\documentclass[twocolumn,trackchanges]{aastex701}

\usepackage{graphicx}
\usepackage{xcolor}
\usepackage{natbib}
\usepackage{amsmath}
\usepackage[subfigure]{tocloft}
\usepackage{subfigure}
\usepackage{booktabs}
\usepackage{CJK}
\usepackage{makecell}
\usepackage{multirow}
\usepackage[utf8]{inputenc}
\usepackage{mhchem}

\newcounter{RomanNumber}

\newcommand{\ue}{\mathrm{e}}

\received{xxx}
\revised{xxx}
\accepted{xxx}

\shorttitle{Multi-messenger view of IMBH-WD TDE}

\begin{document}
\begin{CJK*}{UTF8}{gbsn}

\title{Multi-messenger View of White Dwarf Tidal Disruption Events by Intermediate-Mass Black Holes: I. Gravitational Waves and Disk Photon and Neutrino Emissions}

\author[0000-0002-3525-791X]{Jin-Hong Chen (陈劲鸿)}
\affiliation{Department of Physics, University of Hong Kong, Pokfulam Road, Hong Kong, People's Republic of China}
\affiliation{The Hong Kong Institute for Astronomy and Astrophysics, University of Hong Kong, Hong Kong, China}
\affiliation{Shenzhen Institute of Research and Innovation, The University of Hong Kong, Shenzhen 518057, P. R. China}
\email{chenjh2@hku.hk}
\author[0000-0002-9589-5235]{Lixin Dai (戴丽心)}
\affiliation{Department of Physics, University of Hong Kong, Pokfulam Road, Hong Kong, People's Republic of China}
\affiliation{The Hong Kong Institute for Astronomy and Astrophysics, University of Hong Kong, Hong Kong, China}
\email{lixindai@hku.hk}
\author[0000-0002-9725-2524]{Bing Zhang (张冰)}
\affiliation{Department of Physics, University of Hong Kong, Pokfulam Road, Hong Kong, People's Republic of China}
\affiliation{The Hong Kong Institute for Astronomy and Astrophysics, University of Hong Kong, Hong Kong, China}
\email{bzhang1@hku.hk}

\correspondingauthor{Jin-Hong Chen, Lixin Dai, Bing Zhang}
\email{chenjh2@hku.hk,lixindai@hku.hk,bzhang1@hku.hk}

\begin{abstract}
White dwarf (WD) tidal disruption events (TDEs) provide a unique window onto intermediate-mass black holes (IMBHs). We present a multi-messenger view of these systems in two papers. In this paper, we develop an accretion-disk model for WD--TDEs in which the bound debris accretes at extremely super-Eddington rates, $\sim 10^5$--$10^9$ times higher than in typical (main-sequence) TDEs. The model includes magnetic pressure, nuclear-burning heating, wind mass loss, and neutrino production via $e^{\pm}$ pair annihilation. At such high accretion rates, the gas and radiation temperatures of the inner flow can reach $T\gtrsim 10^9\,\mathrm{K}$, enabling prolific pair production and MeV neutrino emission. We find that the disk is predominantly advection dominated over a broad range of accretion rates, while disk winds can partially cool the flow and reduce the inner temperature. The predicted thermal EM emission is nearly insensitive to the fallback rate in the super-Eddington regime: the luminosity only mildly exceeds the IMBH Eddington luminosity and the spectrum peaks at $\sim 0.1$--$1\,\mathrm{keV}$, implying detectability with current X-ray facilities such as Einstein Probe. For low-mass IMBHs ($\sim 10^3\,M_{\odot}$), the disk can also produce a burst of MeV neutrinos with luminosities up to $\sim 10^{47}\,\mathrm{erg\,s^{-1}}$ for ONeMg WD--TDEs, although detectability with current neutrino detectors (e.g., Super-Kamiokande and JUNO) is limited to Galactic distances. Finally, we estimate the GW burst produced during the final passage prior to disruption, which peaks at $\sim 0.1$--$1\,\mathrm{Hz}$, placing WD--TDEs in the target band of proposed decihertz detectors and motivating coordinated GW+EM+neutrino searches. We also present a first exploration of GWs from a precessing WD--TDE disk; this signal is much weaker, with a detection horizon $\lesssim 1\,\mathrm{Mpc}$ for these missions.
\end{abstract}

\keywords{Neutrinos; Tidal disruption; Black hole physics; Galaxies; Accretion; Gravitational waves}

\section{Introduction}
\label{sec:introduction}
Tidal disruption events (TDEs) provide a powerful probe of intermediate-mass black holes (IMBHs; $M_{\rm h}\sim 10^2$--$10^5\,M_{\odot}$) in the nuclei of stellar clusters and dwarf galaxies. A TDE can produce bright, multi-wavelength emission, revealing an otherwise quiescent IMBH \citep{Ramirez-Ruiz_THE_2009,Chen_Tidal_2018}. As the missing link between stellar-mass black holes and supermassive black holes (SMBHs), IMBHs are central to understanding black-hole growth and black hole--galaxy co-evolution.

A growing sample of IMBH--TDE candidates has been reported, including 3XMM~J215022.4$-$055108 \citep{Lin_A_2018,Chen_Tidal_2018}, AT~2018cqh \citep{wang_tidal_2026}, EP240222a \citep{jin_intermediate-mass_2025}, and the possible candidate EP241021a \citep{shu_ep241021a_2025}. These events may be explained by the disruption of main-sequence stars by IMBHs \citep{Ramirez-Ruiz_THE_2009,Chen_Tidal_2018}.

Because a white dwarf (WD) can be disrupted outside the event horizon only by an IMBH, WD--TDEs provide a particularly clean ``smoking-gun'' signature of IMBHs \citep{Rosswog_Tidal_2009}. Einstein Probe (EP) recently discovered the fast, luminous X-ray transient EP250702a \citep{li_fast_2026} and its associated gamma-ray burst (GRB)~250702BDE \citep{cheng_ep250702a_2025,levan_day-long_2025}. Its day-long X-ray peak, off-nuclear host position, and progressive spectral softening make it a strong candidate for a jetted WD--TDE. Aside from this source, an earlier work reported a possible WD--TDE candidate near M86 \citep{jonker_discovery_2013}.

Motivated by these discoveries, we investigate the central engine of WD--TDEs. Super-Eddington accretion is generically expected in TDEs \citep{ohsuga_supercritical_2005,jiang_global_2014,mckinney_three-dimensional_2014,sadowski_global_2015,dai_unified_2018}, and the resulting flow can drive powerful disk winds through magnetic stresses and radiation pressure. In WD--TDEs, the fallback/accretion rate can far exceed that of main-sequence TDEs, increasing the disk temperature and potentially enabling additional high-energy processes, including nuclear burning, copious $e^{\pm}$ pair creation, and neutrino production.

TDEs are also promising multi-messenger targets. They have long been discussed as potential sources of high-energy cosmic rays and neutrinos: particle acceleration was first emphasized in relativistic jets \citep{wang_probing_2011,wang_tidal_2016,pfeffer_ultrahigh-energy_2017,yuan_revisiting_2025}, and later proposed in the disk, corona, and disk winds \citep{hayasaki_neutrino_2019,murase_high-energy_2020}. Several TDE candidates have been reported in temporal coincidence with high-energy neutrino alerts \citep{stein_tidal_2021,reusch_candidate_2022,yuan_at2021lwx_2024,ji_at2022sxl_2025}; however, refined IceCat-2 reconstructions disfavour spatial associations for some proposed pairs \citep{zegarelli_icecat-2_2025}. Nevertheless, the shared phenomenology of these candidates motivates continued exploration of neutrino--TDE connections \citep{yuan_at2021lwx_2024}.

Gravitational-wave (GW) signals from TDEs have also been discussed as targets for decihertz observatories \citep{pfister_observable_2021,toscani_updated_2025}; see \citet{toscani_updated_2025} for a dedicated study of GW detectability from TDEs. In particular, repeating WD tidal stripping is a promising source class for space-based observatories \citep{Sesana_Observing_2008,Chen_tidal_2023,ye_observing_2024,Wang_A_2022}, such as the Laser Interferometer Space Antenna (LISA) \citep{Amaro_LISA_2017} and TianQin \citep{Luo_TianQin_2016}.

In this paper, we model the central engine of WD--TDEs by computing the disk structure and associated high-energy physics, and we present the first predictions for the resulting multi-messenger emission (EM, neutrinos, and GWs). In a companion paper (Chen et al., in preparation), we will investigate the multi-messenger signatures of jets, coronae, and disk winds, with particular emphasis on high-energy neutrinos and cosmic rays from WD--TDEs.

This paper is organized as follows. In Section~\ref{sec:TDE} we summarize the basic scalings of WD--IMBH disruptions and the expected fallback/accretion rates. In Section~\ref{sec:accretion_disk} we construct a steady, vertically integrated disk model including wind mass loss, photon diffusion/advection, nuclear burning, and neutrino cooling, and derive the resulting disk structure. In Section~\ref{sec:disk_radiation} we compute the associated EM and neutrino signals and discuss their parameter dependence and detectability. The GW signal from the final passage before disruption and from a precessing disk is explored in Section~\ref{sec:GW}. In Section~\ref{sec:Multi_messenger_evolution} we summarize multi-messenger observational characteristics relevant for time-domain searches. We discuss disk formation, evolution, and wind emission in Section~\ref{sec:discussion}, and conclude in Section~\ref{sec:conclusion}.

\section{Basic theory of WD--IMBH TDE}
\label{sec:TDE}
When a WD encounters an IMBH, a TDE occurs if the pericenter distance $R_{\rm p}$ is comparable to (or smaller than) the tidal radius \citep{Rees_Tidal_1988,Phinney_Manifestations_1989}
\begin{equation} \label{eq:Rt}
    R_{\rm t} = R_*\left(\frac{M_{\rm h}}{M_*}\right)^{1/3} \simeq 7 \times 10^9\ r_{*,-2}\, M_{3}^{1/3}\, m_*^{-1/3}\ {\rm cm},
\end{equation}
where $M_{\rm h}$ is the IMBH mass and $M_*$ and $R_*$ are the WD mass and radius. We define $M_{3} \equiv M_{\rm h}/(10^3\,M_{\odot})$, $r_{*,-2} \equiv R_*/(10^{-2}\,R_{\odot})$, and $m_* \equiv M_*/M_{\odot}$.

In WDs, the pressure is dominated by a degenerate electron gas. The zero-temperature mass--radius relation can be written as \citep{nauenberg_analytic_1972,Paczynski_Models_1983,han_determining_2018}
\begin{equation} \label{eq:R_M}
    R_* = 9 \times 10^8 \left[1- \left(\frac{M_*}{M_{\rm ch}}\right)^{4/3}\right]^{1/2} \left(\frac{M_*}{M_{\odot}}\right)^{-1/3}\ {\rm cm},
\end{equation}
where $M_{\rm ch} \simeq 1.44\,M_{\odot}$ is the Chandrasekhar mass. By composition and mass, WDs are commonly grouped into He WDs ($\lesssim 0.5\,M_{\odot}$), CO WDs ($\sim 0.5$--$1\,M_{\odot}$), and ONeMg WDs ($\sim 1$--$1.4\,M_{\odot}$).

More massive WDs are denser and thus have smaller tidal radii. A WD TDE can only be observed if disruption occurs outside the gravitational radius $R_{\rm g} = GM_{\rm h}/c^2$. Defining the impact parameter $\beta \equiv R_{\rm t}/R_{\rm p}$, the pericenter distance in units of $R_{\rm g}$ is
\begin{equation} \label{eq:Rp_Rg}
    R_p \simeq 47\, \beta^{-1}\, M_{3}^{-2/3}\, m_*^{-1/3}\, r_{*,-2}\, R_{\rm g}.
\end{equation}

Equation (\ref{eq:Rp_Rg}) highlights that WD disruptions preferentially occur around lower-mass black holes ($\lesssim 10^6\,M_{\odot}$), i.e., IMBHs. WD TDEs therefore provide a promising channel for IMBH searches.

After disruption, the debris is stretched into a stream with a characteristic spread in specific orbital energy\footnote{This expression applies to full disruptions. For partial TDEs, the energy spread depends on $\beta$, and $R_{\rm t}$ can be replaced by $R_{\rm p}$, as found in simulations \cite{Chen_tidal_2023}.} $\Delta \epsilon \simeq GM_{\rm h} R_*/R_{\rm t}^2$ \citep{Lodato_Stellar_2009,Guillochon_Hydrodynamical_2013,Chen_tidal_2023}. If the WD approaches the IMBH with specific orbital energy $\epsilon_0$, the most-bound debris has
\begin{equation} \label{eq:E_mb}
    \epsilon_{\rm mb} \simeq \epsilon_0 - \Delta \epsilon.
\end{equation}
We focus on the parabolic case ($\epsilon_0 = 0$).

For $\epsilon_0 = 0$, Equation (\ref{eq:E_mb}) gives
\begin{equation} \label{eq:E_mb_para}
    \epsilon_{\rm mb} = -\frac{GM_{\rm h}}{2a_{\rm mb}} \simeq -\frac{GM_{\rm h} R_*}{R_{\rm t}^2},
\end{equation}
where $a_{\rm mb}$ is the semimajor axis of the most-bound debris. Using Kepler's third law, the fallback timescale is $t_{\rm fb} \simeq 2\pi\sqrt{a_{\rm mb}^3/(GM_{\rm h})}$. Substituting Equation (\ref{eq:E_mb_para}) yields
\begin{equation} \label{eq:t_fb}
    t_{\rm fb} \simeq 112\, M_3^{1/2}\, r_{*,-2}^{3/2}\, m_*^{-1}\ {\rm s}.
\end{equation}

The debris subsequently returns at a rate $\dot{M}_{\rm fb}$. For full disruptions, the late-time fallback rate follows $\dot{M}_{\rm fb} \propto t^{-5/3}$ \citep{Rees_Tidal_1988}, while partial disruptions can exhibit a steeper decline $\propto t^{-9/4}$ \citep{Coughlin_Partial_2019,Miles_Fallback_2020}.

The peak fallback rate for WD TDEs is
\begin{equation} \label{eq:dotM_peak}
    \dot{M}_{\rm peak} \sim \frac{M_*}{3t_{\rm fb}} \simeq 5 \times 10^9\, M_3^{-3/2}\, r_{*,-2}^{-3/2}\, m_*^{2}\, \dot{M}_{\rm Edd},
\end{equation}
where $\dot{M}_{\rm Edd} = L_{\rm Edd}/(\eta c^2)$ is the Eddington accretion rate. Here $L_{\rm Edd}$ is the Eddington luminosity and we adopt a nominal accretion efficiency $\eta \simeq 0.12$ \citep{novikov_astrophysics_1973}.

The duration of the super-Eddington fallback phase can be estimated by solving $\dot{M}_{\rm fb}(t_{\rm Edd}) = \dot{M}_{\rm Edd}$, giving
\begin{equation} \label{eq:t_Edd}
    t_{\rm Edd} \simeq 6\, M_3^{-2/5}\, r_{*,-2}^{3/5}\, m_*^{1/5}\ {\rm yr}.
\end{equation}
Thus, super-Eddington accretion can persist for several years in WD TDEs.

The extreme fallback rates in WD TDEs (e.g., $\dot{M}_{\rm peak} \sim 10^{9}\,\dot{M}_{\rm Edd}$) are far higher than those expected for main-sequence TDEs (typically $\sim 10$--$100\,\dot{M}_{\rm Edd}$ for SMBHs, and $\sim 10^4$--$10^5\,\dot{M}_{\rm Edd}$ for IMBHs; \citealt{Chen_Tidal_2018}).

\begin{figure}
\centering
\includegraphics[scale=0.5]{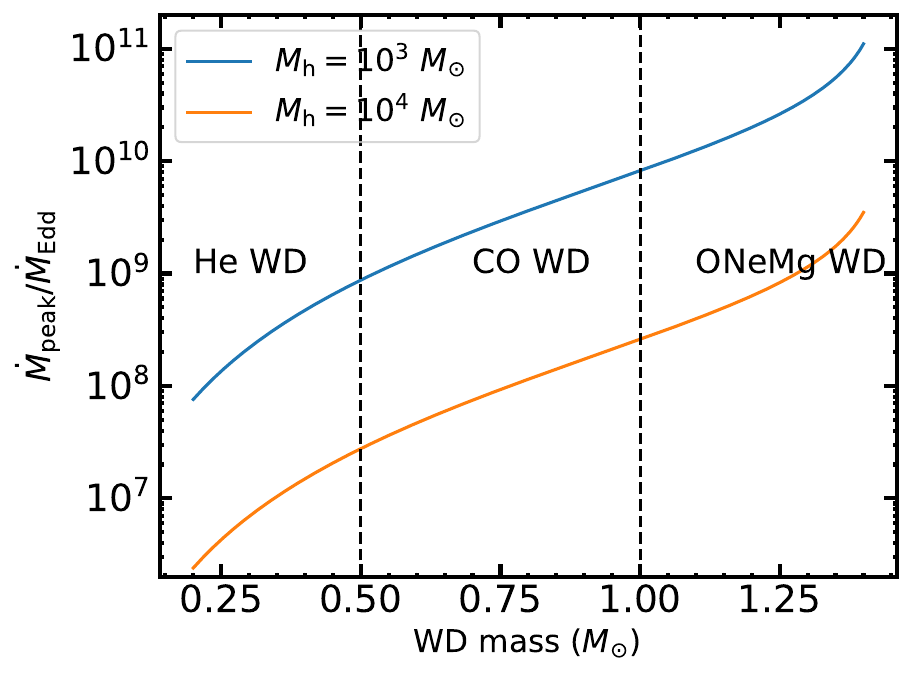}
\caption{Eddington ratio versus WD mass for WD TDEs, derived from Equations (\ref{eq:R_M}) and (\ref{eq:dotM_peak}).}
\label{fig:macc_wd}
\end{figure}

This difference suggests distinct accretion dynamics in WD TDEs, potentially involving non-standard disk physics. Substituting Equation (\ref{eq:R_M}) into Equation (\ref{eq:dotM_peak}) yields the Eddington ratio as a function of WD mass (Figure~\ref{fig:macc_wd}).

If the disk can efficiently process the returning debris, the disk accretion rate should closely track the fallback rate. It is therefore common to assume $\dot{M}_{\rm acc} \simeq \dot{M}_{\rm fb}$, although the mapping between these two rates depends on how rapidly the debris circularizes and forms a disk (see Section~\ref{sec:discussion}).

Disk winds can reduce the mass that ultimately reaches the IMBH. In the absence of outflows, one would expect $\dot{M}_{\rm BH} \approx \dot{M}_{\rm fb}$; however, if a fraction of the inflowing material is lost to winds (Section~\ref{subsec:disk_wind}), then $\dot{M}_{\rm BH} < \dot{M}_{\rm fb}$.

Hereafter, we first assume $\dot{M}_{\rm acc} \simeq \dot{M}_{\rm fb}$ and calculate the disk properties and the corresponding observational features. We will further discuss this aspect in Section \ref{subsec:disk_evolution} to explore the time evolution of disk, and we will see that this assumption would be reasonable for WD-TDE disk with disk wind mediate.

\section{Accretion disk in WD-IMBH TDE}
\label{sec:accretion_disk}
Accretion-disk formation in TDEs is commonly attributed to debris self-intersections (stream--stream and stream--disk collisions) followed by viscous angular-momentum redistribution \citep{bonnerot_long-term_2017,Chen_pTDE_2021}.

The properties and time evolution of disks in ``typical'' (main-sequence) TDEs have been studied extensively \citep{Shen_EVOLUTION_2014}. When the accretion rate is moderately super-Eddington, $\dot{M}_{\rm acc} \sim 10$--$10^2\,\dot{M}_{\rm Edd}$, photon trapping leads to an advection-dominated (``slim-disk'') regime in which a large fraction of the dissipated energy is carried inward with the flow rather than radiated locally. As a result, the emergent luminosity tends to saturate near the Eddington limit.

At the opposite extreme, hyper-accreting disks have been widely discussed in the context of gamma-ray bursts (GRBs). In standard GRB central-engine models, the accretion rate can reach $\sim 10^{11}$--$10^{15}\,\dot{M}_{\rm Edd}$ (equivalently $\sim 10^{-3}$--$10\,M_{\odot}\,\rm s^{-1}$ for a stellar-mass BH) \citep{popham_hyperaccreting_1999,kawanaka_neutrino-cooled_2007}. At these rates, the disk can reach $T\gtrsim 10^{10}\,\rm K$, enabling nuclear photodisintegration and efficient neutrino cooling via $e^{\pm}$ pair annihilation and electron capture on nuclei \citep{di_matteo_neutrino_2002}. Such systems are often referred to as neutrino-dominated accretion flows (NDAFs).

WD--TDEs occupy an intermediate regime in which the accretion rate is far above that of standard TDE disks yet typically below GRB hyper-accretion, making pair production and neutrino emission potentially important without necessarily controlling the overall energetics. Motivated by this, we investigate WD--TDE disks using a steady, vertically integrated model.

Because WD--TDE disks can reach $T\gtrsim 10^8\,\rm K$, thermonuclear burning in the inner flow may also occur, processing the debris into heavier elements. Related ``nuclear-dominated'' disks have been explored in compact-object merger contexts \citep{metzger_nuclear-dominated_2012,margalit_time-dependent_2016,dan_structure_2014}.

\begin{figure*}
\centering
 \includegraphics[scale=1]{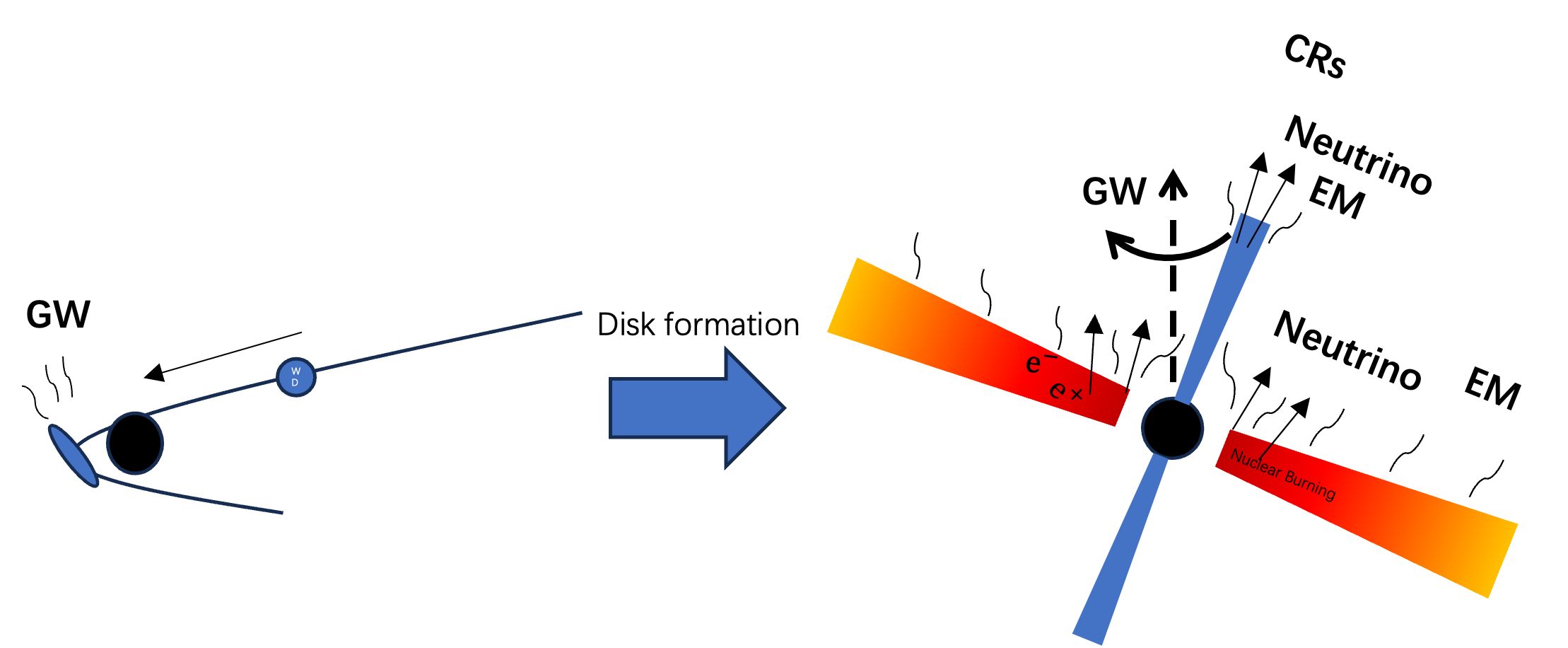}
    \caption{Schematic overview of a WD--TDE and its multi-messenger channels. The pericenter passage can generate a GW burst, while subsequent disruption and circularization form a thick accretion disk fed by rapid fallback. At sufficiently high temperatures, $e^{\pm}$ pairs in the inner disk produce MeV neutrinos via pair annihilation, and nuclear burning ($T\gtrsim 10^8\,\rm K$) provides additional heating. If the disk angular-momentum axis is misaligned with the IMBH spin, Lense--Thirring torques can drive (approximately) rigid-body precession, imprinting variability on the electromagnetic emission and potentially the GW signal. The disk may also power a relativistic jet that accelerates non-thermal particles, producing high-energy cosmic rays and neutrinos through $p\gamma$ and $pp$ interactions.}
\label{fig:sketch}
\end{figure*}

In the following sections we incorporate both nuclear burning and pair/neutrino physics and refer to these flows as pair--nuclear accretion disks (PNADs). Figure~\ref{fig:sketch} provides a schematic overview.

For deep encounters with $\beta\gg 1$, tidal compression near pericenter can ignite the WD and produce a supernova-like transient \citep{rosswog_high-resolution_2003,Rosswog_Atypical_2008,Rosswog_Tidal_2009,macleod_optical_2016,tanikawa_does_2017,Anninos_Relativistic_2018}. Here we focus on ``typical'' WD--TDEs with $\beta\sim 1$, for which nuclear reactions are not triggered during the disruption itself; instead, any burning occurs later in the hot, newly formed disk. Exploring deep-encounter (prompt-ignition) cases and their impact on the subsequent disk is an important topic for future work.

In the Appendix \ref{appendix:ingredients} we present the PNAD model in detail, including prescriptions for viscosity, pressure support, neutrino emission, nuclear burning, disk winds, and radiative cooling. The resulting disk structure is summarized in Section below.

\subsection{Disk Structure} \label{subsec:Disk_structure}
We compute the steady-state structure of the WD--TDE disk over a broad range of radii and accretion rates by explicitly accounting for the relevant heating and cooling channels.

The total heating is the sum of viscous dissipation $Q_{\rm vis}$ (Equation~(\ref{eq:Q_vis})) and nuclear burning $Q_{\rm nuc}$ (Equation~(\ref{eq:Q_nuc})),
\begin{equation}
    Q_+ = Q_{\rm vis} + Q_{\rm nuc}.
\end{equation}
Cooling includes radiative diffusion $Q_{\rm rad}$ (Equation~(\ref{eq:Q_rad})), advection $Q_{\rm adv}$, wind losses $Q_{\rm w}$ (Equation~(\ref{eq:Q_w})), and neutrino emission $Q_{\nu}$ (Equation~(\ref{eq:Q_v})), i.e.,
\begin{equation}
    Q_- = Q_{\rm rad} + Q_{\rm adv} + Q_{\rm w} + Q_{\nu}
\end{equation}
where the standard form for the advective term is
\begin{equation} \label{eq:Q_adv}
    Q_{\rm adv} = \xi \frac{\dot{M}_{\rm in}}{2\pi R^2} \frac{P}{\rho}.
\end{equation}
Here $\xi \sim 1$ \citep{Frank_Accretion_1985}. In steady state the energy equation reads
\begin{equation} \label{eq:balance}
    Q_+ = Q_-.
\end{equation}
Combining Equation~(\ref{eq:balance}) with the total pressure (Equation~(\ref{eq:P_tot})) allows us to solve numerically for the disk structure at fixed $\dot{M}_{\rm acc}$.

\begin{figure}
\centering
 \begin{minipage}[t]{0.5\textwidth}
 \centering
 \includegraphics[scale=0.5]{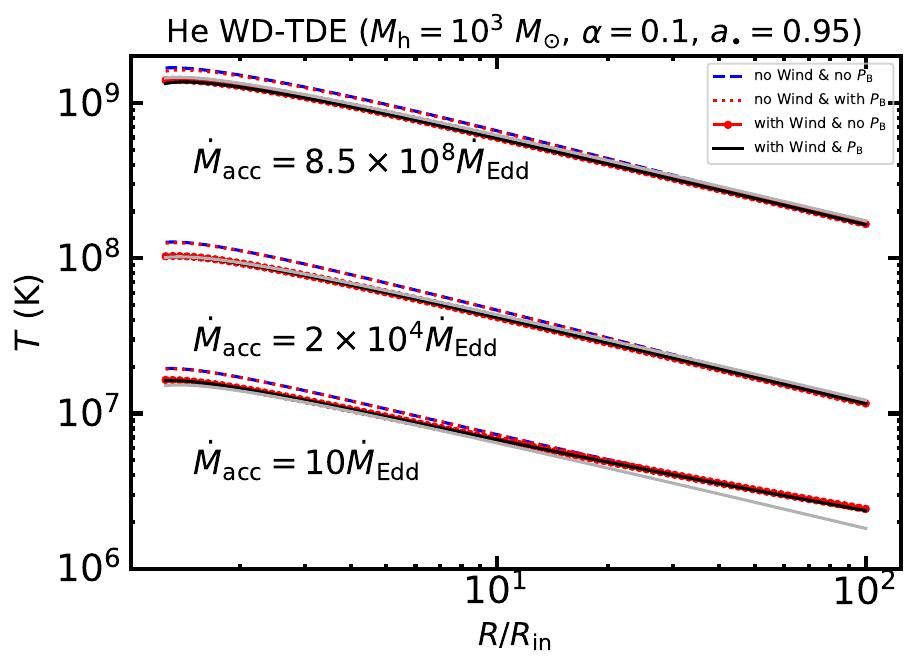}
\end{minipage}
 \begin{minipage}[t]{0.5\textwidth}
 \centering
 \includegraphics[scale=0.5]{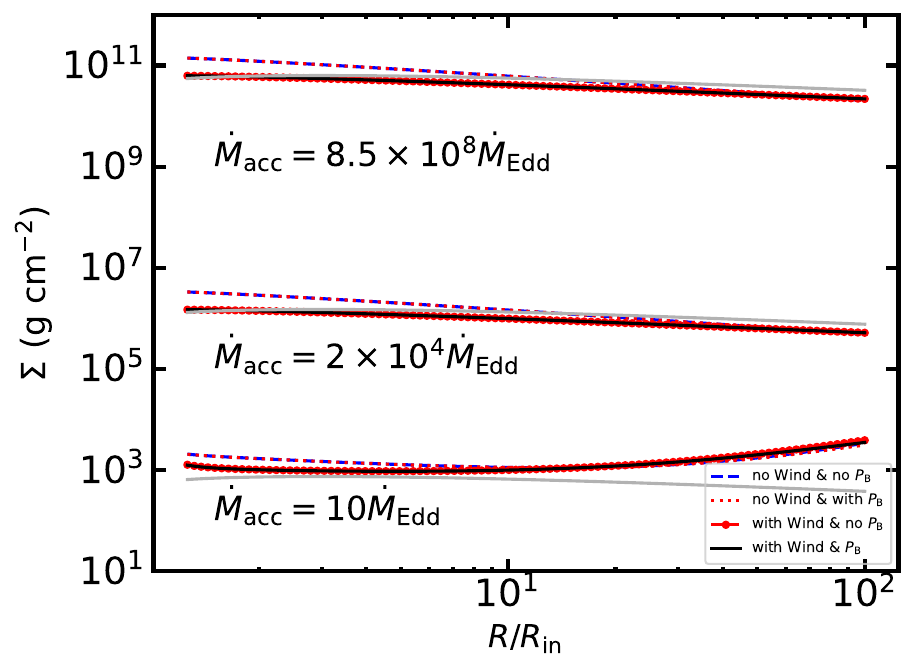}
\end{minipage}
 \begin{minipage}[t]{0.5\textwidth}
 \centering
 \includegraphics[scale=0.5]{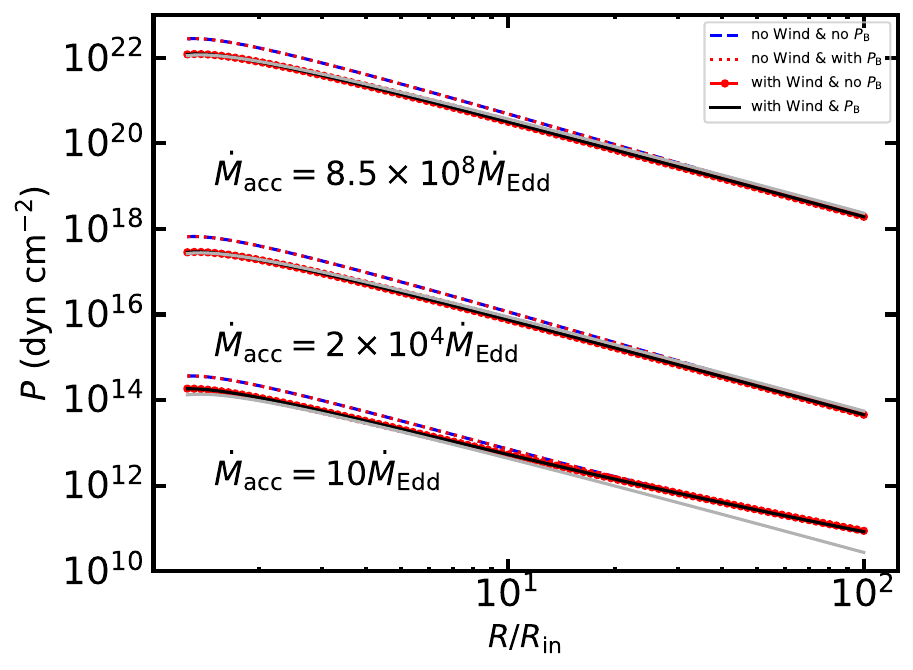}
\end{minipage}
    \caption{He WD--TDE disk profiles for temperature (upper), surface density (middle), and pressure (lower) at three representative fallback rates. We vary magnetic pressure and disk winds independently to assess their impact on the structure at different accretion rates. Gray curves show the analytic scalings from Equations~(\ref{eq:T_adv_w}--\ref{eq:Sigma_adv_w}).}
\label{fig:HeWD_disk_profile}
\end{figure}

To isolate the roles of magnetic pressure, winds, and nuclear burning, we compute models in which these effects are included separately. Figure~\ref{fig:HeWD_disk_profile} shows the He WD--TDE disk profiles at three representative fallback rates, $10\,\dot{M}_{\rm Edd}$, $2\times 10^4\,\dot{M}_{\rm Edd}$, and the peak rate $8.5\times 10^8\,\dot{M}_{\rm Edd}$ (for a $0.5\,M_{\odot}$ He WD).

Magnetic pressure contributes little to the overall support in these models (i.e., $P_{\rm B}$ remains sub-dominant). By contrast, wind mass loss modestly depresses the temperature, density, and pressure in the inner disk by reducing the inflowing mass and associated heating.

Advection dominates the cooling over most radii, particularly in the inner disk. In the radiation-pressure regime ($P\simeq P_{\rm rad}$) one can recover the standard advection-dominated scalings (with and without wind modulation), summarized in Equations~(\ref{eq:T_adv}--\ref{eq:Sigma_adv_w}) and derived in Appendix~\ref{appendix:adv_disk}. The numerical solutions closely follow these analytic scalings; the main deviation occurs for $\dot{M}_{\rm acc}=10\,\dot{M}_{\rm Edd}$ in the outer disk, where radiative cooling becomes efficient.

\begin{figure}
\centering
 \includegraphics[scale=0.35]{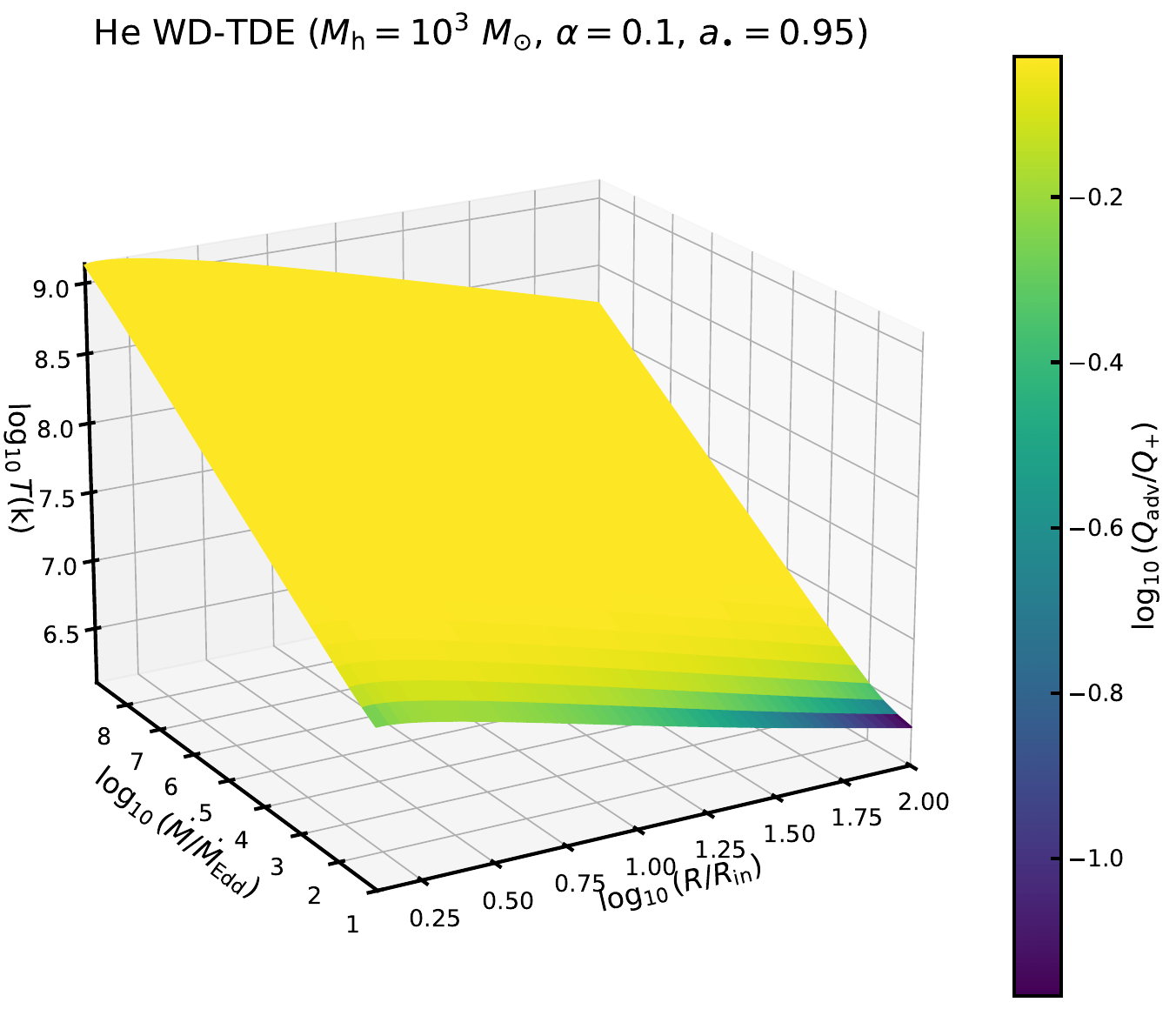}
    \caption{Temperature profile of the He WD--TDE disk over a range of accretion rates. Colors indicate the ratio $Q_{\rm adv}/Q_+$, highlighting the dominance of advective cooling. Radiative cooling becomes competitive only in the outer disk at low accretion rates ($\dot{M}_{\rm acc}\lesssim 10\,\dot{M}_{\rm Edd}$).}
\label{fig:He_adv}
\end{figure}

Figure~\ref{fig:He_adv} further illustrates this behavior by showing the temperature profile over a range of accretion rates: advection remains dominant except in the outer regions at low accretion rates ($\dot{M}_{\rm acc} \lesssim 10\,\dot{M}_{\rm Edd}$).

\begin{figure}
\centering
 \begin{minipage}[t]{0.5\textwidth}
 \centering
 \includegraphics[scale=0.5]{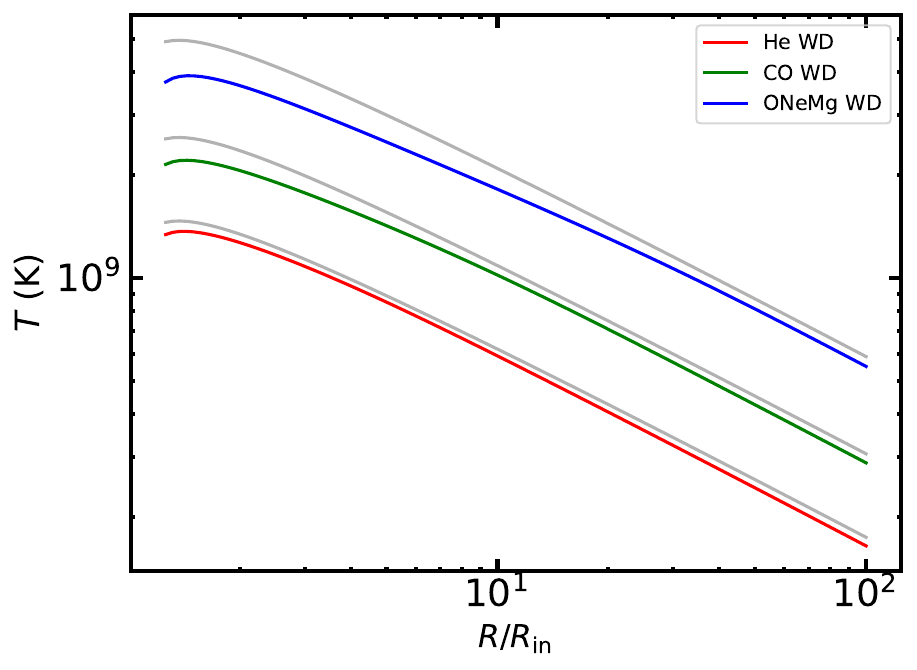}
\end{minipage}
 \begin{minipage}[t]{0.5\textwidth}
 \centering
 \includegraphics[scale=0.5]{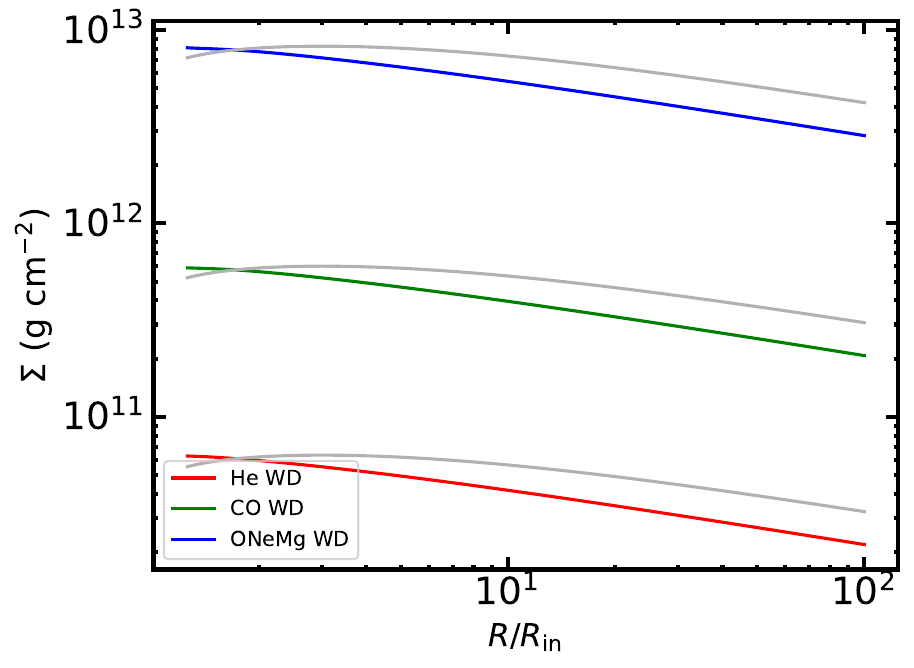}
\end{minipage}
 \begin{minipage}[t]{0.5\textwidth}
 \centering
 \includegraphics[scale=0.5]{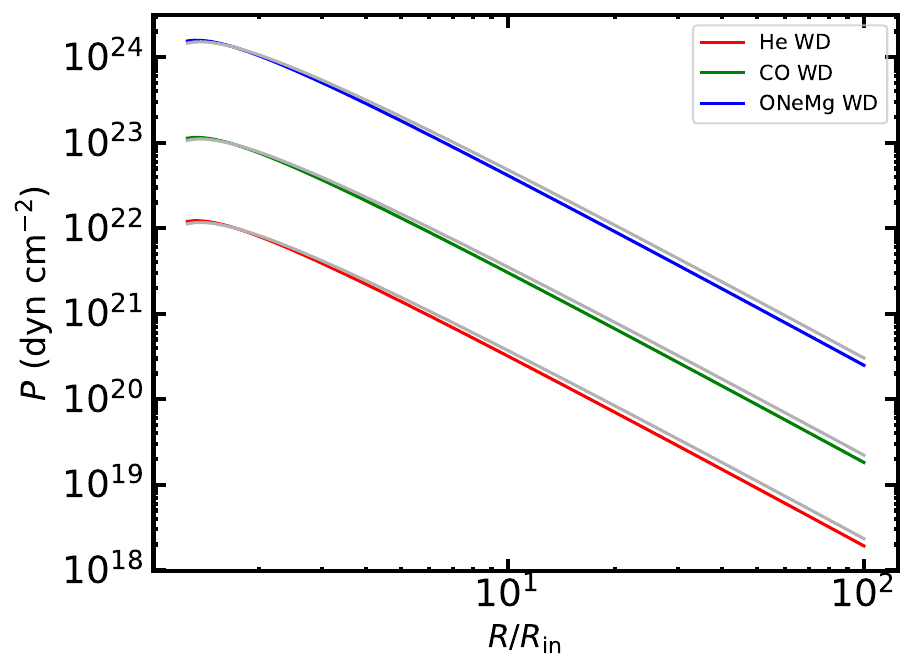}
\end{minipage}
    \caption{Disk profiles for He, CO, and ONeMg WD--TDEs: temperature (upper), surface density (middle), and pressure (lower), evaluated at representative peak fallback rates: $\simeq 8.5\times 10^8\,\dot{M}_{\rm Edd}$ (He), $\simeq 8.1\times 10^9\,\dot{M}_{\rm Edd}$ (CO), and $\simeq 10^{11}\,\dot{M}_{\rm Edd}$ (ONeMg), with a$10^3\ M_{\odot}$ IMBH. Gray curves show the analytic scalings from Equations~(\ref{eq:T_adv_w}--\ref{eq:Sigma_adv_w}).}
\label{fig:WD_disk_profile}
\end{figure}

For a direct comparison among WD compositions, Figure~\ref{fig:WD_disk_profile} shows the temperature, surface density, and total pressure profiles evaluated at the corresponding peak fallback rates. The CO and ONeMg disks are systematically cooler than the purely analytic, radiation-dominated expectation at small radii because gas pressure and magnetic pressure become non-negligible in the inner disk at high $\dot{M}_{\rm acc}$. Neutrino cooling can also modify the structure, although its effect is small for the parameters explored here.

Including both nuclear heating and neutrino cooling, we find that $Q_{\rm nuc}$ is at least $\sim 10$ orders of magnitude smaller than $Q_{\rm vis}$ across the parameter space considered.

\begin{figure}
\centering
 \includegraphics[scale=0.35]{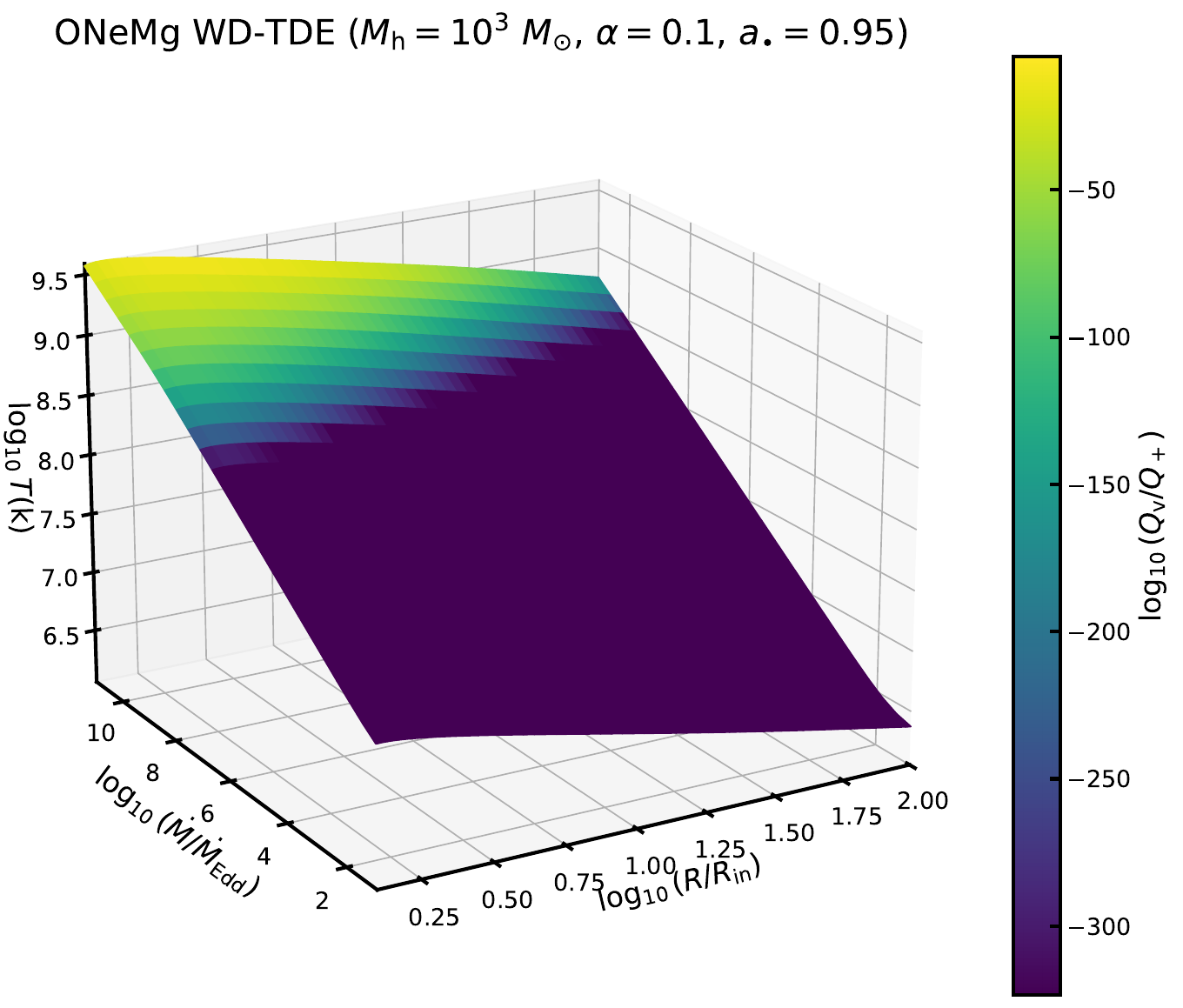}
    \caption{Temperature profile of the ONeMg WD--TDE disk over a range of accretion rates. Colors indicate the ratio $Q_{\nu}/Q_+$, demonstrating that neutrino cooling is negligible.}
\label{fig:ONeMg_neutrino}
\end{figure}

Figure~\ref{fig:ONeMg_neutrino} shows the relative importance of neutrino losses for the ONeMg WD--TDE. Even at the highest accretion rates, neutrino emission remains negligible in the disk energy budget (in contrast to the neutrino-dominated accretion flows invoked for GRB engines). Nevertheless, neutrino production may still be relevant for multi-messenger searches for WD--TDEs.

\section{Parameter Dependence of Disk Emissions}
\label{sec:disk_radiation}
In this section, we calculate the electromagnetic (EM) and neutrino emissions from the disk and examine how these outputs depend on the disk properties.

\subsection{EM Radiation}
\label{subsec:EM}
\begin{figure}
\centering
 \includegraphics[scale=0.5]{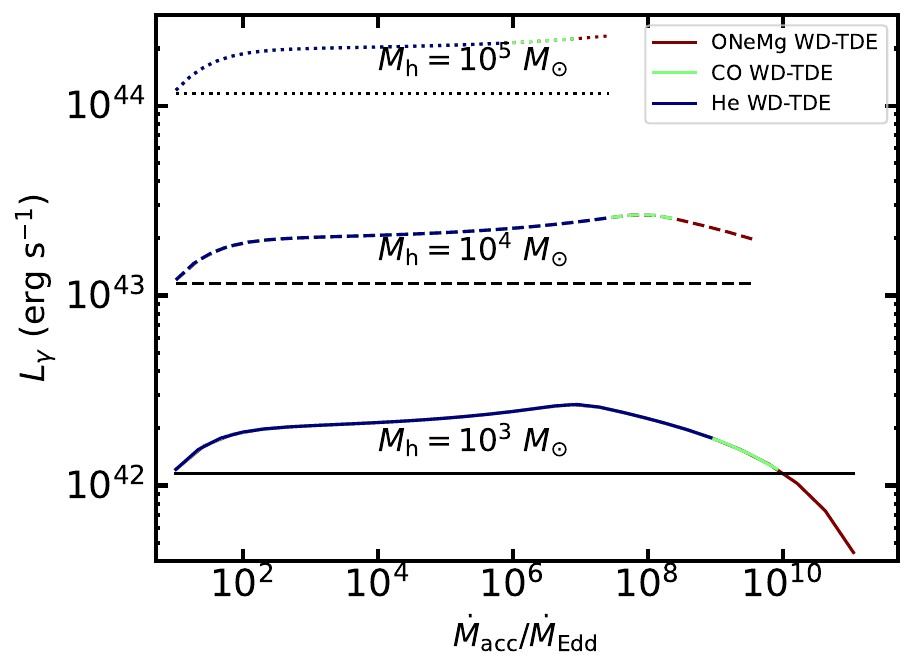}
    \caption{EM luminosity of WD--TDE disks as a function of mass fallback rate, from $10\,\dot{M}_{\rm Edd}$ up to the peak fallback rate for each WD composition. The horizontal black line marks the analytic luminosity of an advection- and radiation-pressure-dominated disk, $2\ln(R_{\rm out}/R_{\rm in})\,L_{\rm Edd}\simeq 9\,L_{\rm Edd}$ (Equation~(\ref{eq:L_adv})). The luminosity varies weakly with $\dot{M}_{\rm acc}$ and peaks near $\dot{M}_{\rm acc}\simeq 10^8\ \dot{M}_{\rm Edd}$, where $e^{\pm}$ pairs contribute to the pressure and neutrino cooling becomes non-negligible.}
\label{fig:dotE_rad}
\end{figure}

We first compute the EM luminosity radiated by the disk,
\begin{equation} \label{eq:dotE_rad}
    L_{\gamma} = 4\pi \int^{R_{\rm out}}_{R_{\rm in}} Q_{\rm rad} R\, dR.
\end{equation}
Figure~\ref{fig:dotE_rad} shows the results for $R_{\rm out}=100\,R_{\rm in}$\footnote{Our fiducial choice of $R_{\rm out}$ is close to $R_{\rm c}=2R_{\rm p}$ for typical parameters; however, as discussed in Section \ref{subsec:disk_evolution}, viscous spreading can increase the disk size beyond $R_{\rm c}$.}. The luminosity depends only weakly on the fallback rate as it increases from $10\,\dot{M}_{\rm Edd}$ up to the peak rate of each WD--TDE. The luminosity reaches a maximum near $\dot{M}_{\rm acc}\simeq 10^8\,\dot{M}_{\rm Edd}$, where $e^{\pm}$ pairs contribute appreciably to the pressure and neutrino cooling becomes non-negligible; for $\dot{M}_{\rm acc}\gtrsim 10^8\,\dot{M}_{\rm Edd}$, the EM luminosity decreases at the highest accretion rates.

The peak luminosity of a WD-composition disk is typically a factor of $\sim 2$ higher than that of an otherwise similar H-rich disk, primarily because the electron-scattering opacity is lower for WD materials (see Appendix~\ref{appendix:adv_disk}).

Because the disk is optically thick, each surface element radiates approximately as a blackbody with effective temperature $T_{\rm eff}(R)$ satisfying $\sigma T_{\rm eff}^4 = Q_{\rm rad}$. This implies
\begin{equation} \label{eq:Teff4}
    T_{\rm eff}^4 \simeq \frac{16}{3\kappa_{\rm R} \Sigma}T^4.
\end{equation}
For an observer at luminosity distance $D_{\rm L}$, the face-on specific flux can be estimated using \citep{Frank_Accretion_1985}
\begin{equation} \label{eq:Fv}
    F_{\nu} = \frac{4\pi h \nu^3}{c^2 D_{\rm L}^2}\int^{R_{\rm out}}_{R_{\rm in}} \frac{R\, dR}{\ue^{h\nu/k_{\rm b}T_{\rm eff}(R)}-1}.
\end{equation}
This is a simplified treatment; the geometry of a thick disk and general-relativistic effects can modify the observed spectrum, but we defer such modeling to future work.

At high inclination, emission from the inner disk may be obscured by the outer disk and/or reprocessed by disk winds \citep{dai_unified_2018,qiao_early_2025}. Emission along the polar direction ($i\simeq 0$) is least affected by reprocessing; therefore, we focus on the face-on case ($i=0$).

\begin{figure}
 \begin{minipage}[t]{0.5\textwidth}
 \centering
 \includegraphics[scale=0.5]{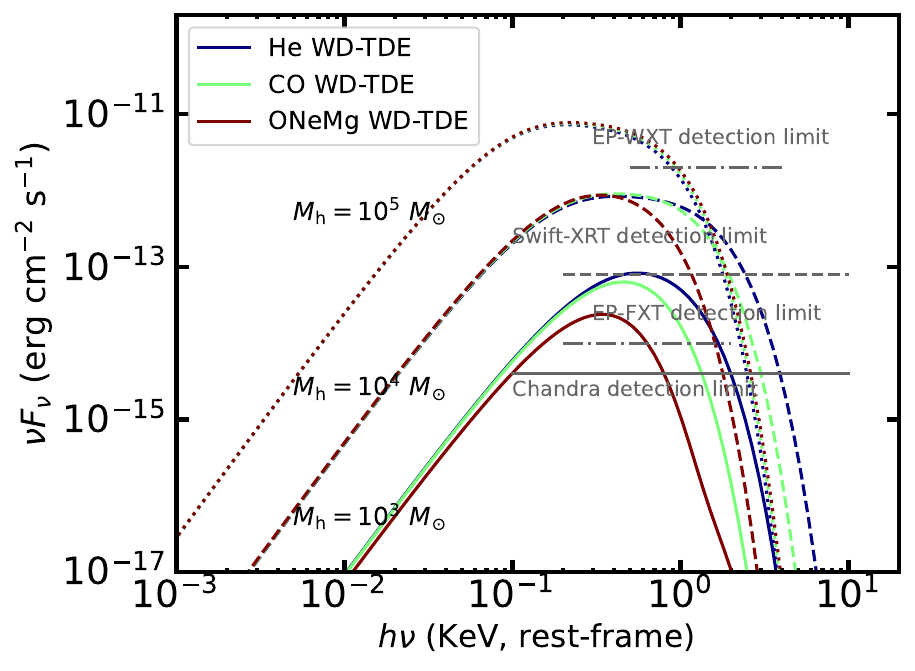}
\end{minipage}
 \begin{minipage}[t]{0.5\textwidth}
 \centering
 \includegraphics[scale=0.5]{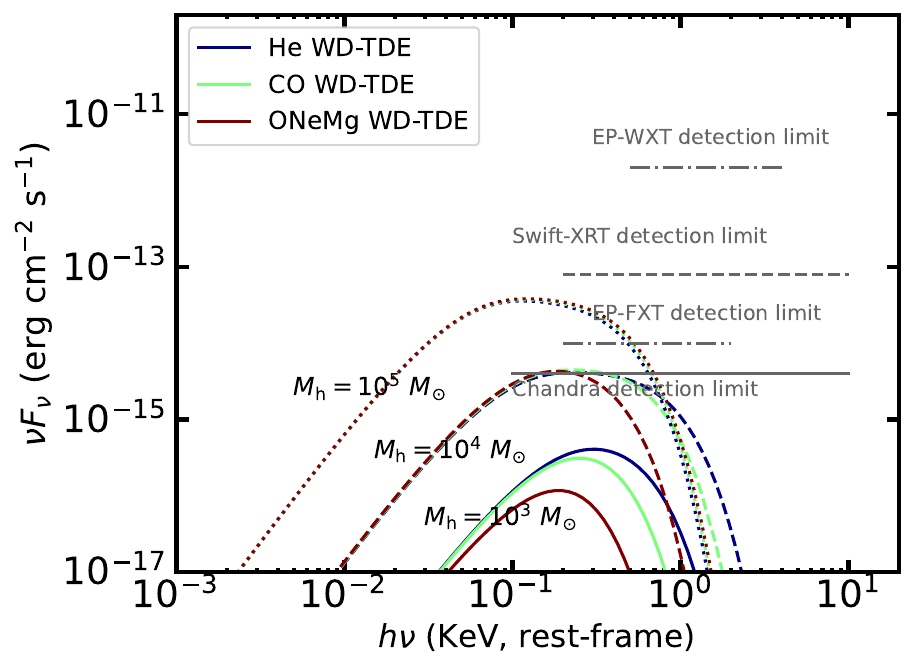}
\end{minipage}
    \caption{Observed EM spectra at the peak fallback rate for each WD--TDE disk, assuming redshift $z=0.1$ (upper) and $z=1$ (lower). Solid, dashed, and dot-dashed curves correspond to IMBH masses $10^3$, $10^4$, and $10^5\,M_{\odot}$, respectively. Horizontal gray lines indicate representative flux limits of soft X-ray instruments (e.g., EP-WXT/FXT, Swift-XRT, Chandra).}
\label{fig:diskEM_spec}
\end{figure}

The numerically computed spectra for each WD--TDE are shown in Figure~\ref{fig:diskEM_spec}. The spectra typically peak at $\sim 0.1$--$1\,\mathrm{keV}$. For CO and ONeMg disruptions onto a $10^3\,M_{\odot}$ IMBH, the spectra shift to slightly lower energies and fluxes because neutrino cooling becomes more important at the corresponding higher accretion rates.


\begin{figure*}
 \begin{minipage}[t]{0.5\textwidth}
 \centering
 \includegraphics[scale=0.5]{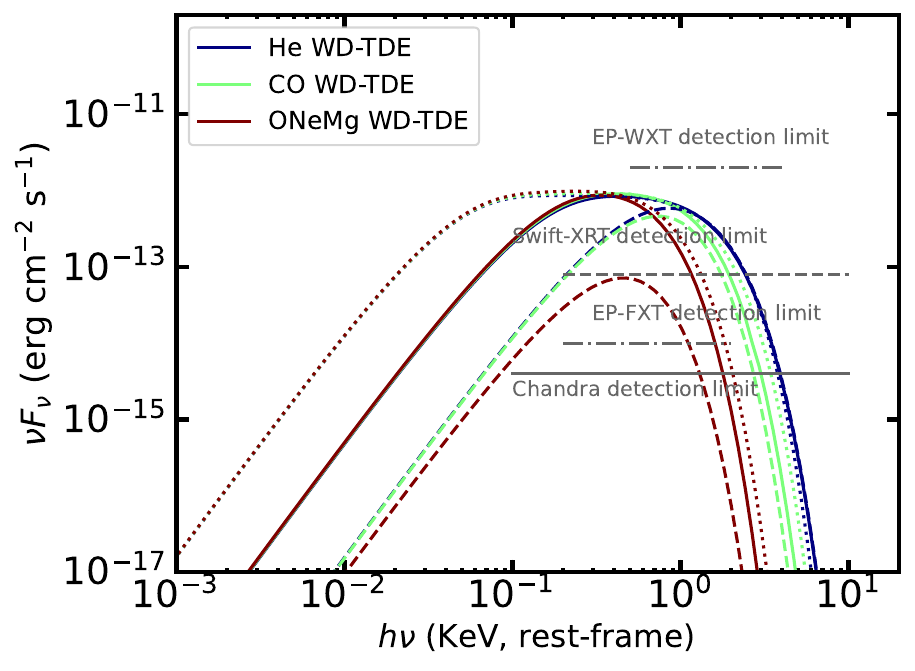}
\end{minipage}
 \begin{minipage}[t]{0.5\textwidth}
 \centering
 \includegraphics[scale=0.5]{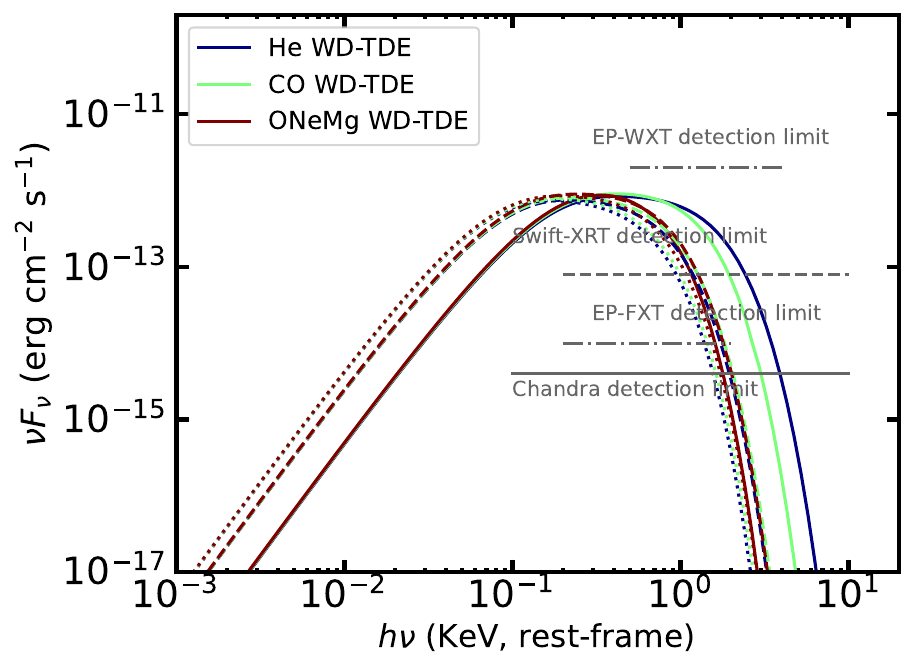}
\end{minipage}
 \begin{minipage}[t]{0.5\textwidth}
 \centering
 \includegraphics[scale=0.5]{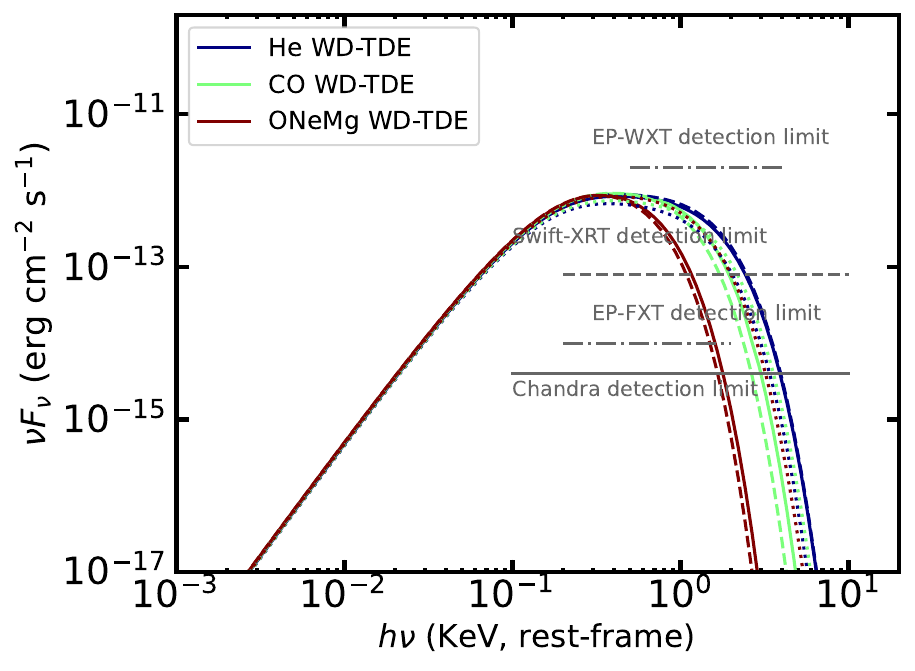}
\end{minipage}
 \begin{minipage}[t]{0.5\textwidth}
 \centering
 \includegraphics[scale=0.5]{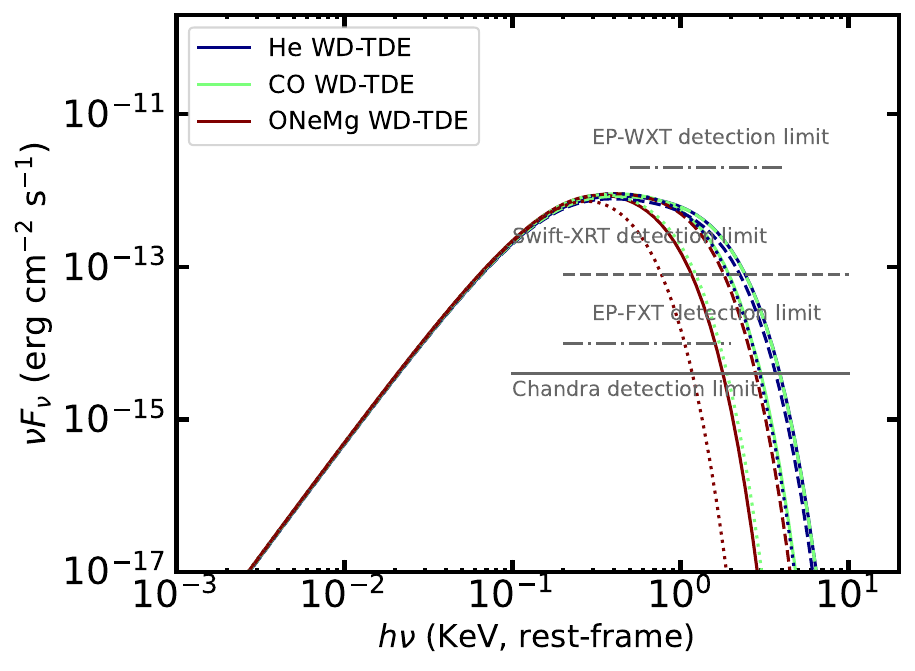}
\end{minipage}
    \caption{Dependence of the observed EM spectra (at the peak fallback rate) on different parameters, assuming $z=0.1$ and $M_{\rm h}=10^4\,M_{\odot}$. Upper left (disk size $R_{\rm out}$): solid, dashed, and dotted curves show $R_{\rm out}=100\,R_{\rm in}$, $R_{\rm out}=10\,R_{\rm in}$, and $1000\,R_{\rm in}$, respectively. Upper right (IMBH spin $a_{\bullet}$): solid, dashed, and dotted curves show $a_{\bullet}=0.95$, $a_{\bullet}=0$, and $a_{\bullet}=-0.95$, respectively. Lower left (wind $s$-index): solid, dashed, and dotted curves show $s=0.2$, $s=0$, and $s=1$, respectively. Lower right (viscosity $\alpha$): solid, dashed, and dotted curves show $\alpha=0.1$, $\alpha = 1$, and $\alpha = 0.01$, respectively.}
\label{fig:diskEM_spec_R_a}
\end{figure*}

This thermal disk component can fall within the sensitivity of current X-ray facilities, both in all-sky surveys and in targeted follow-up. For example, the lobster-eye telescope WXT on board the Einstein Probe (EP) may discover nearby events ($z\lesssim 0.1$) and potentially capture the early emission shortly after disk formation. Continued monitoring with EP-FXT, Swift-XRT, Chandra, and XMM-Newton can probe the thermal component at later times and, for sufficiently luminous events, out to $z\sim 1$ \citep{Zhang_EPsensitivity_2022}.

The parameter dependence of the disk spectrum (outer radius $R_{\rm out}$, IMBH spin $a_{\bullet}$, wind $s$-index, and viscosity $\alpha$) is shown in Figure~\ref{fig:diskEM_spec_R_a}. The spectrum is most sensitive to $R_{\rm out}$: increasing $R_{\rm out}$ raises the flux and enhances the low-energy tail because a larger fraction of low-energy photons originates from the outer disk. Decreasing $a_{\bullet}$ increases $R_{\rm in}$ and shifts the spectrum to softer energies. The wind $s$-index has a negligible effect on the EM spectrum in our models, while viscosity affects primarily the ONeMg WD--TDE case: for fixed accretion rate, smaller $\alpha$ leads to stronger neutrino cooling and thus lower EM output. Soft X-ray observations therefore have the potential to constrain the disk size and, in combination with other diagnostics, the IMBH mass and spin.

\subsection{Neutrino Radiation}
\label{subsec:Neutrino_radiation}
\begin{figure}
\centering
 \includegraphics[scale=0.5]{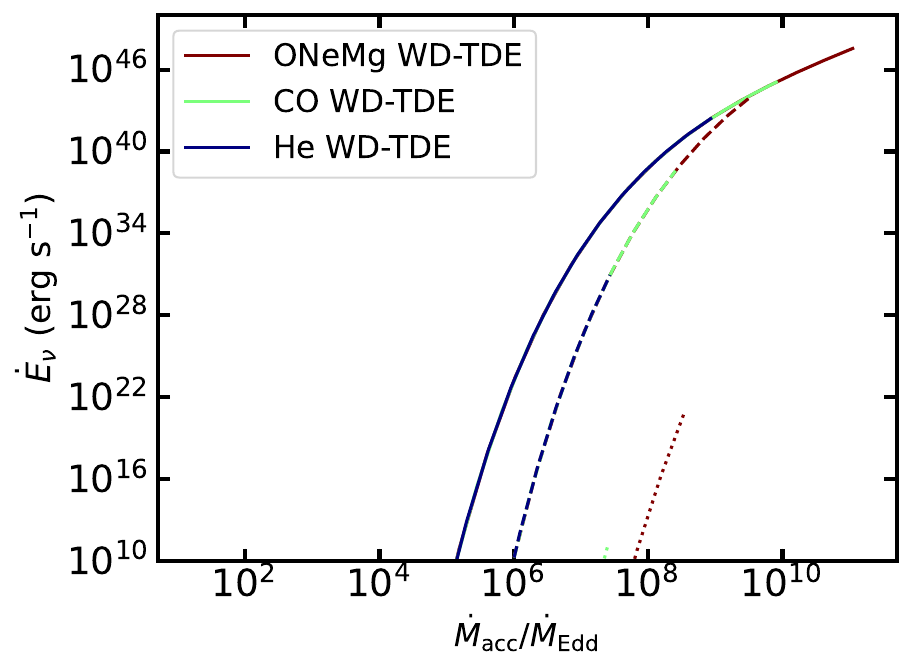}
    \caption{Neutrino luminosity of WD--TDE disks from electron--positron pair annihilation, as a function of mass fallback rate (from $10\,\dot{M}_{\rm Edd}$ to the peak fallback rate for each WD composition). Neutrino production is negligible for $\dot{M}_{\rm acc}\lesssim 10^8\,\dot{M}_{\rm Edd}$ and rises steeply with $\dot{M}_{\rm acc}$. Solid, dashed and dotted lines are $\dot{E}_{\nu}$ for $M_{\rm h} = 10^{3}\ M_{\odot}$, $10^4\ M_{\odot}$ and $10^5\ M_{\odot}$, respectively. For $M_{\rm h}=10^3\,M_{\odot}$, the peak values are $\dot{E}_{\nu}\simeq 3\times 10^{38}$, $2\times 10^{42}$, and $2.5\times 10^{45}\,\mathrm{erg\ s^{-1}}$ for He, CO, and ONeMg WD--TDEs, respectively.}
\label{fig:dotE_v}
\end{figure}

Thermal neutrinos and antineutrinos produced in the disk escape essentially freely because the neutrino optical depth is very small. The total neutrino power is
\begin{equation} \label{eq:dotE_v}
    \dot{E}_{\nu} = 4\pi \int^{R_{\rm out}}_{R_{\rm in}} Q_{\nu} R\, dR.
\end{equation}
Figure~\ref{fig:dotE_v} shows that WD--TDE disks can reach $\dot{E}_{\nu}\sim 10^{40}$--$10^{47}\,\mathrm{erg\ s^{-1}}$, with smaller IMBHs producing higher disk temperatures and hence higher neutrino luminosities.

For $\dot{M}\lesssim 10^8\,\dot{M}_{\rm Edd}$, neutrino emission is negligible, but it increases rapidly at higher $\dot{M}_{\rm acc}$. Even for ONeMg WD--TDEs, the neutrino luminosity is far below that of core-collapse supernovae \citep[$\sim 10^{53}\,\mathrm{erg\ s^{-1}}$][]{janka_explosion_2012} and neutrino-dominated accretion flows around stellar-mass black holes \citep[$\sim 10^{50}$--$10^{51}\,\mathrm{erg\ s^{-1}}$][]{liu_detectable_2016}.

\begin{figure}
\centering
 \begin{minipage}[t]{0.5\textwidth}
 \centering
 \includegraphics[scale=0.5]{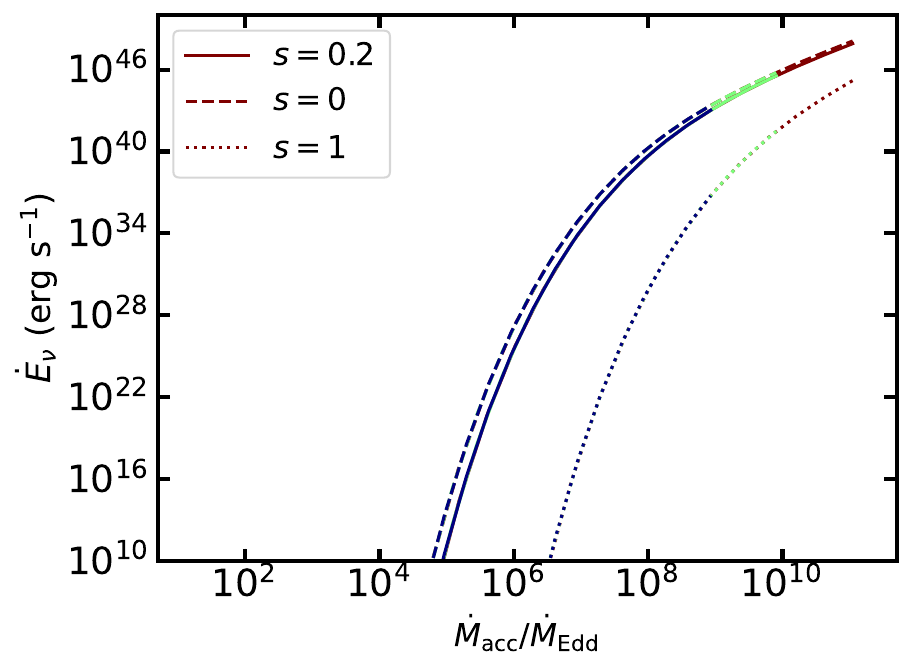}
\end{minipage}
 \begin{minipage}[t]{0.5\textwidth}
 \centering
 \includegraphics[scale=0.5]{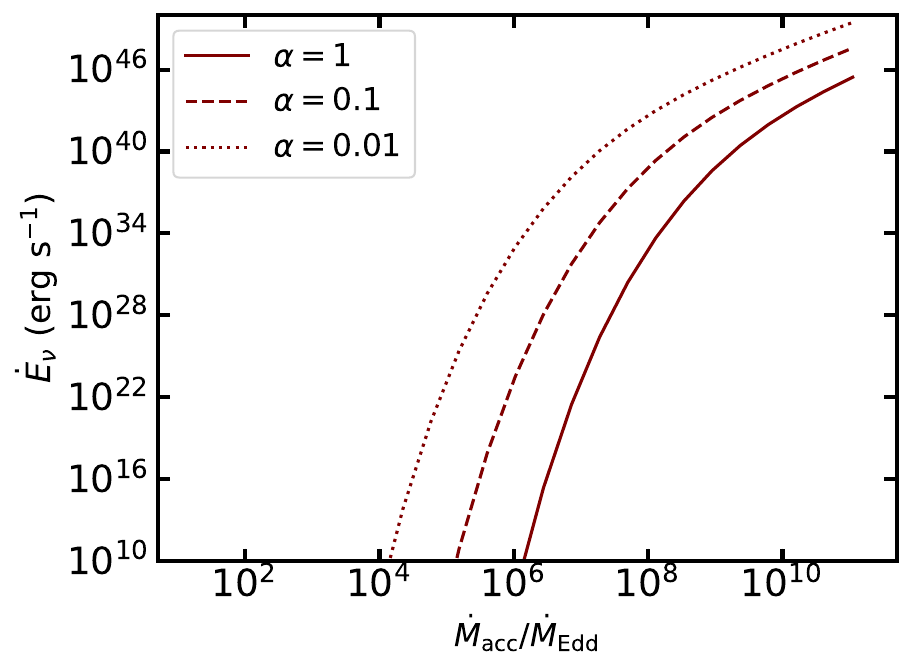}
\end{minipage}
    \caption{Dependence of the neutrino luminosity of WD-TDE disks on parameter $s$-index ($\dot{M}_{\rm in} \propto R^s$, upper panel) and viscosity $\alpha$ (lower panel). $s = 0$ represents no mass loss through wind. As disk wind provides an additional cooling on disk, it can reduce the neutrino luminosity. For the same accretion rate, lower $\alpha$ disk has higher temperature, thus produce higher neutrino luminosity.}
\label{fig:dotE_v_s_alpha}
\end{figure}

The dependence of $\dot{E}_{\nu}$ on the wind $s$-index and viscosity is shown in Figure~\ref{fig:dotE_v_s_alpha}. Disk winds provide an additional cooling channel and therefore reduce the neutrino luminosity. For the same accretion rate, smaller $\alpha$ yields a hotter disk and hence a higher neutrino luminosity.

The spectral shape is set by the pair-annihilation process. We adopt the fitting form of \citet{misiaszek_neutrino_2006},
\begin{equation}
    \phi(E_{\nu},T) = \frac{A_{\nu}}{k_{\rm b}T}\left(\frac{E_{\nu}}{k_{\rm b}T}\right)^{3.2} \ue^{-E_{\nu}/k_{\rm b}T},
\end{equation}
where $\phi$ is normalized such that $\int \phi\, dE_{\nu}=1$, and $A_{\nu}\simeq 0.14$.

\begin{figure}
\centering
 \includegraphics[scale=0.5]{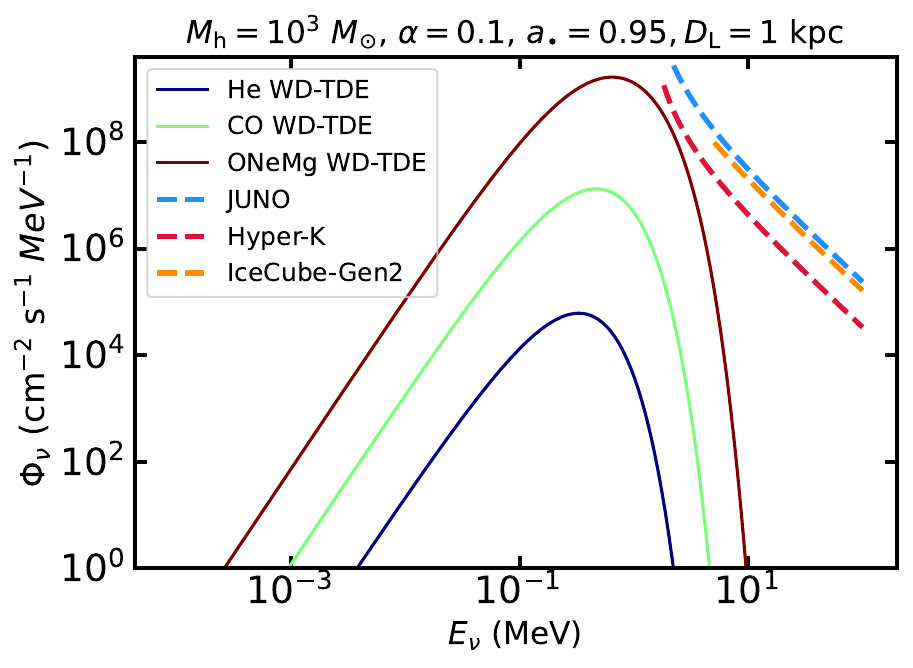}
    \caption{Differential neutrino spectra from WD--TDE disks. The three dashes lines represent the detection flux limit for JUNO, Hyper-K and IceCube-Gen2 with $T_{\rm expo} \simeq 10\ {\rm s}$ exposure time. The emission peaks at $\sim 0.1$--$1$ MeV.}
\label{fig:neutrino_spectrum}
\end{figure}

The neutrino flux ${\rm (cm^{-2}\ s^{-1}\ MeV^{-1})}$ is then
\begin{equation} \label{eq:Phi_v}
    \Phi_{\nu} = \frac{1}{D_{\rm L}^2}\int^{R_{\rm out}}_{R_{\rm in}} \phi(E_{\nu},T)\,\frac{Q_{\nu}}{\overline{E}_{\nu}}\, R\, dR.
\end{equation}
Here $\overline{E}_{\nu}=\int \phi E_{\nu}\, dE_{\nu}\simeq 4.6\,k_{\rm b}T$ is the mean neutrino energy at temperature $T$. As shown in Figure~\ref{fig:neutrino_spectrum}, the spectrum peaks at $\sim 0.1$--$1\,\mathrm{MeV}$. The most optimistic case is the ONeMg WD--TDE, which has the highest neutrino luminosity.

These low-energy neutrinos are challenging to detect with current facilities because of weak interactions and large backgrounds in MeV neutrino detectors. To estimate detectability, we compare the model spectra with the background-limited flux sensitivities for Jiangmen Underground Neutrino Observatory (JUNO), Hyper-Kamiokande (Hyper-K), and IceCube-Gen2 in Figure~\ref{fig:neutrino_spectrum}; the calculation of these sensitivity limits is described in Appendix~\ref{sec:flux_limit_framework}. Even for the most promising case, the detectable distance is $\lesssim 1\,\mathrm{kpc}$.


\section{GW Emission}
\label{sec:GW}

\subsection{Gravitational wave emission during passage}
\label{subsec:GW}
As a WD approaches an IMBH from large separation down to pericenter, the rapidly varying quadrupole moment generates gravitational waves (GWs). Here we focus on ``one-off'' WD--TDEs on parabolic orbits, in which the WD is fully disrupted during a single passage with $R_{\rm p}=R_{\rm t}$. If the pericenter is slightly larger than the tidal radius ($R_{\rm p}\gtrsim R_{\rm t}$), the WD may instead undergo repeated tidal stripping on an eccentric orbit; the GW signal in that scenario has been explored in \citet{chen_repeated_2024}.

The total instantaneous power radiated in GWs is given by \citep{Peters_Gravitational_1963}:
\small{
\begin{equation} \label{eq:P_GW}
    P_{\rm GW}(\Psi) = \frac{1}{60}\frac{G^4}{c^5}\frac{M_{\rm h}^3M_*^2}{R_{\rm p}^5} (1+\cos{\Psi})^4 [12(1+\cos{\Psi})^2+\sin^2{\Psi}],
\end{equation}}
where $\Psi$ is the true anomaly of the WD's orbit. $\Psi = 0$ corresponds to the pericenter position, where the GW power reaches the maximum:
\begin{equation} \label{eq:P_GW_max}
    P_{\rm GW,max} = 2 \times 10^{46} M_3^{4/3} r_{*,-2}^{-5} m_*^{11/3}\ {\rm erg\ s^{-1}} .
\end{equation}

To estimate the detectability in the parabolic case, we compute the signal-to-noise ratio (SNR) by combining the characteristic strain of the GW burst with the detector noise power spectral density. Because the encounter is unbound, the GW signal is typically treated as a single burst.

The SNR of a single-burst GW event is \citep{Barack_Confusion_2004,berry_gravitational_2010}
\begin{equation} \label{eq:SNR}
    {\rm SNR}^2 \simeq \int^\infty_0 \frac{|h_{\rm eff}(f)|^2}{|h_{\rm n}(f)|^2} \, d\ln{f},
\end{equation}
where $h_{\rm eff}(f)$ is the effective characteristic strain of the signal and $h_{\rm n}(f)$ is the characteristic strain of the detector noise. Roughly, the signal becomes detectable when $h_{\rm eff}\gtrsim h_{\rm n}$ over the frequency range containing most of the radiated power.

The characteristic strain can be written as
\begin{equation} \label{eq:h_eff}
    |h_{\rm eff}(f)|^2 \simeq \frac{4G}{c^3\pi^2 D_{\rm L}^2} \frac{dE_{\rm GW}}{df},
\end{equation}
where $dE_{\rm GW}/df$ is the GW energy spectrum. For a parabolic encounter \citep{berry_gravitational_2010},
\begin{equation} \label{eq:dE_df}
    \frac{dE_{\rm GW}}{df} \simeq \frac{4\pi^2}{5}\frac{G^3 M_*^2 M_{\rm h}^2}{c^5 R_{\rm p}^2} \, l\!\left(\frac{f}{f_{\rm c}}\right),
\end{equation}
where
\begin{equation} \label{eq:f_c}
    f_{\rm c} = \frac{1}{2\pi}\sqrt{\frac{GM_{\rm h}}{R_{\rm p}^3}}
\end{equation}
is the orbital frequency of a circular orbit at $R_{\rm p}$. For encounters with $R_{\rm p}=R_{\rm t}$, the scaling $R_{\rm t}\propto M_{\rm h}^{1/3}$ implies that $f_{\rm c}$ is only weakly dependent on $M_{\rm h}$. The dimensionless function $l(f/f_{\rm c})$ is given by Equation~(24) of \citet{berry_gravitational_2010}; Equation~(\ref{eq:dE_df}) follows from the parabolic limit of \citet{Peters_Gravitational_1963}. The spectrum peaks at $f\simeq 1.6 f_{\rm c}$.

\begin{figure*}
\centering
 \begin{minipage}[t]{1\textwidth}
 \centering
 \includegraphics[scale=0.5]{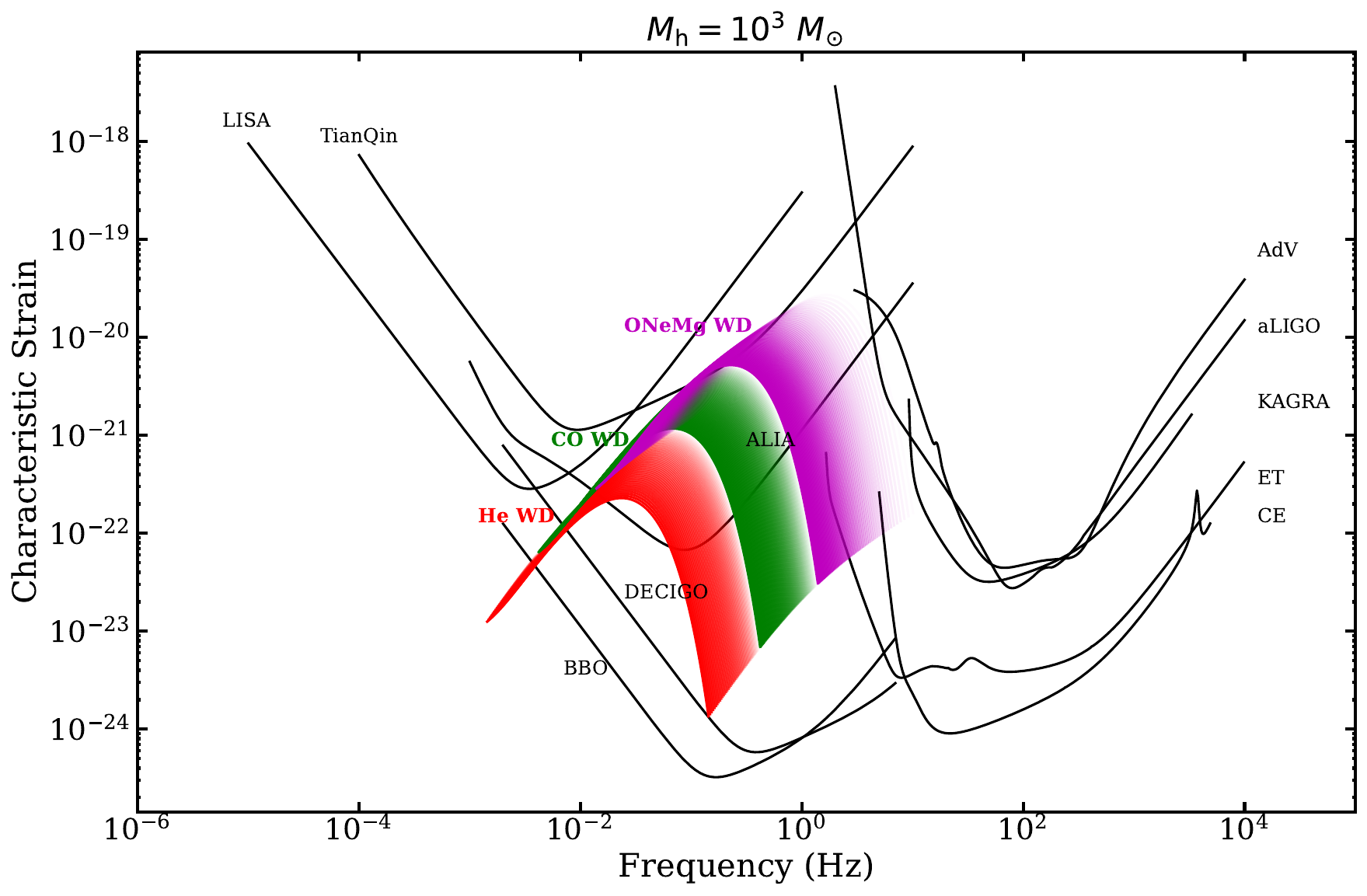}
\end{minipage}
 \begin{minipage}[t]{1\textwidth}
 \centering
 \includegraphics[scale=0.5]{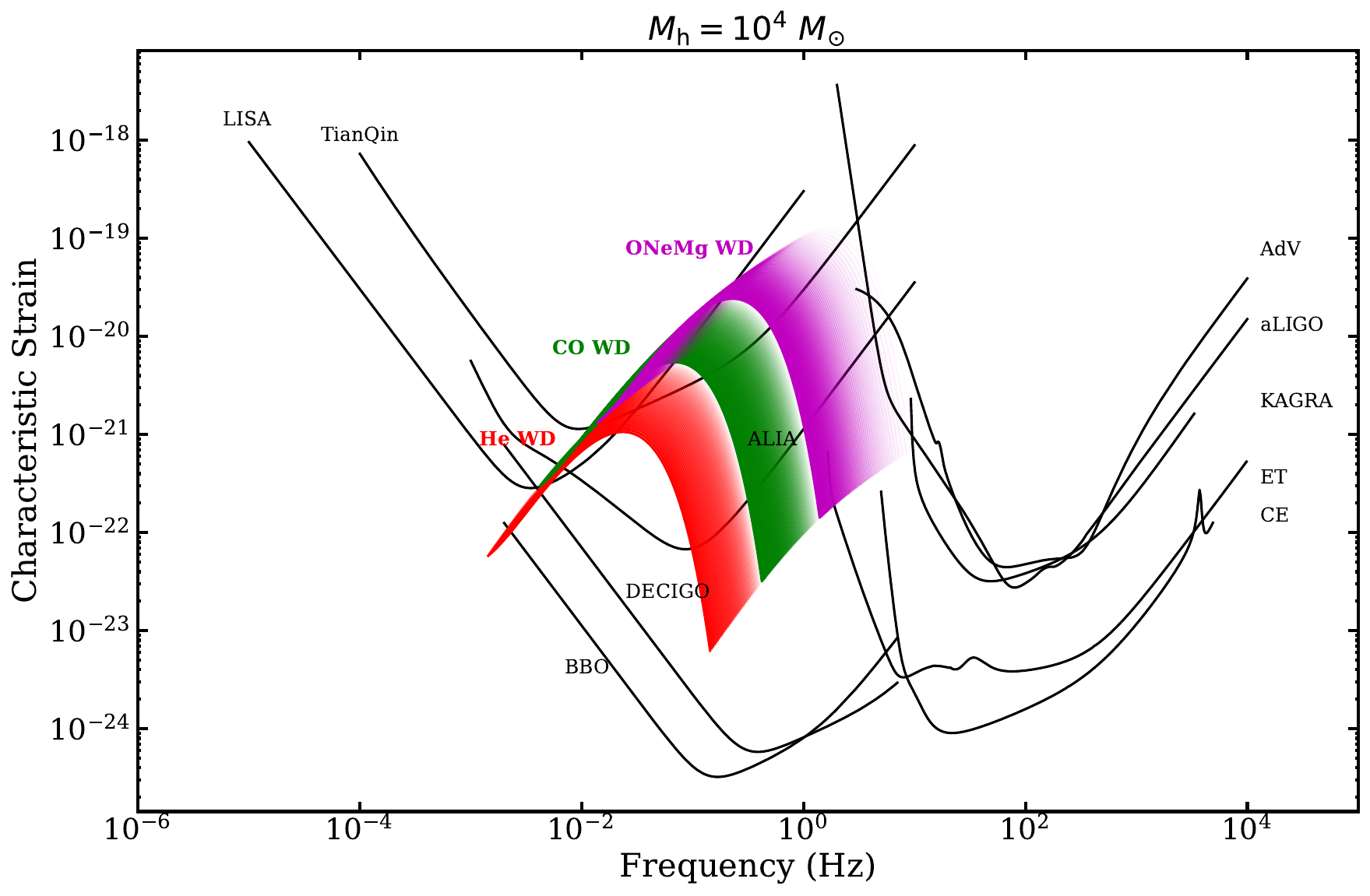}
\end{minipage}
    \caption{Effective GW strain for a WD--TDE on a parabolic orbit with $R_{\rm p}=R_{\rm t}$ at a distance of $D_{\rm L}=1$~Mpc, compared with the sensitivity curves of different GW detectors. Different WD compositions are shown with different colors. The extended shapes indicate different $M_*$; more massive WDs generally yield higher peak frequencies and larger strains. The characteristic GW frequency peaks at $\sim 0.1$--$1$~Hz. Next-generation space-based detectors such as ALIA, DECIGO, and BBO are therefore ideal for one-off WD--TDEs, while LISA and TianQin may detect the nearest sources.}
\label{fig:GW_spec}
\end{figure*}

In Figure~\ref{fig:GW_spec}, we compare $h_{\rm eff}$ with detector sensitivity curves $h_{\rm n}$ for a fiducial distance $D_{\rm L}=1$~Mpc. The sensitivity curves are taken from \url{http://gwplotter.com}. The characteristic GW frequency typically falls in the $\sim 0.1$--$1$~Hz band. Next-generation space-based detectors operating in this band---such as the Advanced Laser Interferometer Antenna (ALIA), the DECI-hertz Interferometer Gravitational Wave Observatory (DECIGO; \citealt{sato_decigo_2009}), and the Big Bang Observer (BBO; \citealt{harry_laser_2006})---are therefore well suited to detecting one-off WD--TDEs. The Laser Interferometer Space Antenna (LISA; \citealt{Amaro_LISA_2017}) and TianQin \citep{Luo_TianQin_2016} may also detect the nearest events.

\begin{figure}
\centering
 \includegraphics[scale=0.4]{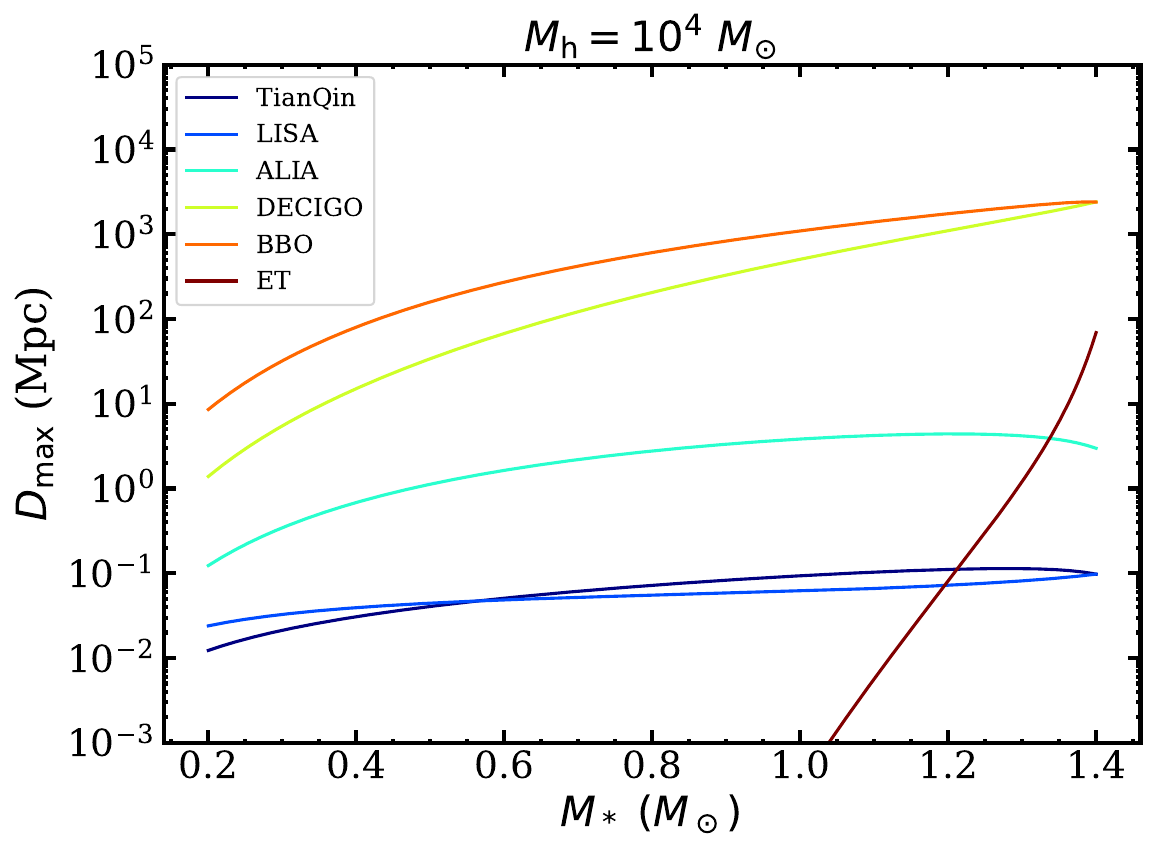}
    \caption{GW horizon distance for the burst emitted during the disruptive passage. Different lines represent different detectors.}
\label{fig:GW_dmax}
\end{figure}

To estimate detection horizons, we adopt a threshold ${\rm SNR}=15$ following \citet{ye_observing_2024}. Figure~\ref{fig:GW_dmax} shows the GW horizon distance $D_{\rm max}$ for full disruptions on parabolic orbits with $R_{\rm p}=R_{\rm t}$. DECIGO and BBO could detect WD--TDEs involving $10^3\,M_{\odot}$ IMBHs out to $\sim 10^3$--$10^4$~Mpc. For $10^4\,M_{\odot}$ IMBHs, the horizon increases by roughly an order of magnitude due to ${\rm SNR}\propto D_{\rm L}^{-1} M_{\rm h}^{2/3}$.

GW detections of WD--TDEs would enable precise measurements of key system parameters (e.g., $M_{\rm h}$, $M_*$, and orbital configuration). They could also provide an advance trigger for rapid multiwavelength follow-up, potentially capturing the earliest phases of debris circularization and disk formation.

\subsection{GWs from a precessing disk}
\label{subsec:GW_disk}
A nascent accretion disk in a TDE is typically misaligned with the BH spin axis because the stellar orbital orientation is essentially random \citep{chen_diverse_2026}. A tilted, geometrically thick disk can therefore precess approximately as a rigid body due to global Lense--Thirring torques \citep{lense_uber_1918}.

The repeating GRB~250702B from a WD--TDE candidate has been suggested to arise from co-precession of the disk--jet system \citep{levan_day-long_2025}. Disk-jet co-precession has also been observed in the TDE candidate AT2020afhd \citep{wang_detection_2025}.

The precession frequency depends on disk properties (e.g., surface density profile and radial extent) as well as the IMBH mass and spin \citep{stone_observing_2012,Chen_AT2019avd_2022}. It can be obtained by integrating the Lense--Thirring torque over the precessing disk \citep{fragile_global_2007}:
\begin{equation} \label{eq:f_prec}
    f_{\rm prec} = \frac{G^2M_{\rm h}^2 |a_{\bullet}|}{\pi c^3} \frac{\int^{R_{\rm out}}_{R_{\rm in}} R'^{-3/2} \Sigma(R')\,dR'}{\int^{R_{\rm out}}_{R_{\rm in}} R'^{3/2}\Sigma(R')\,dR'}.
\end{equation}
For the WD--TDE disk, we find $\Sigma\propto R^{\zeta}$ with $\zeta\sim -0.25$ to $-0.2$, so that most of the mass resides at large radii. In this limit, Equation~(\ref{eq:f_prec}) reduces to
\begin{equation} \label{eq:f_prec2}
    \begin{split}
    f_{\rm prec} &\simeq \frac{G^2M_{\rm h}^2 |a_{\bullet}|}{\pi c^3} R_{\rm out}^{-3} \\
    &\simeq 0.06\,|a_{\bullet}|\,M_3^{-1} \left(\frac{R_{\rm out}}{10R_{\rm g}}\right)^{-3}\ {\rm Hz}.
    \end{split}
\end{equation}

We assume that alignment is negligible during the super-Eddington phase, so the disk tilt remains approximately constant. As the accretion rate declines into the sub-Eddington regime, the disk cools and eventually enters the diffusive-warp regime (roughly when $\alpha \gtrsim H/R$), in which warps are rapidly damped and the disk aligns with the BH spin \citep{bardeen_lense-thirring_1975,nelson_hydrodynamic_1999,franchini_lense-thirring_2016}. Viscous shear also dissipates the precession energy at a rate $\propto \alpha$ \citep{bate_observational_2000}, implying an alignment timescale $t_{\rm Bate} \simeq \alpha^{-1} (H/R)^2 \Omega(R_{\rm out})/f_{\rm prec}^2$ and thus a number of precession cycles before alignment of $t_{\rm Bate} f_{\rm prec} \simeq 34\,(\alpha/0.1)^{-1}|a_{\bullet}|^{-1}(R_{\rm out}/10R_{\rm g})^{3/2}$. In reality, continued injection of misaligned angular momentum by the debris stream could prolong the alignment time.

A precessing, axisymmetric body emits GWs (e.g., in GRB disk models; \citealt{romero_gravitational_2010}). Here we apply this framework to WD--TDE disks and estimate detectability.

The GW strain can be written as the sum of the two polarization states,
\begin{equation} \label{eq:h_prec}
    h_{\rm prec}(t) = h_{+}(t) + h_{\times}(t),
\end{equation}
with
\begin{equation} \label{eq:h_+}
    h_+(t) = F_{+,1} \cos{(2\pi f_{\rm prec} t)} + F_{+,2}\cos{(4\pi f_{\rm prec} t)},
\end{equation}
\begin{equation} \label{eq:h_times}
    h_{\times}(t) = F_{\times,1} \sin{(2\pi f_{\rm prec} t)} + F_{\times,2}\sin{(4\pi f_{\rm prec} t)}.
\end{equation}
Assuming the disk precesses as an axisymmetric rigid body, the emission occurs at the fundamental frequency $f_{\rm prec}$ and the second harmonic $2f_{\rm prec}$.

The directional amplitude coefficients depend on the viewing geometry and tilt angle $\theta$:
\begin{equation} \label{eq:F_+1}
    F_{+,1} = h'_0 \sin{(2\theta)} \sin{\iota} \cos{\iota},
\end{equation}
\begin{equation} \label{eq:F_+2}
    F_{+,2} = 2h'_0 \sin^2{\theta} (1+\cos^2{\iota}),
\end{equation}
\begin{equation} \label{eq:F_times1}
    F_{\times,1} = h'_0 \sin{(2\theta)} \sin{\iota},
\end{equation}
\begin{equation} \label{eq:F_times2}
    F_{\times,2} = 4h'_0 \sin^2{\theta}\cos{\iota},
\end{equation}
where $\iota$ is the angle between the BH spin axis and the line of sight, and
\begin{equation} \label{eq:h_0}
    h'_0=-\frac{4\pi^2 G}{c^4} \frac{(I_3-I_1)f_{\rm prec}^2}{d_{\rm L}}
\end{equation}
sets the overall amplitude.

The moments of inertia satisfy $I_1 = I_2\neq I_3$ in the disk rest frame. For the WD--TDE disk, using $\Sigma\propto R^{\zeta}$ with $\zeta\sim -0.25$ to $-0.2$, we obtain $I_3-I_1 \simeq (1/24) M_{\rm d} R_{\rm out}^2$, where $M_{\rm d}$ is the precessing disk mass.

The total instantaneous power radiated in GWs is the integral of the energy flux over a sphere:
\small{
\begin{equation} \label{eq:P_prec}
    P_{\rm prec}(t) = \int^{2\pi}_0 d\phi \int^\pi_0 \left[\frac{c^3}{16\pi G} \left(\dot{h}_+^2+\dot{h}_{\times}^2\right)\right] d_{\rm L}^2 \sin{\iota}\,d\iota,
\end{equation}
}
Assuming the disk properties and $f_{\rm prec}$ do not vary significantly over a cycle, one obtains the time-average power over one precession cycle:
\begin{equation} \label{eq:P_prec2}
    \begin{split}
    P_{\rm prec} &= \frac{2}{5} (2\pi)^6 \frac{G}{c^5}(I_3-I_1)^2f_{\rm prec}^6 \sin^2{\theta}(1+15 \sin^2{\theta}) \\
    &\simeq 10^{39}|a_{\bullet}|^6 \left(\frac{M_{\rm d}}{0.1M_{\odot}}\right)^2 M_3^{-2} \left(\frac{R_{\rm out}}{10R_{\rm g}}\right)^{-14} \\
    &\times \sin^2{\theta}(1+15 \sin^2{\theta})\ {\rm erg\ s^{-1}}.
    \end{split}
\end{equation}
This scaling shows that the GW power is extremely sensitive to the size of the precessing disk. We typically take the initial disk size to be $R_{\rm out}\simeq R_{\rm c}$. In Section~\ref{subsec:disk_evolution}, we will explore the disk size evolution. Initially, the disk size is close to $R_{\rm c}$ with slow outward expansion.

\begin{figure}
\centering
 \includegraphics[scale=0.35]{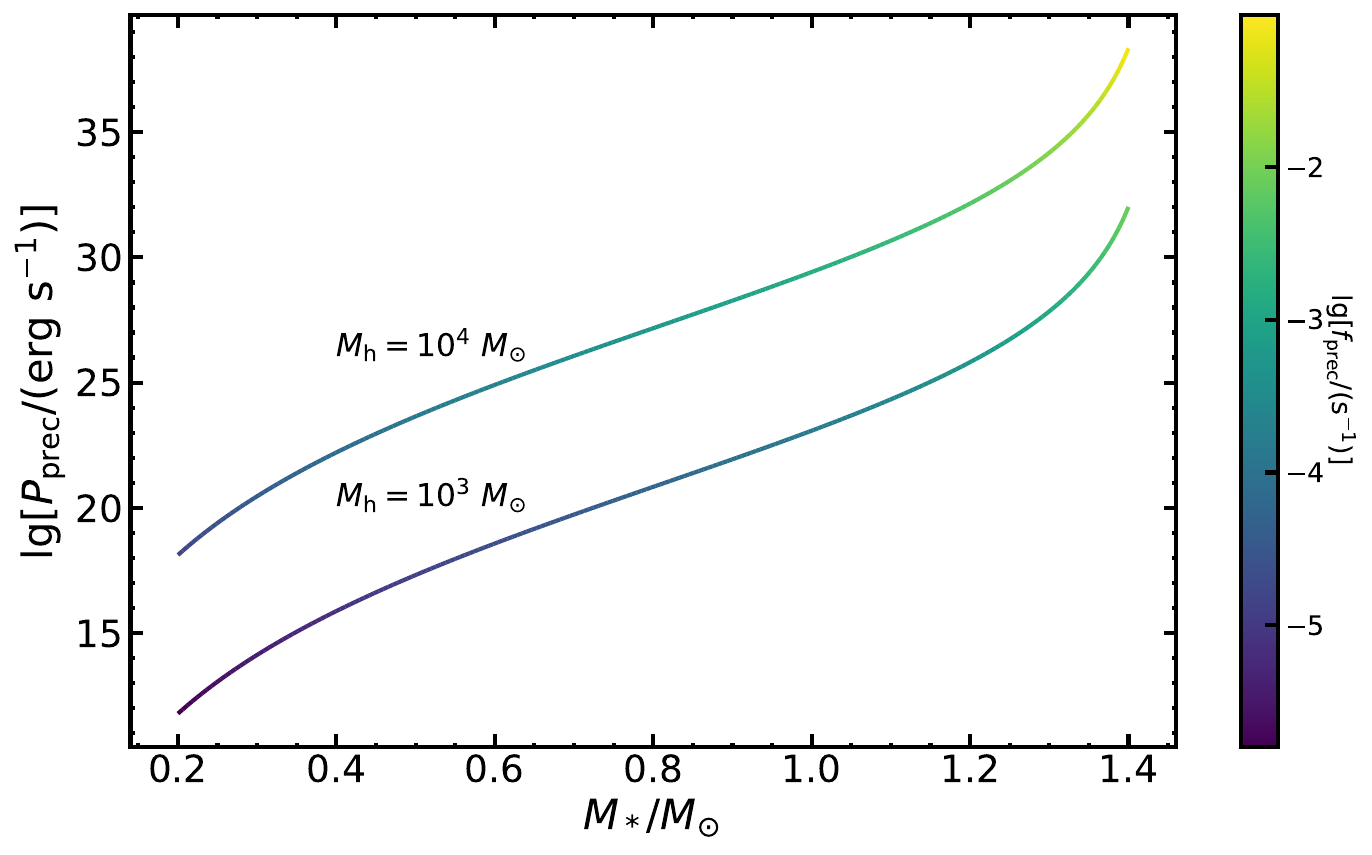}
    \caption{GW power $P_{\rm prec}$ from a precessing disk versus WD mass. The color indicates the precession frequency $f_{\rm prec}$. The two curves correspond to different IMBH masses. The disk tilt angle and disk size are fixed to $\theta=\pi/4$ and $R_{\rm out} \simeq R_{\rm c}$.}
\label{fig:P_prec}
\end{figure}

The disk mass is related to the accretion rate via $M_{\rm d} \simeq \dot{M}_{\rm acc} t_{\nu}(R_{\rm out}) \simeq \dot{M}_{\rm acc}/\bigl(\alpha \Omega(R_{\rm out})\bigr)$, where $t_{\nu}(R_{\rm out})$ and $\Omega(R_{\rm out})$ are the viscous timescale and Keplerian angular frequency at $R_{\rm out}$. Near peak accretion ($\dot{M}_{\rm acc} \simeq \dot{M}_{\rm peak}$) and for $R_{\rm out} \simeq R_{\rm c}$, Equation~(\ref{eq:P_prec2}) can be rewritten as
\begin{equation} \label{eq:P_prec3}
    \begin{split}
    P_{\rm prec} &\simeq 10^{23} \alpha^{-2} |a_{\bullet}|^6 M_3^{19/3} r_{*,-2}^{-14} m_*^{23/3}\left(\frac{R_{\rm out}}{R_{\rm c}}\right)^{-11} \\
    &\times \sin^2{\theta}(1+15 \sin^2{\theta})\ {\rm erg\ s^{-1}}.
    \end{split}
\end{equation}
The GW power $P_{\rm prec}$ and corresponding precession frequency $f_{\rm prec}$ are shown in Figure~\ref{fig:P_prec}. More massive WDs produce denser, more compact disks, yielding higher $f_{\rm prec}$ and larger $P_{\rm prec}$. For the ONeMg WD case ($M_*\gtrsim M_{\odot}$), we find $f_{\rm prec}\sim 10^{-3}$--$10^{-2}\,\mathrm{Hz}$. For a larger IMBH ($10^4\,M_{\odot}$), the frequency can reach $\sim 0.1\,\mathrm{Hz}$.

In all cases, the GW power is far below the accretion power $\sim 0.1\dot{M}_{\rm acc}c^2$, so GW energy losses have a negligible impact on disk evolution.

Next, we estimate the detectability of the GW signal from a precessing WD--TDE disk by comparing it with detector sensitivity curves.

The frequency-domain amplitude follows from the Fourier transformation $\tilde{h}(f)=\mathcal{F}\{h_{\rm prec}(t)\}$:
\begin{equation} \label{eq:h(f)^2}
    \begin{split}
    |\tilde{h}(f)|^2 &= \frac14 F_{+,1}^2 [\delta(f-f_{\rm prec})]^2+\frac14 F_{+,2}^2 [\delta(f-2f_{\rm prec})]^2 \\
    &+\frac14 F_{\times,1}^2 [\delta(f-f_{\rm prec})]^2+\frac14 F_{\times,2}^2 [\delta(f-2f_{\rm prec})]^2.
    \end{split}
\end{equation}

Sky averaging gives
\begin{equation} \label{eq:<h(f)^2>}
    \begin{split}
    \langle|\tilde{h}(f)|^2\rangle_{\rm sky} &= \frac{1}{4\pi}\int^{2\pi}_0 d\phi \int^{\pi}_0 |\tilde{h}(f)|^2 \sin{\iota} \, d\iota\\
    &= \frac{16}{5}{h'_0}^2 \sin^2{\theta} \cos^2{\theta} \delta(f-f_{\rm prec})T_{\rm obs} \\
    &+ \frac{64}{5} {h_0'}^2 \sin^4{\theta}\delta(f-2f_{\rm prec})T_{\rm obs},
    \end{split}
\end{equation}
where we regularize the ill-defined $\delta(f-f_{\rm prec})^2$ by assuming a finite observation time $T_{\rm obs}$, i.e., $\delta(f-f_{\rm prec})^2\rightarrow \delta(f-f_{\rm prec})T_{\rm obs}$. Here $T_{\rm obs}$ is the detector mission lifetime, or the source duration if shorter. We take $T_{\rm obs}\sim t_{\rm fb}$, since the disk evolves on a timescale $t_{\rm fb}$; afterwards, the source fades and spreads viscously, reducing detectability.

The SNR can be written as
\begin{equation} \label{eq:SNR_prec}
    \begin{split}
    {\rm SNR}^2 &= 4 \int^{\infty}_0 \frac{\langle|\tilde{h}(f)|^2\rangle_{\rm sky}}{S_n(f)}\ df \\
    &= \frac{16}{5}{h_0'}^2T_{\rm obs} \left[\frac{\sin^2{\theta}\cos^2{\theta}f_{\rm prec}}{|h_n(f_{\rm prec})|^2}+\frac{8\sin^4{\theta}f_{\rm prec}}{|h_n(2f_{\rm prec})|^2}\right],
    \end{split}
\end{equation}
where $h_{\rm n}=\sqrt{fS_{\rm n}}$ is the characteristic strain of the detector noise.

\begin{figure}
\centering
 \includegraphics[scale=0.4]{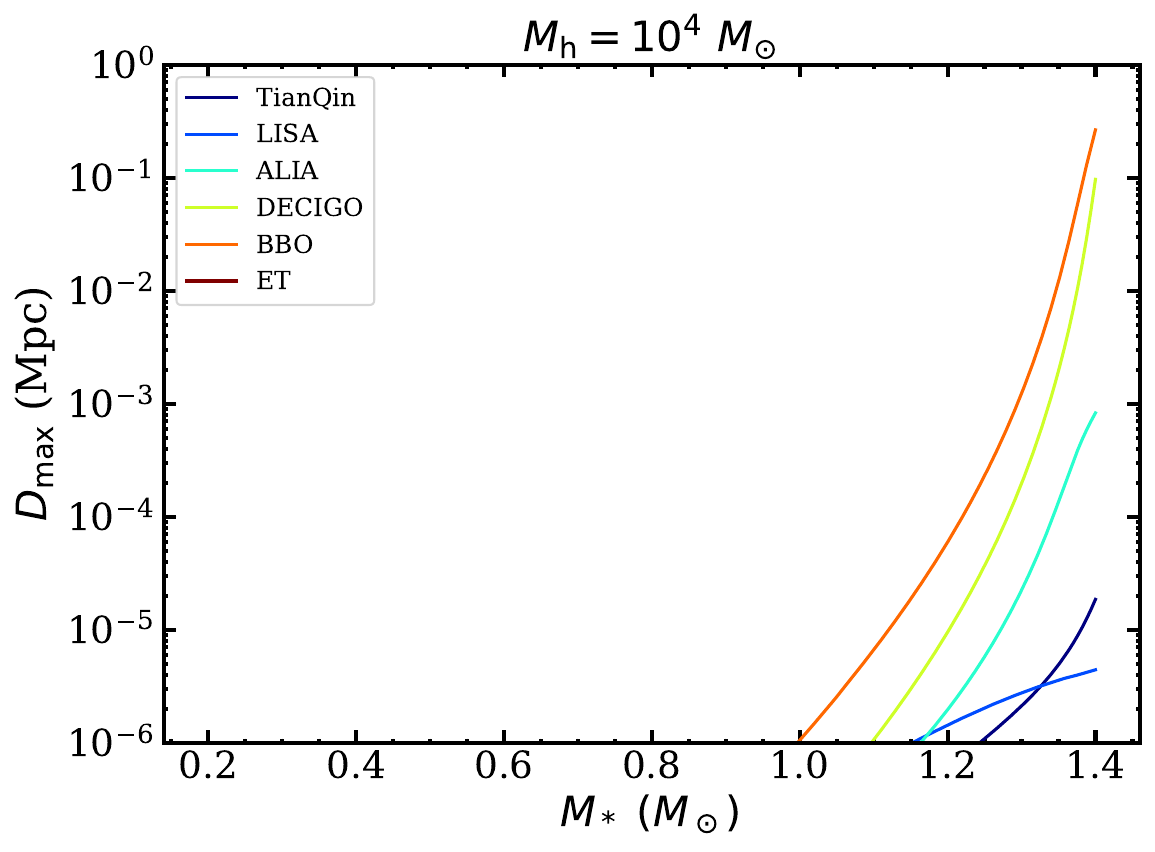}
    \caption{GW horizon distance for precessing disks in WD--TDEs. Different lines represent different detectors. The disk tilt angle and disk size are fixed to $\theta=\pi/4$ and $R_{\rm out} \simeq R_{\rm c}$.}
\label{fig:GW_dmax_prec}
\end{figure}

We adopt a threshold ${\rm SNR}=15$ to estimate detection horizons. Figure~\ref{fig:GW_dmax_prec} shows the GW horizon distance for a precessing-disk signal in WD--TDEs. This signal is significantly weaker than the burst emitted during the disruptive passage, and the detectable distance is limited to $\sim 1$~Mpc even for decihertz detectors.




\section{Multi-messenger signals of WD TDEs: a summary}
\label{sec:Multi_messenger_evolution}
\begin{figure*}
\centering
 \begin{minipage}[t]{0.45\textwidth}
 \centering
 \includegraphics[scale=0.32]{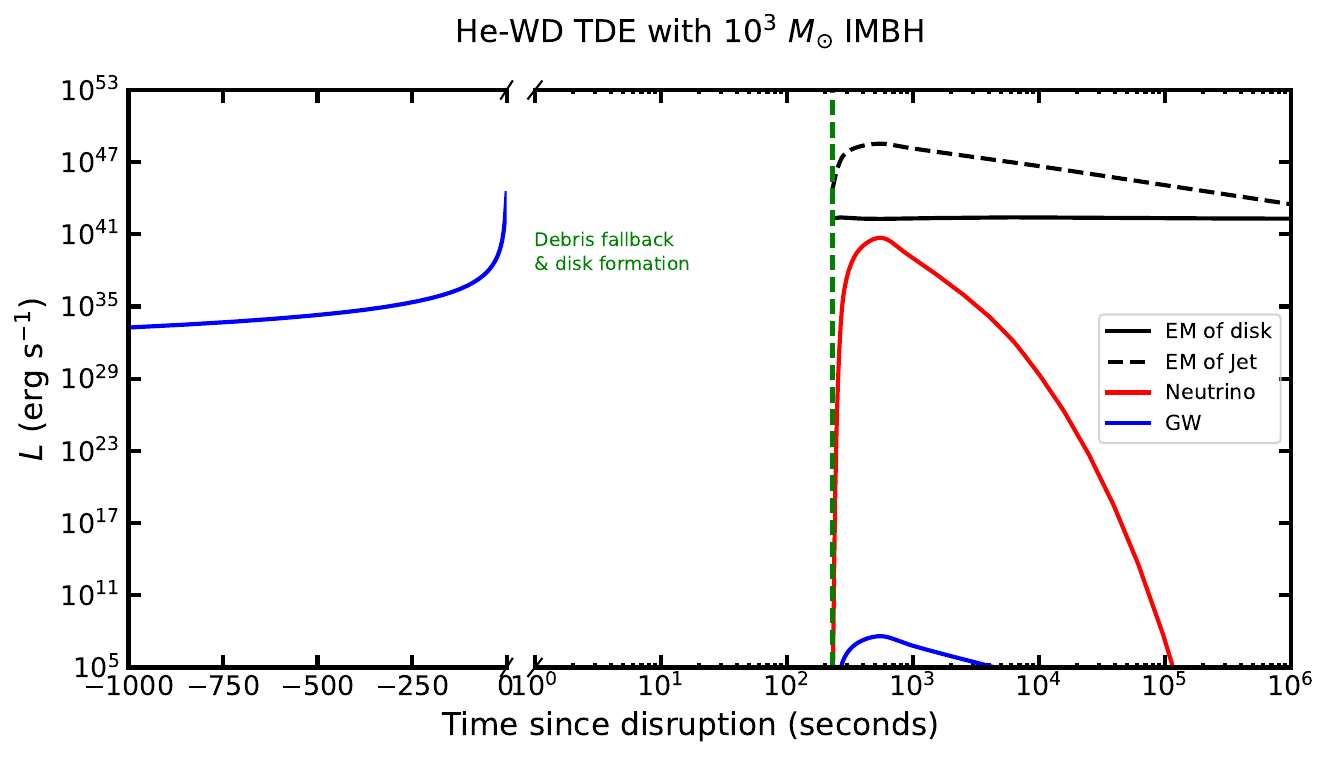}
\end{minipage}
 \begin{minipage}[t]{0.45\textwidth}
 \centering
 \includegraphics[scale=0.32]{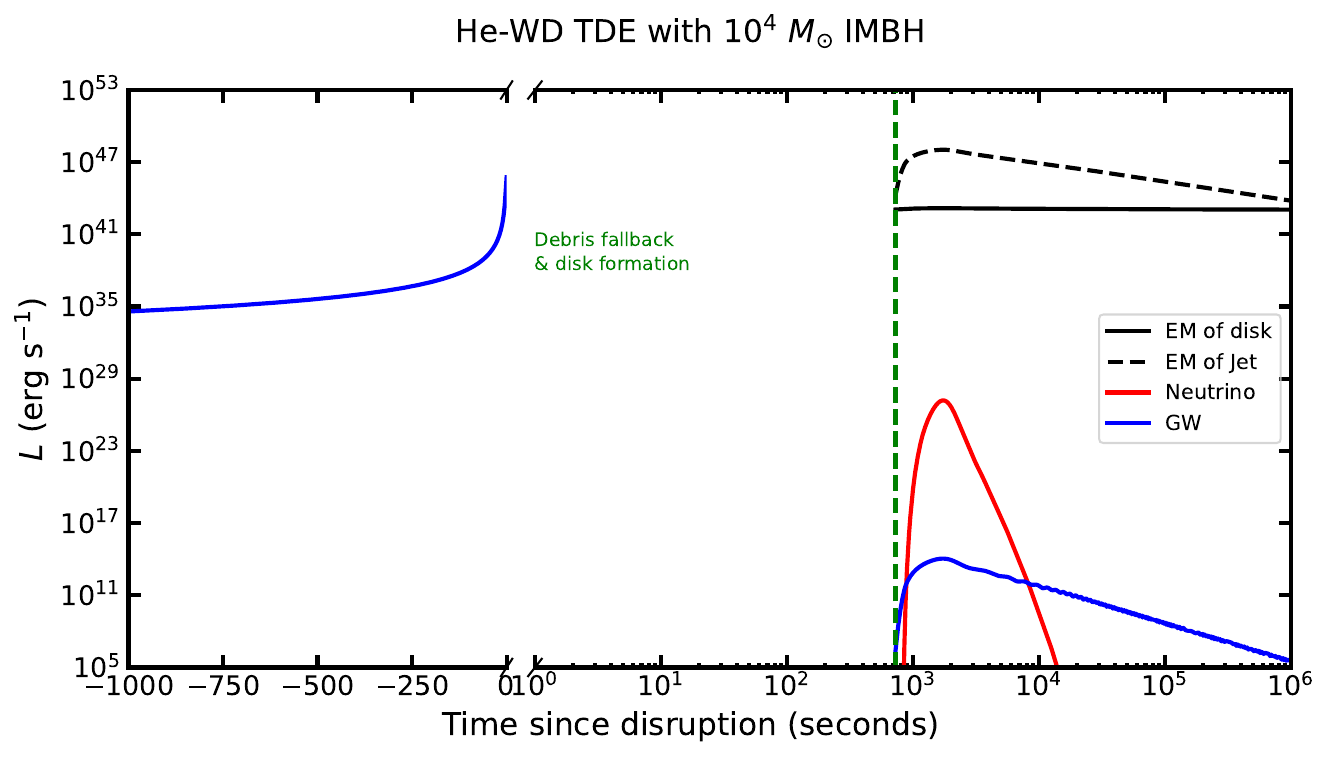}
\end{minipage}
 \begin{minipage}[t]{0.45\textwidth}
 \centering
 \includegraphics[scale=0.32]{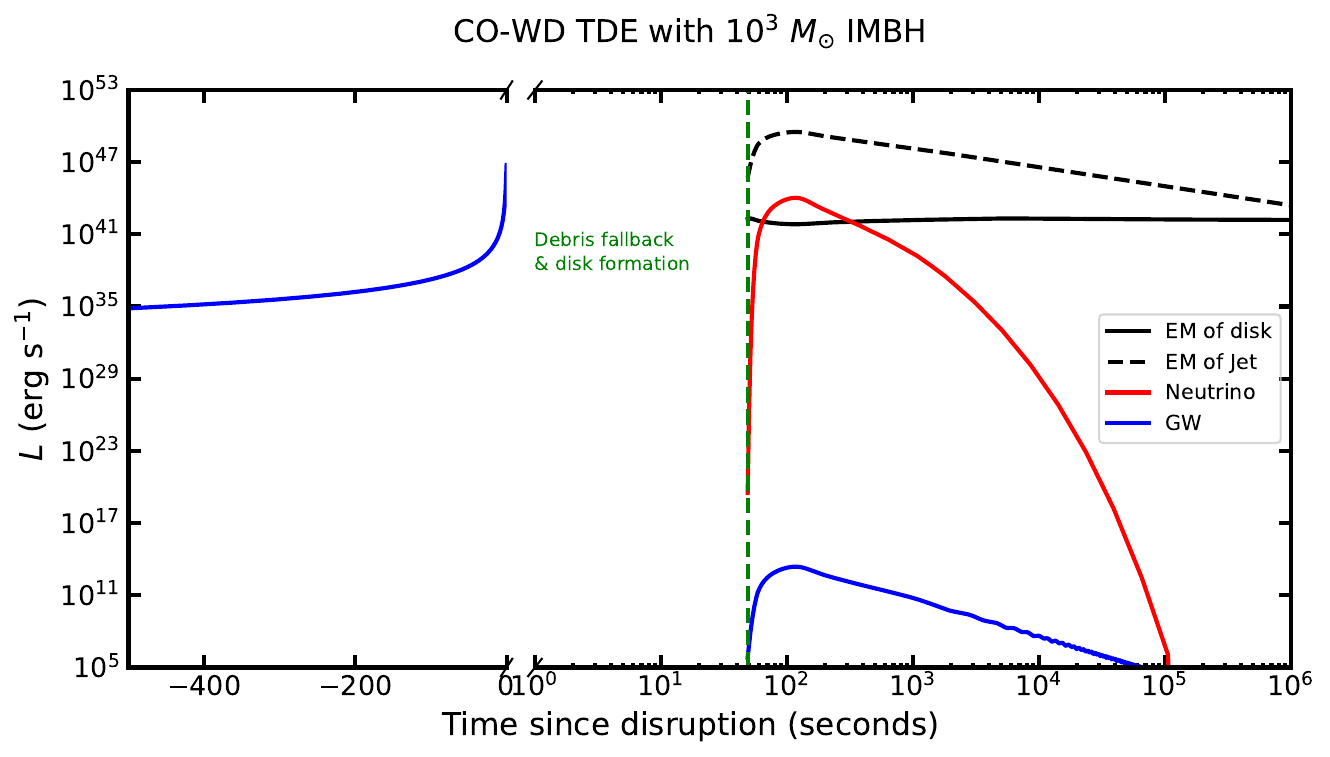}
\end{minipage}
 \begin{minipage}[t]{0.45\textwidth}
 \centering
 \includegraphics[scale=0.32]{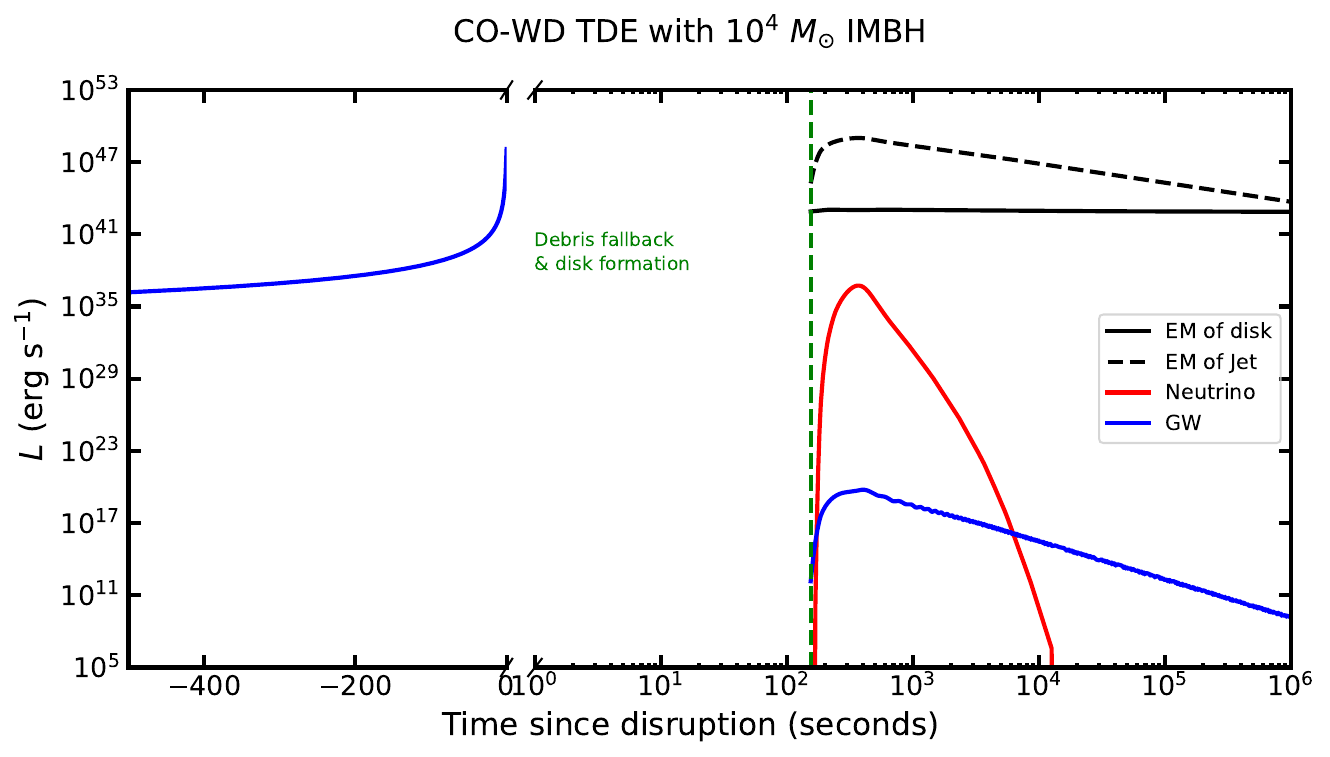}
\end{minipage}
 \begin{minipage}[t]{0.45\textwidth}
 \centering
 \includegraphics[scale=0.32]{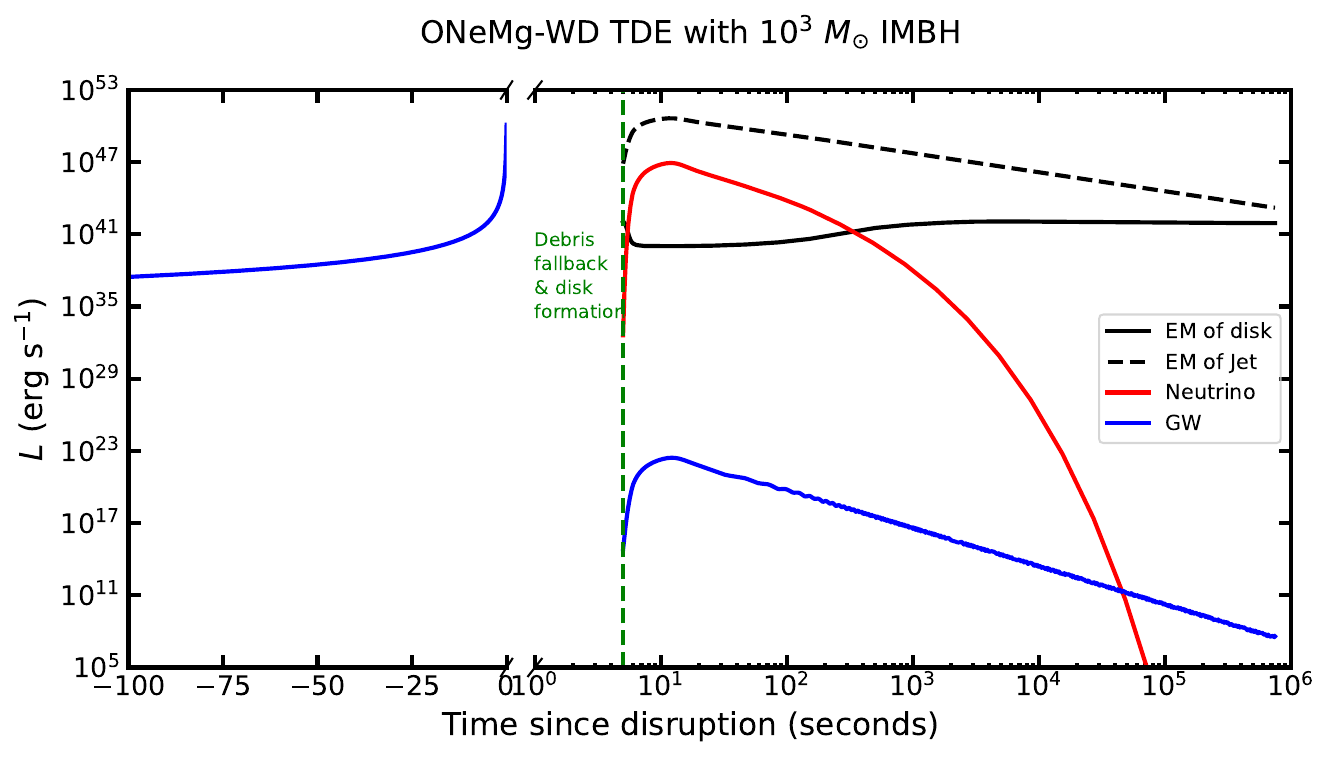}
\end{minipage}
 \begin{minipage}[t]{0.45\textwidth}
 \centering
 \includegraphics[scale=0.32]{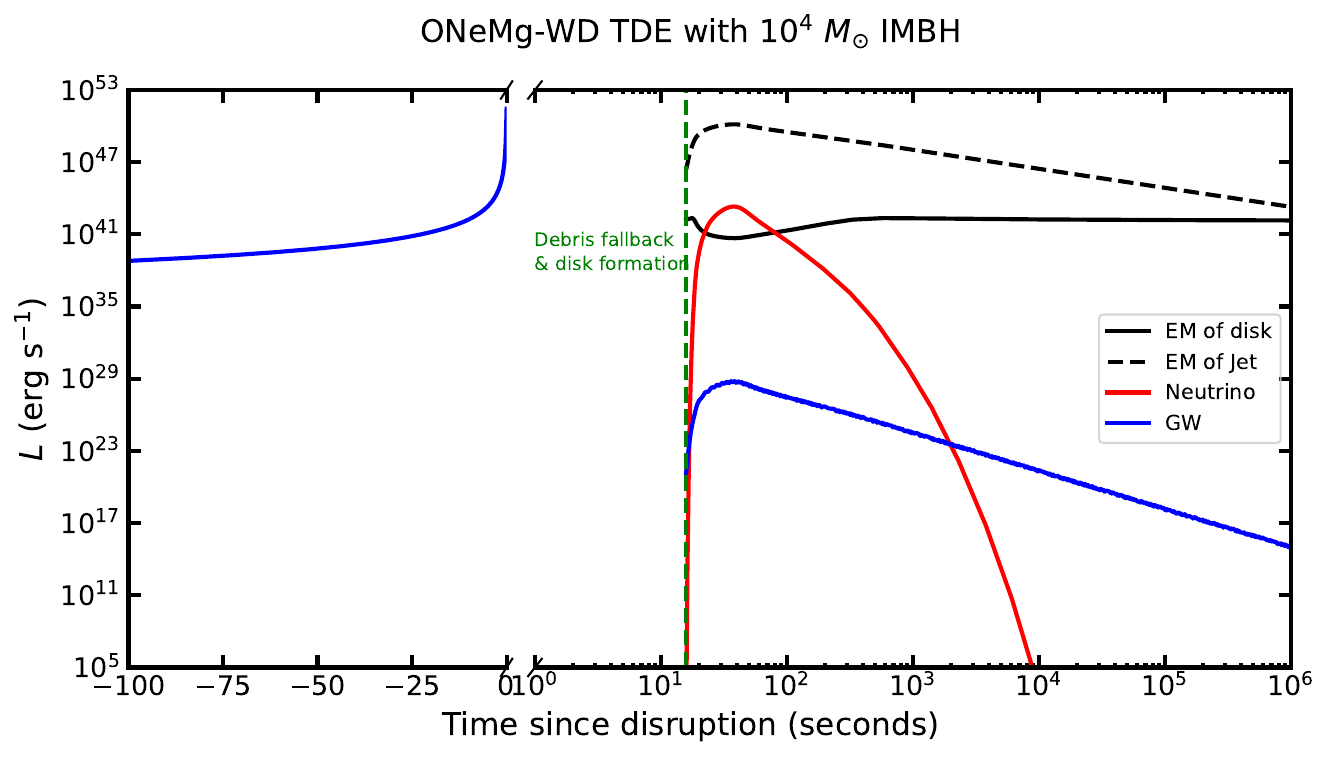}
\end{minipage}
    \caption{EM, GW, and neutrino power histories for WD TDEs with different IMBH masses and WD types, for representative peak fallback rates of $\simeq 8.5\times 10^8\,\dot{M}_{\rm Edd}$ (He), $\simeq 8.1\times 10^9\,\dot{M}_{\rm Edd}$ (CO), and $\simeq 10^{11}\,\dot{M}_{\rm Edd}$ (ONeMg). During pericenter passage, the time-varying quadrupole moment produces a GW burst. Subsequent debris fallback can form a compact accretion disk, whose thermal emission mildly exceeds the Eddington luminosity. For ONeMg-WD TDEs, the high accretion rate can trigger substantial neutrino production, slightly reducing the EM output while producing appreciable MeV neutrino emission. If the disk is misaligned, disk precession may also generate a GW signal. The dashed lines indicate a possible jet power $\simeq 0.01\dot{M}_{\rm fb}c^2$ assuming a radiative efficiency of $\sim 0.01$.}
\label{fig:multi_LC}
\end{figure*}

WD TDEs are potential multi-messenger transients, producing gravitational waves (GWs), electromagnetic (EM) radiation, and (in some cases) neutrinos. Here we summarize the main observational channels relevant to time-domain searches. Figure~\ref{fig:multi_LC} shows representative EM, GW, and neutrino power histories for WD TDEs across different IMBH masses and WD types.

As the WD approaches pericenter and is disrupted by the IMBH, the rapidly varying quadrupole moment generates a GW burst. Such signals may be detectable by next-generation detectors (e.g., ALIA, BBO, DECIGO, and ET), with detectability improving for larger $M_{\rm h}$ and $M_*$. 

After disruption, roughly half of the debris remains bound and returns to pericenter. Stream self-intersections and stream--disk collisions dissipate orbital energy and enable disk formation (Section~\ref{subsec:Disk_Formation}); here we assume rapid and efficient circularization.

If the disk forms promptly ($\sim 10$--$1000\,\rm s$ after disruption), accretion powers the EM emission. Continued fallback feeds the disk; with efficient viscous transport and wind mass loss, the disk accretion rate can closely track the fallback rate (Section~\ref{subsec:disk_evolution}).

The thermal disk emission only mildly exceeds the Eddington luminosity and peaks at $0.1$--$1\,\rm keV$, making it readily detectable in X-rays with instruments such as Swift/XRT, EP-FXT, Chandra, and XMM-Newton. In particular, the wide-field EP-WXT survey could capture this emission for sufficiently nearby events ($z \lesssim 0.1$). 

At extreme accretion rates ($\gtrsim 10^9\,\dot{M}_{\rm Edd}$), $e^{\pm}$ pair annihilation in the disk can power neutrino emission. This channel is most promising for $M_{\rm h}\simeq 10^3\,M_{\odot}$ and a high-mass ONeMg WD, yielding $\dot{E}_{\nu} \simeq 10^{46}\,{\rm erg\,s^{-1}}$.

However, these neutrinos are at MeV energies and are therefore difficult to detect: IBD in JUNO and Super-Kamiokande would be sensitive only to sources within $\sim 0.1\,\rm kpc$.

Higher-energy neutrinos may offer better prospects. In Chen et al.
(in preparation), we will explore high-energy neutrinos and non-thermal particles produced by acceleration in disk and jet/wind systems.

A potentially more luminous EM component is jet emission, which can dominate when the jet is aligned with the observer, as suggested by the WD-TDE candidate EP250702a \citep{li_fast_2026}. Assuming a jet power $\simeq 0.01\,\dot{M}_{\rm fb}c^2$ and a radiative efficiency of $\sim 0.01$, we plot the jet-power history in Figure~\ref{fig:multi_LC}. For on-axis observers, relativistic beaming boosts the isotropic-equivalent luminosity by a factor of $\sim \Gamma^2$ (where $\Gamma$ is the bulk Lorentz factor) and shifts the spectrum to higher energies, enabling detection by $\gamma$-ray missions such as \textit{Fermi} and SVOM. We will explore the jet emission in detail Chen et al. (in preparation).

Furthermore, as the jet/wind propagates outward and interacts with the ambient medium, it can produce late-time synchrotron emission from radio to X-ray bands.

\section{Discussion}
\label{sec:discussion}
\subsection{Disk Formation}
\label{subsec:Disk_Formation}
After disruption, the debris streams can self-interact; such collisions are widely regarded as the primary pathway to circularization and disk formation \citep{Dai_Soft_2015,bonnerot_long-term_2017}. Below we briefly summarize the expected circularization picture for WD-TDEs.

After pericenter passage, relativistic apsidal precession by an angle $\phi \sim 3\pi R_{\rm g}/R_{\rm p}$ causes the outgoing and infalling streams to collide near apocenter. The corresponding specific energy dissipation can be estimated as
\begin{equation} \label{eq:dE_I}
    \begin{split}
    \Delta \epsilon_{\rm I} &= \left(\frac{9\pi^2}{16c^4}\right) e_{\rm mb}^2 \left(\frac{GM_{\rm h}}{R_{\rm p}}\right)^3 \\ 
    &\simeq 10^{16} \beta^3 M_3^2 m_{*,0} r_{*,-2}^{-3}\ {\rm erg/g},
    \end{split}
\end{equation}
which is only a small fraction of the dissipation required to circularize at $R_{\rm c}$, i.e., $GM_{\rm h}/(2R_{\rm c}) \simeq 10^{18} \beta M_3^{2/3} m_{*,0}^{1/3} r_{*,-2}^{-1}\ {\rm erg/g}$. Thus, a single self-intersection is unlikely to complete circularization. Subsequent stream--stream and stream--disk interactions may instead provide sustained dissipation, powering radiation and driving both inflows and outflows \citep{Lu_Self_2020,steinberg_stream-disk_2024,meza_radiation-magnetohydrodynamic_2025}.

For a misaligned stellar orbit, Lense--Thirring precession of the debris stream can induce an offset self-intersection, reducing the dissipated energy \citep{Jiang_PROMPT_2016} and potentially prolonging disk formation. In more extreme cases, if the stream is sufficiently thin and the offset is large, the stream segments may miss each other at the first self-intersection, delaying stream--stream collisions and hence disk formation \citep{dai_impact_2013,Guillochon_A_2015}.

During circularization, the dissipated orbital energy is partitioned into thermal and kinetic components. A fraction of the gas may be unbound in an outflow, while diffusing radiation from the shocked material can contribute to the early-time TDE light curve \citep{steinberg_stream-disk_2024,meza_radiation-magnetohydrodynamic_2025}.

The remaining bound material near the IMBH can assemble into a nascent accretion disk, with continued dissipation heating the disk. Disk formation in TDEs remains an open problem, and robust radiation-(magneto)hydrodynamic simulations are needed to establish when and how circularization proceeds.

\subsection{Disk Evolution}
\label{subsec:disk_evolution}
In the previous section, we studied the disk structure under the simplifying assumption that the accretion rate follows the fallback rate. Here we model the global disk evolution to quantify the relation between the accretion rate and the fallback supply. We show that $\dot{M}_{\rm acc}$ closely tracks $\dot{M}_{\rm fb}$ for the parameters of interest. We also follow the evolution of the disk outer radius and discuss how it affects the emergent EM spectrum.

Motivated by \citet{Shen_EVOLUTION_2014}, we compute the global evolution by solving coupled conservation equations for the disk mass $M_{\rm d}$ and angular momentum $J_{\rm d} \simeq M_{\rm d} \sqrt{GM_{\rm h} R_{\rm out}}$:
\begin{equation} \label{eq:Md_Jd}
    \begin{cases}
        \frac{dM_{\rm d}}{dt} = \dot{M}_{\rm fb} - \dot{M}_{\rm acc}, \\
        \frac{dJ_{\rm d}}{dt} = j_{\rm fb} \dot{M}_{\rm fb} - \dot{J}_{\rm w}.
    \end{cases}
\end{equation}
Here $\dot{M}_{\rm acc} \simeq M_{\rm d}/t_{\nu}(R_{\rm out})$ represents the net mass loss from the disk through BH accretion and winds. The wind also carries away angular momentum; we approximate the loss rate as
$\dot{J}_{\rm w} \simeq \int \sqrt{GM_{\rm h}R} \, (d\dot{M}_{\rm in}/dS)\,dS \simeq \frac{s}{s+1/2} \dot{M} \sqrt{GM_{\rm h}R_{\rm out}} \simeq \frac{s}{s+1/2} \frac{J_{\rm d}}{t_{\nu}(R_{\rm out})}$.
The viscous timescale is $t_{\nu}=R^2/\nu \simeq \alpha^{-1} (GM_{\rm h}/R^3)^{-1/2} f^{-1}$. The disk is fed by fallback material with specific angular momentum $j_{\rm fb} \simeq \sqrt{GM_{\rm h}R_{\rm c}}$.

We adopt the simulated fallback rate of a polytropic star with index $5/3$ from \citet{Guillochon_Hydrodynamical_2013} and rescale it to the IMBH--WD TDE case using their Equations (A1) and (A2).

\begin{figure}
\centering
 \includegraphics[scale=0.5]{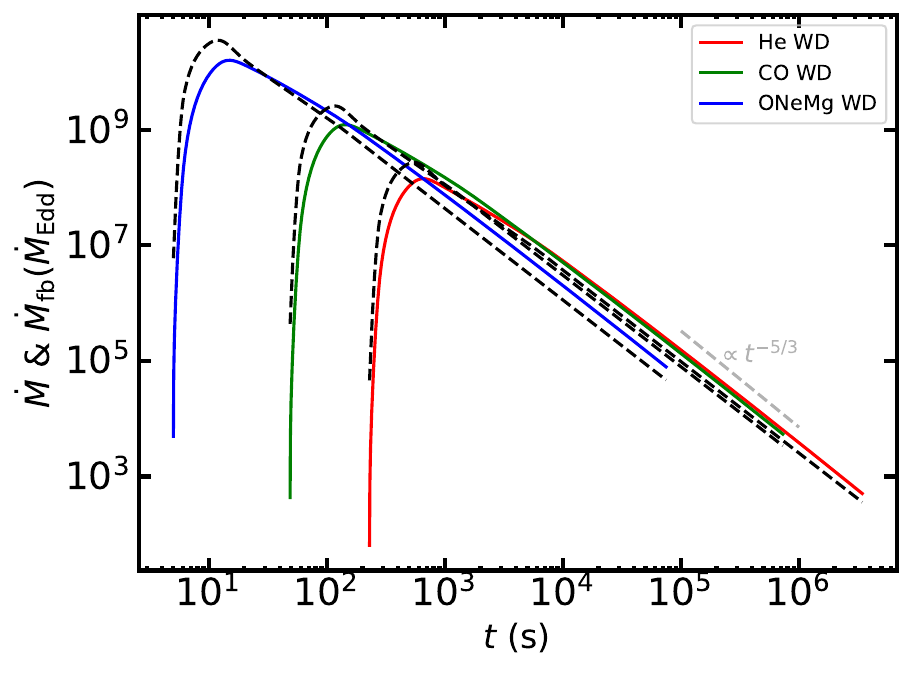}
    \caption{Accretion rate $\dot{M}_{\rm acc}$ (solid) compared with the fallback rate $\dot{M}_{\rm fb}$ (dashed). The accretion rate is calculated by Equation (\ref{eq:Md_Jd}) with $M_{\rm h} = 10^3\ M_{\odot}$ and $\alpha = 0.1$. The two closely coincide because the disk rapidly adjusts through viscous transport while winds regulate the net mass flow.}
\label{fig:Md}
\end{figure}

\begin{figure}
\centering
 \includegraphics[scale=0.5]{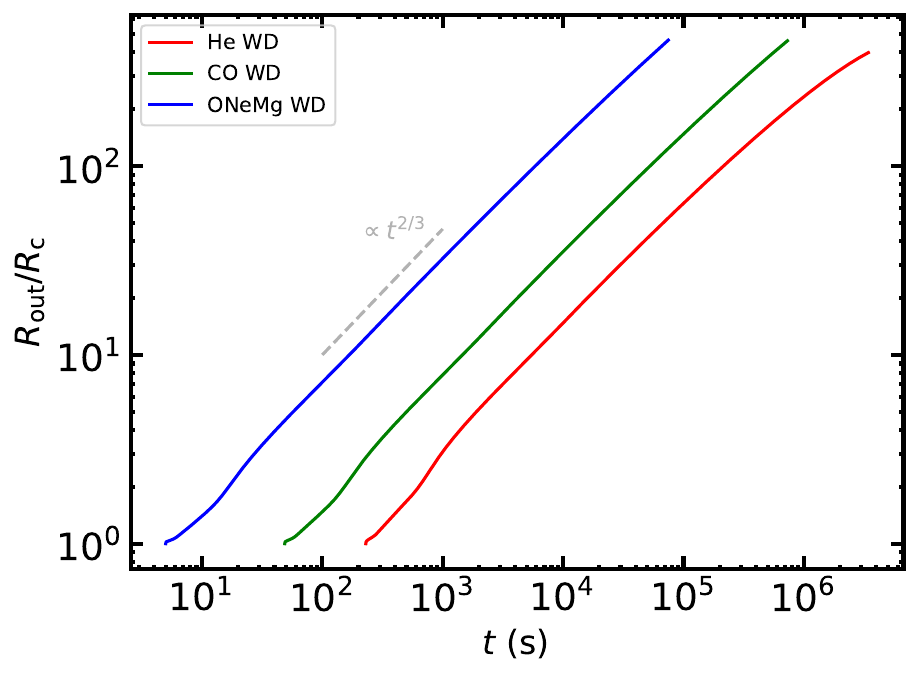}
    \caption{Evolution of the disk outer radius, calculated by Equation (\ref{eq:Md_Jd}) with $M_{\rm h} = 10^3\ M_{\odot}$ and $\alpha = 0.1$.. The disk initially forms at $R_{\rm c}$ and then spreads viscously and slowly, following a similarity solution $R_{\rm out} \propto t^{2/3}$.}
\label{fig:Rout}
\end{figure}

The evolution of the accretion rate $\dot{M}_{\rm acc}$ and the disk size $R_{\rm out}$ is shown in Figures~\ref{fig:Md} and \ref{fig:Rout}, respectively. Owing to efficient viscous transport and substantial wind mass loss, $\dot{M}_{\rm acc}$ closely tracks the fallback rate.

The disk initially forms near $R_{\rm c}$ and subsequently spreads viscously. The radial expansion is modest---even at the peak accretion rate, $R_{\rm out}$ remains close to $R_{\rm c}$---but its growth is nevertheless broadly consistent with the similarity solution, $R_{\rm out} \propto t^{2/3}$ for $\nu \propto R^{1/2}$ \citep{Lynden-Bell_The_1974,hartmann_accretion_1998}.

In Section~\ref{subsec:EM}, we found that a larger disk shifts the thermal spectrum to lower energies. We therefore expect the EM spectrum to soften at late times as the disk spreads.

The secular evolution of the BH (mass and spin) can, in principle, feed back on the disk structure and emission---an effect that can be important for GRB engines, where the accreted mass may be comparable to the BH mass. By contrast, for IMBH--WD TDEs with $M_{\rm h} \gg M_*$, accretion changes the BH mass and spin only weakly and can be neglected.

\subsection{EM emission from Disk Winds}
\label{subsec:wind_emission}
In this work we focus on the EM emission from the accretion disk. In practice, disk winds can substantially reprocess this radiation. Indeed, optical/UV emission in many TDEs has been attributed to an extended wind photosphere \citep{Strubbe_Optical_2009,Metzger_A_2016,dai_unified_2018}. Here we briefly comment on this possibility.

As discussed in Section~\ref{subsec:disk_wind}, a fraction of the inflowing gas can be driven into a disk wind, which can modify the observed emission in two main ways.

First, depending on its geometry, the wind can obscure the disk and intercept a large fraction of the disk luminosity. Simulations suggest a funnel-shaped outflow with lower density toward the polar region, allowing some disk radiation to escape along the poles \citep{ohsuga_supercritical_2005,mckinney_three-dimensional_2014,jiang_global_2014,sadowski_global_2015,dai_unified_2018}.

Second, the wind can carry substantial internal energy at launch. Its electron-scattering optical depth is
\begin{equation} \label{eq:tau}
    \tau \sim \int^{\infty}_{R_{\rm edge}} \kappa_{\rm es}\rho\, dR \sim \frac{\kappa_{\rm es} M_{\rm w}}{4\pi v_{\rm w} R_{\rm edge}t},
\end{equation}
which is initially $\gg 1$. Here $R_{\rm edge} \sim v_{\rm w}t$ is the outer edge of the outflow and $M_{\rm w}$ is the total wind mass. Photons are then trapped and the wind cools primarily via adiabatic expansion. As $\tau$ decreases to $\sim c/v_{\rm w}$, radiation begins to diffuse out \citep{Strubbe_Optical_2009}, defining an approximate photospheric radius
\begin{equation} \label{eq:R_edge}
    R_{\rm edge} \simeq \sqrt{\frac{\kappa_{\rm es} M_{\rm w}v_{\rm w}}{4\pi c}} \simeq 10^{15}\left(\frac{v_{\rm w}}{0.3c}\frac{M_{\rm w}}{0.6\,M_{\odot}}\right)^{1/2}{\rm cm}.
\end{equation}
The corresponding timescale is 
\begin{equation}
    t_{\rm edge} \simeq R_{\rm edge}/v_{\rm w} \sim 3\,{\rm day}.
\end{equation}
By this stage, most of the initial thermal energy has been converted into bulk kinetic energy, and the diffusive component could contribute to late-time optical emission.

As a result, wind emission in WD--TDEs can differ qualitatively from that in typical (main-sequence) TDEs, which evolve on month-long timescales: in the latter, photons can diffuse out of the outflow relatively quickly. In WD--TDEs, by contrast, the photon escape time $t_{\rm edge}$ can be much longer than the fallback time $t_{\rm fb}$, so the wind is unlikely to dominate the early emission and may instead contribute at late times, potentially alongside jet afterglow emission.

\subsection{Disk emission of EP250702a}
\label{subsec:EP250702a}
Recently, Einstein Probe (EP) reported a WD--TDE candidate, EP250702a \citep{li_fast_2026}, associated with an ultra-long GRB \citep[GRB250702BDE][]{cheng_ep250702a_2025,levan_day-long_2025}. Here we briefly discuss this source, focusing on the possible emergence of a thermal disk component.

At early times the emission is dominated by the jet. After $\sim 1$ day, the X-ray light curve begins to fade and the spectrum softens. At later times ($\sim 10$ days), the decay becomes shallower, which may indicate the emergence and gradual dominance of a thermal component that fades more slowly than the jet.

A joint fit to the simultaneous Chandra and EP-FXT observations obtained at $\sim 17$ days yields a power-law plus blackbody component, with $kT \simeq 0.17\,\mathrm{keV}$ and luminosity $\sim 9.8\times 10^{44}\,\mathrm{erg\,s^{-1}}$.

For an inferred IMBH mass of $\sim 7.5\times 10^{4}\,M_{\odot}$, the temperature is consistent with our model (Figure~\ref{fig:diskEM_spec}, where the spectrum peaks near $\sim 0.2\,\mathrm{keV}$). The high luminosity suggests a relatively extended disk, $R_{\rm out} \gtrsim 100\,R_{\rm in}$, which in our calculations yields $L_{\gamma}\sim 10^{44}\,\mathrm{erg\,s^{-1}}$.

The inferred luminosity is still somewhat higher than our model prediction. A likely explanation is that, at the reported redshift ($z\sim 1$), the thermal component is difficult to isolate cleanly and may be partially contaminated by residual jet emission.

Future detections of closer WD--TDEs, and/or improved soft X-ray coverage with more sensitive instruments, should enable more precise measurements of the thermal component.

\section{Conclusion}
\label{sec:conclusion}
In this paper we developed a baseline central-engine model for WD--TDEs by IMBHs, focusing on the steady, vertically integrated structure of the newly formed accretion disk and the associated multi-messenger signals (EM, GWs, and neutrinos). Our main results are as follows:
\begin{itemize}
    \item \textbf{Disk physics.} We compute steady disk solutions including viscous heating, advection, photon diffusion, wind-driven mass loss, nuclear heating, and magnetic pressure support. For the extremely super-Eddington fallback rates expected in WD--TDEs, the inner disk can reach $T\gtrsim 10^9\,\mathrm{K}$, where abundant $e^{\pm}$ pairs form and optically thin neutrino cooling from pair annihilation becomes relevant; nevertheless, the global disk remains predominantly advection dominated across a broad range of accretion rates.

    \item \textbf{Thermal EM emission.} The predicted thermal disk emission is only weakly dependent on the fallback rate: the luminosity typically exceeds the IMBH Eddington luminosity by at most a factor of a few, and the spectrum peaks at $\sim 0.1$--$1\,\mathrm{keV}$. Nearby events ($z\lesssim 1$) should therefore be detectable in soft X-rays with current wide-field and pointed facilities. The spectrum is sensitive to the disk outer radius $R_{\mathrm{out}}$, with larger disks producing softer emission and slightly higher luminosities. For the most extreme ONeMg WD disruptions, neutrino cooling can suppress the high-energy tail of the thermal spectrum.

    \item \textbf{Neutrino emission.} Although neutrino cooling does not dominate the global energy budget, the hottest inner regions can generate a short-lived burst of MeV neutrinos, reaching $\dot{E}_{\nu}\sim 10^{47}\,\mathrm{erg\,s^{-1}}$ for ONeMg debris at the highest accretion rates around a $10^3\,M_{\odot}$ IMBH. Comparing the predicted spectra to background-limited sensitivities implies detectability only for very nearby (Galactic) events (typical horizons $\lesssim 1\,\mathrm{kpc}$) with JUNO, Hyper-K, and IceCube-Gen2. CO and He WD--TDEs can also produce neutrinos, but at much lower luminosities.

    \item \textbf{Gravitational-wave signals.} We estimate the GW burst from one-off WD--TDEs on parabolic orbits and find characteristic frequencies in the $\sim 0.1$--$1\,\mathrm{Hz}$ band, placing them near the most sensitive range of proposed decihertz missions (e.g., DECIGO/BBO/ALIA), with the potential for cosmological detection and advance triggers for EM follow-up. We also explore GWs from a precessing WD--TDE disk; this signal is much weaker, with a detection horizon $\lesssim 1\,\mathrm{Mpc}$ for the most compact, massive (ONeMg) disks.

    \item \textbf{Connection to EP250702a.} The reported thermal component in EP250702a is broadly consistent with disk temperatures ($\sim 0.1$--$1\,\mathrm{keV}$) expected for IMBH masses $\sim 10^{4}$--$10^{5}\,M_{\odot}$, while its high inferred luminosity may indicate a relatively extended disk and/or residual jet contamination at $z\sim 1$. Future detections of closer WD--TDEs, and improved soft X-ray coverage, will enable more robust isolation of the thermal disk component.
\end{itemize}

Looking ahead, a time-domain multi-messenger view of WD--TDEs provides a powerful route to constraining both the disruption dynamics and the physics of super-Eddington accretion. Decihertz GW detections could deliver early-warning localizations and enable rapid, multi-epoch soft X-ray follow-up to track the expected evolution from an early jet-dominated phase (when present) to a longer-lived thermal disk component on day--week timescales, while contemporaneous optical/UV and radio monitoring can constrain reprocessing, outflows, and late-time afterglow. Although MeV neutrinos are likely limited to rare Galactic events, non-detections for extragalactic candidates can still place upper limits on the hottest inner-disk conditions; moreover, a detectable GW signal from disk precession would provide an additional, independent diagnostic of disk geometry. In a companion paper (Chen et al., in preparation), we will explore the multi-messenger signatures of jets, coronae, and disk winds, extending WD--TDE studies toward the ultra-high-energy regime.

\appendix
\counterwithin{figure}{section}

\twocolumngrid

\section{Disk ingredients}
\label{appendix:ingredients}
\subsection{Viscosity} \label{subsec:viscosity}
We summarize the viscosity prescription and its connection to the mass accretion rate. For a steady-state disk, the relationship between the kinematic viscosity $\nu$ and the local mass inflow rate $\dot{M}_{\rm in}(R)$ at radius $R$ is
\begin{equation} \label{eq:nu_macc}
    \nu\Sigma = \frac{\dot{M}_{\rm in}}{3\pi}\, f,
\end{equation}
where $\Sigma$ is the surface density and $f$ is a general-relativistic correction factor for a Kerr spacetime \citep{page_disk-accretion_1974,riffert_relativistic_1995}. Following \cite{lei_hyperaccreting_2017}, we write $f=D\sqrt{BC/A^3}$, where $A$, $B$, $C$, and $D$ are given by their Equations~(5--9). The factor $f$ depends on radius, the BH mass $M_{\rm h}$, and the dimensionless spin parameter $a_{\bullet}$. The inner disk edge $R_{\rm in}$ is taken to be the marginally stable orbit, where $f=0$ (the standard zero-torque boundary condition). Far from the inner edge ($R\gg R_{\rm in}$), one recovers the Newtonian limit $f\simeq 1-\sqrt{R_{\rm in}/R}$ \citep{Frank_Accretion_1985}.

It is now widely recognized that the zero-torque inner-boundary condition can break down in super-Eddington disks. Global MHD simulations show that magnetic field lines can connect the disk to the plunging region, yielding a non-zero magnetic torque \citep{mckinney_three-dimensional_2014,jiang_global_2014,jiang_super-eddington_2019}. Such a torque can enhance the X-ray emission and extend the spectrum to higher energies. Nevertheless, for simplicity we adopt the standard zero-torque condition in our analytical calculations.

We adopt the standard $\alpha$-viscosity prescription \cite{Shakura_viscosity_1973},
\begin{equation} \label{eq:nu}
    \nu = \frac{2}{3}\,\alpha\,\frac{P}{\rho\,\Omega},
\end{equation}
where $P$ and $\rho$ are the midplane total pressure and density, respectively, and
$\Omega = \sqrt{GM_{\rm h}C/(BR^3)}$ is the relativistic Keplerian angular velocity at radius $R$. We assume $\alpha$ is constant with $\alpha\simeq 0.1$--$1$; hereafter we set $\alpha=0.1$ and $a_{\bullet}=0.95$.

The midplane density is related to the surface density via $\rho\simeq \Sigma/(2H)$, where the disk scale height $H$ follows from vertical hydrostatic equilibrium:
\begin{equation} \label{eq:H}
    H \simeq \Omega^{-1}\left(\frac{P}{\rho}\right)^{1/2}.
\end{equation}

The viscous heating rate per unit surface area is
\begin{equation} \label{eq:Q_vis}
    Q_{\rm vis} = \frac94\,\nu\Sigma\Omega^2
    = \frac32\,\alpha\,\Omega\,\Sigma\,\frac{P}{\rho}
    = \frac{3}{4\pi}\,\Omega^2\,\dot{M}_{\rm in}\,f,
\end{equation}
where the final equality follows by combining Equation~(\ref{eq:nu_macc}) with the steady-state relation
$\dot{M}_{\rm in}=2\pi R\,v_{\rm R}\,\Sigma = 3\pi\nu\Sigma\,f^{-1}$ under the zero-torque inner boundary condition. Here $v_{\rm R}$ is the radial drift velocity.

\subsection{Pressure} \label{subsec:pressure}
We adopt the following expression for the total midplane pressure in WD--TDE disks:
\begin{equation} \label{eq:P_tot}
    P_{\rm tot} = P_{\rm rad} + P_{\rm gas} + P_{\rm B} = P_{\rm rad} + P_{e^-} + P_{e^+} + P_{\rm B}.
\end{equation}
Here 
\begin{equation}
    P_{\rm rad}=(1/3)aT^4,
\end{equation}
where $T$ is the midplane temperature and $a$ is the radiation constant. $P_{\rm gas}$ and $P_{\rm B}$ are the gas and magnetic pressures, respectively. Under the conditions relevant to our models, the gas pressure is dominated by electrons and positrons, so that $P_{\rm gas}\simeq P_{e^-}+P_{e^+}$. Below we motivate each term.

The relative importance of these contributions depends on composition and temperature. White dwarfs span He, CO, and ONeMg compositions. In a fully ionized plasma the pressure is supplied by ions, electrons, and radiation; at sufficiently high temperatures ($\gtrsim 10^8$ K), electron--positron pairs also contribute. Neutrino (and antineutrino) production could in principle add a pressure component if neutrinos were trapped and thermalized. In our parameter space, however, the neutrino optical depth is $\ll 1$, so neutrinos escape promptly and do not contribute appreciably to the pressure (Section~\ref{subsec:neutrino_generation}).

At $T\gtrsim 10^{7}$--$10^{8}$ K the material is fully ionized, and the electron pressure exceeds the ion pressure. The electron/positron pressure depends on both number density and temperature; an exact treatment requires integrating the Fermi--Dirac distribution. Electrons arise from ionization, and at $T\gtrsim 10^9$ K in the inner disk, high-energy photons produce abundant pairs through the Breit--Wheeler process ($\gamma+\gamma\to e^-+e^+$). We present the detailed calculation in Appendix~\ref{appendix:Pressure_e}.

In summary, the electron/positron component is non-degenerate in our models and may be written as a sum of contributions from free electrons ($n_{e0}k_{\rm B}T$) and pair-produced leptons ($n_{e,\psi=0}k_{\rm B}T$):
\begin{equation} \label{eq:elec_pressure}
    \begin{split}
    &P_{e^-} \simeq (n_{e0}+n_{e,\psi=0})k_{\rm B} T, \\
    &P_{e^+} \simeq n_{e,\psi=0}k_{\rm B} T.
    \end{split}
\end{equation}
Equation~(\ref{eq:elec_pressure}) is valid in both the non-relativistic and relativistic regimes, and therefore applies across the range of WD--TDE disk conditions considered here.

Magnetic fields can be amplified by the dynamo action associated with the magnetorotational instability (MRI; \citealt{balbus_instability_1998}). The field strength and geometry in WD--TDE disks remain uncertain. We follow the common expectation that the field saturates when the Alfv\'{e}n speed $v_{\rm A}$ approaches $\sqrt{v_{\rm K}c_{\rm gas}}$ \citep{pessah_stability_2005,begelman_accretion_2007}, where $c_{\rm gas}=\sqrt{k_{\rm B}T/(\mu m_{\rm p})}$ is the gas sound speed. The corresponding magnetic pressure is then
\begin{equation} \label{eq:P_B}
    P_{\rm B} = \rho v_{\rm A}^2 \simeq \rho v_{\rm K}c_{\rm gas}.
\end{equation}
\citet{huang_black_2023} found that magnetic pressure can play an important role in shaping the disk structure at low accretion rates.

In WD--TDE disks the high temperatures typically make radiation pressure dominant. Using vertical support $H\sim R$ (Equation~\ref{eq:H}) implies $P/\rho\sim v_{\rm K}^2$, so $P_{\rm B}/P\sim c_{\rm gas}/v_{\rm K}\ll 1$. For $T\lesssim 10^9\,\mathrm{K}$, gas pressure is mainly due to free electrons ($P_{\rm gas}\simeq \rho c_{\rm gas}^2$), giving $P_{\rm rad}>P_{\rm B}>P_{\rm gas}$. When $T\gtrsim 10^9\,\mathrm{K}$, pair production increases the lepton density and can raise $P_{\rm gas}$ to be comparable to $P_{\rm rad}$. In both regimes the magnetic pressure remains sub-dominant and does not significantly modify the disk structure (Section~\ref{subsec:Disk_structure}); we nonetheless retain $P_{\rm B}$ for completeness.

Finally, as accretion proceeds, magnetic flux can be advected inward and accumulate near the BH, potentially producing a magnetically arrested disk (MAD) in the innermost region \citep{narayan_magnetically_2003,tchekhovskoy_efficient_2011,mckinney_efficiency_2015}. A MAD can efficiently power relativistic jets via the Blandford--Znajek mechanism \citep{blandford_electromagnetic_1977}. We do not model the MAD state here, focusing instead on a standard disk with moderate magnetization; we defer an exploration of MAD disks to Paper~II (Chen et al., in preparation).

\subsection{Neutrino Generation and Optical Depth in Disk} \label{subsec:neutrino_generation}
Electron pairs within the disk can undergo annihilation, producing neutrinos and anti-neutrinos through weak interactions ($e^- + e^+ \to \nu_i + \overline{\nu}_i$), where $i$ denotes electron-type neutrinos ($\nu_e$, $\overline{\nu}_e$) and heavy-lepton neutrinos ($\nu_\mu$, $\overline{\nu}_{\mu}$, $\nu_\tau$, $\overline{\nu}_\tau$) \citep{itoh_neutrino_1989,popham_hyperaccreting_1999,di_matteo_neutrino_2002}.

Additionally, nuclear burning processes such as electron capture, proton-proton chain reactions, and $\beta$ decay, contribute minor amounts of $\nu_e$/$\overline{\nu}_e$. Given that neutrinos are predominantly produced through pair annihilation, we disregard these minor contributions in this study. The detailed calculation of pair annihilation is outlined in Appendix \ref{appendix:Neutrino_pair}.

In the typical central engines of GRBs, characterized by high accretion rates and extreme temperatures $\gtrsim 10^{10}\ {\rm K}$, nearly all electron pairs are relativistic, resulting in a neutrino production rate $\dot{q}_{\nu \overline{\nu}} \propto T^9$ \citep[see the application in ][]{popham_hyperaccreting_1999}. However, the WD-TDE disk under consideration spans a wide temperature range, encompassing both non-relativistic and relativistic regimes. Consequently, a generalized formula is necessary to describe the neutrino production rate across this broad parameter space.

\cite{itoh_neutrino_1989} computed the neutrino production rate over a wide density-temperature range (their Equation (18)). We have derived a simplified formula to approximate their results for the density-temperature regime pertinent to WD-TDE scenarios:
\begin{equation} \label{eq:q_ee}
    \begin{split}
    \dot{q}_{\nu \overline{\nu}} &\simeq 3.8\times 10^{22} \left[\left(\frac{T}{6 \times 10^9\ \rm K}\right)^{4.5}+\left(\frac{T}{6 \times 10^9\ \rm K}\right)^9\right] \\
    &\times \ue^{-2\frac{m_ec^2}{k_{\rm b}T}}\ {\rm erg\ s^{-1}\ cm^{-3}}.
    \end{split}
\end{equation}
Here, $T^{4.5}$ and $T^9$ correspond to the non-relativistic and relativistic regimes, respectively (see Appendix \ref{appendix:Neutrino_pair}). We also incorporate an exponential cutoff for low temperatures, which arises from the pair number density cutoff at low temperatures (see Equation (\ref{eq:n_epm})).

The generated neutrinos and anti-neutrinos may interact with other particles within the disk. We now estimate the optical depth for neutrinos.

The inverse process of electron-positron pair annihilation corresponds to absorption. The interaction of neutrinos with one another results in an optical depth \citep{di_matteo_neutrino_2002,kohri_neutrino-dominated_2005} $\tau_{a,\nu \overline{\nu}} \simeq \dot{q}_{\nu \overline{\nu}}H/(7/2\sigma T^4)$, which is approximately $\sim10^{-4}$ for $T \simeq 6 \times 10^9$ K and $H \sim R_{\rm in}$. Other processes, such as neutrino/anti-neutrino absorption via bremsstrahlung, the inverse process of plasmon decay, and $\nu_e/\overline{\nu}_e$ absorption via neutron \citep{kohri_neutrino-dominated_2005}, can also absorb neutrinos. However, in the WD-TDE disk, the scarcity of free nucleons and relatively low density render these processes negligible.

For disks dominated by heavy element compositions in WD-TDEs, another potential source of opacity is Coherent Elastic Neutrino-Nucleus Scattering, which is significant for neutrinos with energies smaller than the atomic energy. The corresponding optical depth is $\tau_{s,\nu A} \sim n_{\rm A} \sigma_{\nu A} H$, where $n_{\rm A}\simeq \rho/(10m_{\rm p})$ is the atomic number density, and the cross section is given by $\sigma_{\nu A} \simeq 4 \times 10^{-44} (T/6\times 10^9\ {\rm K})^2\ {\rm cm^2}$. Substituting typical values $T \simeq 6 \times 10^9\ {\rm K}$, $\rho \simeq 10^4\ {\rm g\ cm^{-3}}$ and $H\sim R_{\rm in}$, we obtain $\tau_{s, \nu A} \sim 10^{-8}$.

Therefore, the opacity for neutrinos in the disk is exceedingly low. This contrasts with GRB disks, which feature high temperatures ($\gtrsim 10^{10}$ K) and an abundance of free nucleons, leading to frequent neutrino interactions with other particles, such as other neutrinos and free nucleons, and resulting in a thermal distribution of neutrinos in the plasma. In the WD-TDE disk, generated neutrinos can escape immediately without absorption or scattering, implying that the emitted neutrinos should be non-thermal. We will calculate the neutrino spectrum in Section \ref{subsec:Neutrino_radiation}.

The corresponding neutrino cooling rate per surface area for an optically thin scenario is given by
\begin{equation} \label{eq:Q_v}
    Q_{\nu} \simeq \dot{q}_{\nu \overline{\nu}} H.
\end{equation}
This term is less significant than advection cooling for WD-TDEs, unlike in NDAFs. However, this process can generate thermal neutrino emission (refer to Section \ref{subsec:Neutrino_radiation}) and contributes a small amount of energy to the relativistic jet.

\subsection{Nuclear Burning} \label{subsec:nuclear}
In the high temperature environment ($\sim 10^9$ K), the material of the disk undergoes a sequence of nuclear reactions. These reactions lead to the formation of heavier elements and the release of energy. Specifically, in the disks composed of non-degenerate gas, nuclear reactions proceed in a stable manner, without thermonuclear runaway. 

The concept of a nuclear accretion disk has been previously explored in the context of the merger between a white dwarf and another compact object, such as neutron star, or a stellar-mass black hole \citep{metzger_nuclear-dominated_2012,margalit_time-dependent_2016,dan_structure_2014}, and accreting BH in AGN disk \citep{tang_nuclear_2024}. Understanding these reactions is crucial for determining the elemental composition distribution within the disk, as well as the contamination of the disk environment by disk winds. Furthermore, thermonuclear reactions can inject a substantial amount of energy into the disk.

In this paper, we incorporate the energy generated by nuclear reactions into our investigation of the properties of WD-TDE disks.

We consider three WD compositions and discuss the dominant nuclear reactions expected in each WD--TDE disk. Because the peak fallback (and hence accretion) rate depends on WD type (Eq.~\ref{eq:dotM_peak}), the inner-disk temperature also differs, leading to distinct thermonuclear burning regimes.

\begin{itemize} 
\item \textbf{He WD-TDE}: The gas in these disks mainly consists of $\rm ^4He$. The temperature of their inner disk is $\sim 10^8 -10^9$ K. The primary and dominant energy-generating reaction is the triple-alpha fusion of helium into carbon ($\rm ^4He + ^4He \rightleftharpoons ^8Be$, $\rm ^8Be + ^4He \to ^{12}C + \gamma$). Immediately after carbon formation, helium nuclei capture onto freshly synthesized carbon ($\rm ^{12}C + ^4He \to ^{16}O + \gamma$) producing oxygen and forming a carbon-oxygen mixture. At higher temperature near $10^9$ K, sequential $\alpha$-captures proceed at moderate rates ($\rm ^{16}O + ^4He \to ^{20}Ne + \gamma$, $\rm ^{20}Ne + ^4He \to ^{24}Mg + \gamma$) and generate neon and magnesium. In the calculation of energy generation, we only consider the triple-alpha fusion ($\rm ^4He + ^4He + ^4He \to ^{12}C + 7.275\ MeV$) for He WD-TDE.

\item \textbf{CO WD-TDE}: The plasma in these disks is composed of roughly equal mass fractions of $\rm ^{12}C$ and $\rm ^{16}O$. The temperature of their inner disk is $\sim 1-3 \times 10^9$ K. The dominant thermonuclear reaction is carbon-carbon fusion ($\rm ^{12}C + ^{12}C \to ^{23}Na + p + \gamma$, $\rm ^{12}C + ^{12}C \to ^{20}Ne + ^4He + \gamma$). This reaction can fully consume carbon steadily. Free protons and $\alpha$-particles from carbon fusion immediately undergo captures and generate other elements (e.g., Mg, O, F). At higher temperature $\gtrsim 2 \times 10^9$ K, oxygen fusion ($\rm ^{16}O + ^{16}O \to ^{28}Si + ^4He$, $\rm ^{16}O + ^{16}O \to ^{31}P + p$, $\rm ^{16}O + ^{16}O \to ^{32}S + \gamma$) is fully ignited and produce Si, S, P. In the calculation of energy generation, we only consider the carbon fusion ($\rm ^{12}C + ^{12}C \to ^{24}Mg + 13.933\ MeV$) and oxygen fusion ($\rm ^{16}O + ^{16}O \to ^{32}S + 16.542\ MeV$)for CO WD-TDE.

\item \textbf{ONeMg WD-TDE}: The plasma in these disks primarily consists of $\rm ^{16}O$, with lesser amounts of $\rm ^{20}Ne$ and $\rm ^{24}Mg$. The temperature of their inner disk is $\sim 3-6 \times 10^9$ K. The dominant reaction is oxygen fusion. In this temperature, the oxygen can be largely consumed and processed into Si, P, S. Photodisintegration becomes significant for Ne and Mg nuclei ($\rm \gamma + ^{20}Ne \to ^{16}O + ^4He$, $\rm \gamma + ^{24}Mg \to ^{20}Ne + ^4He$), and can, to some extent, cool down the disk. However, since photodisintegration occurs rapidly and is confined to a very small radius, we do not consider it in this paper. For the calculation of energy generation, we only consider the oxygen fusion ($\rm ^{16}O + ^{16}O \to ^{32}S + 16.542\ MeV$) for ONeMg WD-TDE.
\end{itemize}

The energy generation rate via nuclear reaction is calculated for different temperatures and densities using the following equation:
\begin{equation} \label{eq:Q_nuc}
    Q_{\rm nuc} \simeq \sum_i n_{1,i} n_{2,i}f_{\rm nuc, i} \tilde{Q}_i H
\end{equation}
where $n_{1,i}$ and $n_{2,i}$ represent the number densities of two isotope particles involved in the $i$-th reaction, and $f_{\rm nuc,i}$ and $\tilde{Q}_i$ are the nuclear reaction rate coefficient and the Q value for the $i$-th reaction, respectively. They are determined using the analytic expressions provided in \citep{caughlan_thermonuclear_1988} \footnote{See the webpage at \url{http://www.nuclear.csdb.cn/data/CF88/}. For the three isotope particles involved in the reaction, $Q_{\rm nuc} \simeq \sum_i n_{1,i} n_{2,i} n_{3,i} f_{\rm nuc, i} \tilde{Q}_i H$}.

In this paper, we do not incorporate the element isotope distribution resulting from nuclear burning. In reality, nuclear burning may result in an element isotope distribution within the disk. As matter accretes to smaller radii, it experiences higher temperatures and burns into increasingly heavier elements. Consequently, as matter accretes toward the inner disk, much of the material may already be exhausted. 

Another effect of isotope distribution is its impact on the composition of the disk wind. Elements synthesized near the mid-plane are transported to the surface, such that the local composition of the wind matches that of the mid-plane. The disk wind composition varies across different radial ranges, potentially imprinting signatures on the outflow's emission, such as the continuum and emission lines in the wind.

The intricate calculations involved in nuclear accretion processes merit further investigation in future research endeavors. For the present study, we have merely considered nuclear burning as an additional source of heating energy. Nevertheless, our findings indicate that the energy generation rate associated with nuclear burning in WD-TDE scenarios is relatively insignificant compared to viscous heating.

\subsection{Disk Wind} \label{subsec:disk_wind}
A highly super-Eddington accretion disk is capable of generating disk winds through the combined effects of radiation and magnetic pressure within the disk structure \citep{stone_hydrodynamical_1999,igumenshchev_two-dimensional_2000}. These winds effectively carry material away from the disk, serving as a mechanism for disk cooling.

At each radial position within the disk, a certain fraction of the accreting mass is ejected via the wind. Consequently, the net mass inflow rate diminishes as we move towards smaller radii. The precise functional form of this decrease remains uncertain, but it is commonly anticipated that the mass inflow rate decreases according to the relation $\dot{M}_{\rm in} \propto R^s$. This relationship has been employed in models of gamma-ray burst (GRB) disks \citep{kumar_mass_2008} and active galactic nucleus (AGN) disks, and its validity has been corroborated by simulations \citep{yuan_numerical_2015,yang_numerical_2021}. Typically, the value of the $s$-index falls within the range $s \sim 0.2- 1$ \citep{shi_comparing_2025}. The terminal velocity of the wind ejected from a radius $R$ is approximated by the local Keplerian velocity, given by $v_{\rm k}\simeq \Omega R$.

The cooling effect of the disk through wind emission can be estimated using the following equation:
\begin{equation} \label{eq:Q_w}
    Q_{\rm w} \simeq \frac12 K v_{\rm k}^2 \frac{d\dot{M}_{\rm in}}{dS}f \simeq \frac{1}{4\pi} sK\Omega^2 \dot{M}_{\rm in}f,
\end{equation}
where $dS = 2\pi RdR$. By comparing this with Equation (\ref{eq:Q_vis}), we find that $Q_{\rm w} \simeq (sK/3) Q_{\rm vis}$, indicating that the energy loss rate due to disk winds is a fraction of the viscous heating rate. For instance, if we assume $s \simeq 0.2$ and $K = 1$, then $Q_{\rm w} \simeq 0.1 Q_{\rm vis}$. 

Despite the seemingly modest wind cooling rate, the reduction in mass inflow rate can significantly decrease the corresponding viscous and nuclear heating rates at smaller radii. The final mass accretion rate onto the IMBH is expressed as $\dot{M}_{\rm BH} \simeq \dot{M} (R_{\rm in}/R_{\rm out})^s$, where $\dot{M} \equiv \dot{M}_{\rm in}(R_{\rm out})$, and $R_{\rm out}$ represents the outer radius of the disk.

The precise disk formation in TDE remains uncertain. It is generally proposed that the debris's orbit will circularize at a radius of approximately $R_{\rm c} \simeq 2R_{\rm p}$ following one or more stream-stream collisions, in accordance with angular momentum conservation principles \citep{Shen_EVOLUTION_2014,bonnerot_long-term_2017}. Then the first return mass will form a small disk at $R_{\rm c}$. Subsequent mass fallback would supply from the disk outer edge. If the outer radius is $R_{\rm out}$ at $R_{\rm out} = 100 R_{\rm in}$, the accretion rate becomes $\dot{M}_{\rm acc} \simeq 0.4 \dot{M}$, and the corresponding mass loss rate via wind is $\dot{M}_{\rm w} \simeq 0.6 \dot{M}$ for $s=0.2$.

The thick disk wind expands adiabatically. Due to its high optical depth, it can initially block emissions from the disk itself. However, if the line of sight is near the polar direction, where the wind is thinner, more photons from the inner disk can escape \citep{dai_unified_2018}.

This type of super-Eddington wind is considered to be the origin of the optical flares observed in TDEs \citep{Strubbe_Optical_2009}. As the heated outflow expands outward, at which point the optical depth becomes thin, the trap photons can escape from the photosphere. We will delve deeper into this topic in Section \ref{subsec:wind_emission}.

In addition, disk winds can remove substantial angular momentum from the flow, which can slow the outward expansion of the disk radius.

\subsection{Photon Opacity and Radiative Cooling} \label{subsec:Photon_opacity}
In a high-temperature plasma, photon opacity is primarily governed by electron Compton scattering. Within the non-relativistic regime, opacity can be approximated using the Thomson electron scattering formula: $\kappa_{\rm es} \simeq \sigma_{\rm T}/(\mu m_{\rm p}) \simeq 0.2\ {\rm cm^2\ g^{-1}}$, where $\sigma_{\rm T}$ represents the Thomson scattering cross-section, and $\mu \simeq 2$ denotes the mean molecular weight per electron for the composition of a WD. Conversely, when electrons attain relativistic velocities, the electron scattering opacity has to be corrected. Furthermore, the creation of electrons and positrons through pair production processes can significantly contribute to opacity, thereby enhancing it.

For precise opacity calculations, it is imperative to incorporate the specific Klein-Nishina correction and account for the energy distribution of electrons and positrons \citep{chin_opacity_1965,svensson_non-thermal_1987}. A thorough analysis and methodology for computing the Rosseland mean opacity were introduced by \cite{buchler_compton_1976}, with \cite{poutanen_rosseland_2017} subsequently providing a more refined fitting formula. We adopt the Rosseland mean opacity for Compton scattering as outlined in the fitting formula presented by \cite{poutanen_rosseland_2017}:
\begin{equation}
    \kappa_{\rm R} \simeq \frac{\sigma_{\rm T}(n_{e^-}+n_{e^+})}{\rho}/\left[1+\left(\frac{12k_{\rm b}T}{m_{\rm e}c^2}\right)^{0.885}\right].
\end{equation}
When $k_{\rm b}T \ll m_{\rm e} c^2$, electrons are non-relativistic, and pair production is negligible, leading to $n_{e^-} \simeq \rho/(\mu m_{\rm p})$. In this scenario, the opacity reverts to the standard case: $\kappa_{\rm R} \simeq \kappa_{\rm es}$.

Our findings indicate that this correction is only significant within a narrow range of $\rho-T$ parameters for ONeMg WD-TDE systems, where the inner disk temperature is sufficiently high to generate a substantial amount of pairs, thereby enhancing opacity to one order of magnitude.

Utilizing the Rosseland mean opacity, we can calculate the radiative cooling rate due to photon escape from the disk surface as follows: 
\begin{equation} \label{eq:Q_rad}
    \begin{split}
        Q_{\rm rad} &= \frac{4acT^4}{3\kappa_{\rm R} \Sigma}.
    \end{split}
\end{equation}
Given the extremely high optical depth in WD-TDE disks, radiative cooling plays a relatively minor role compared to other cooling mechanisms. Nonetheless, it contributes to thermal emission observed in the spectrum.

\section{Electron and positron pressure}
\label{appendix:Pressure_e}
A precise evaluation of the electron/positron pressure requires integrating the full Fermi--Dirac distribution. The number densities of $e^-$ and $e^+$ are
\begin{equation} \label{eq:n_e}
    n_{e^\pm} = \frac{4\pi g_e}{h^3} \int^{\infty}_0 \frac{p^2\, dp}{\ue^{(E-\psi_{e^{\pm}})/k_{\rm b}T}+1},
\end{equation}
where $\psi_{e^{\pm}}$ is the chemical potential, $h$ is Planck's constant, $k_{\rm b}$ is Boltzmann's constant, and $g_e=2$ is the spin degeneracy. The particle energy is
$E=\sqrt{p^2c^2+m_e^2c^4}$. Because $e^{\pm}$ are in chemical equilibrium with radiation through $\gamma+\gamma\to e^-+e^+$ and photons have zero chemical potential, we have $\psi_{e^-}=-\psi_{e^+}$ \citep[see also][]{chen_neutrino-cooled_2007}.

Free (ionization) electrons suppress pair production. For neutral disk material, the exact $e^-$ and $e^+$ number densities are obtained by solving Equation~(\ref{eq:n_e}) together with the charge-neutrality condition
\begin{equation} \label{eq:ne_ne}
    n_{e^-} - n_{e^+} = n_{e0} = \frac{\rho}{\mu m_{\rm p}},
\end{equation}
where $n_{e0}$ is the free-electron number density, $m_{\rm p}$ is the proton mass, and $\mu\simeq 2$ is the mean molecular weight per electron for WD compositions.

\begin{figure}
\centering
 \includegraphics[scale=0.35]{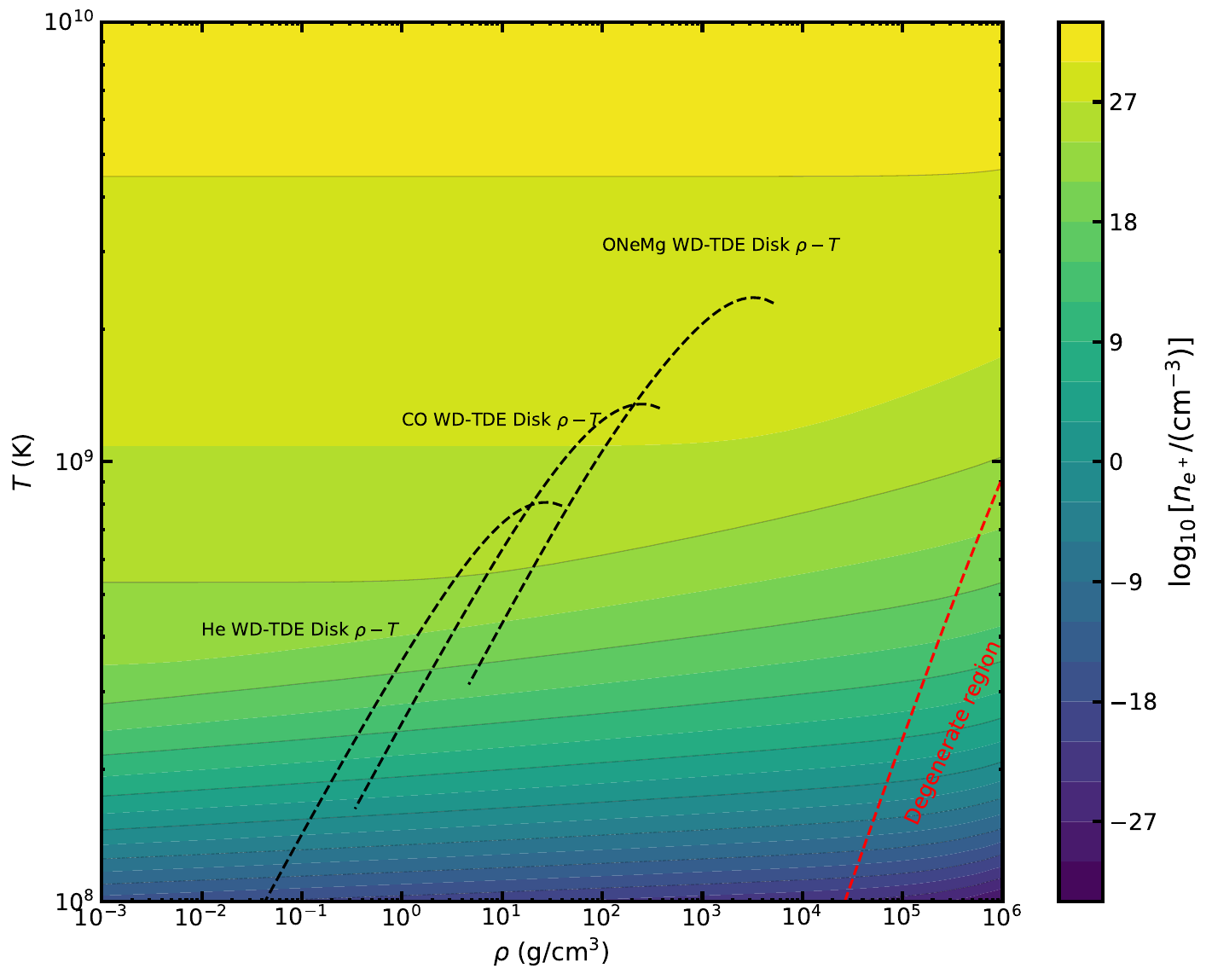}
    \caption{Positron number density $n_{e^+}$ in the $\rho$--$T$ plane. $n_{e^+}$ is highly temperature sensitive and decreases slightly with increasing density due to suppression of pair production by free electrons. The red dashed contour marks the degenerate region, defined by $\psi_{e^-} \gtrsim k_{\rm b}T + m_{\rm e}c^2$. Dashed black lines show representative $\rho$--$T$ profiles for different WD-TDE disks, which remain hot and non-degenerate.}
\label{fig:ne_ne0}
\end{figure}

Figure~\ref{fig:ne_ne0} shows $n_{e^+}$ in the $\rho$--$T$ plane. In the temperature range $10^9$--$10^{10}\,\rm K$ relevant to WD-TDE disks, pair production can generate a substantial number of $e^{\pm}$, and the gas remains hot and non-degenerate.

When the pair-produced electrons dominate over the free electrons ($n_{e,\psi=0}\gg n_{e0}$), the chemical potential approaches $\psi_{e^{\pm}}\simeq 0$. In the non-degenerate limit, the pair number density can then be approximated with a Boltzmann distribution:
{\footnotesize
\begin{equation} \label{eq:n_epm}
    \begin{split}
    &n_{e,\psi=0} \simeq \frac{4\pi g_e}{h^3} \int^{\infty}_0 p^2\, \ue^{-E/(k_{\rm b}T)}\, dp \\
    &\sim \begin{cases}
    2(2\pi m_e k_{\rm b} T/h^2)^{3/2} \ue^{-m_ec^2/(k_{\rm b} T)}, \quad k_{\rm b}T<m_e c^2 \\
    2\pi(2k_{\rm b} T/hc)^3 \ue^{-m_ec^2/(k_{\rm b} T)}, \quad k_{\rm b}T>m_e c^2
    \end{cases} \\
    &\sim \begin{cases}
    1.5 \times 10^{29} \left(\frac{T}{10^9\,\rm K}\right)^{3/2} \ue^{-(6 \times 10^9\,\rm K)/T} \,{\rm cm^{-3}}, \quad T<6\times 10^9\,\rm K \\
    1.7 \times 10^{28} \left(\frac{T}{10^9\,\rm K}\right)^3 \ue^{-(6 \times 10^9\,\rm K)/T} \,{\rm cm^{-3}}, \quad T>6\times 10^9\,\rm K
    \end{cases}
    \end{split}.
\end{equation}
}
Here $k_{\rm b}T<m_e c^2$ and $k_{\rm b}T>m_e c^2$ correspond to the non-relativistic and relativistic limits, respectively. The pair number density is exponentially suppressed at low temperatures: when $T$ decreases from $10^9$ to $10^8\,\rm K$, $n_{e,\psi=0}$ drops by a factor of $\sim 10^{-28}$.

\begin{figure}
\centering
 \includegraphics[scale=0.35]{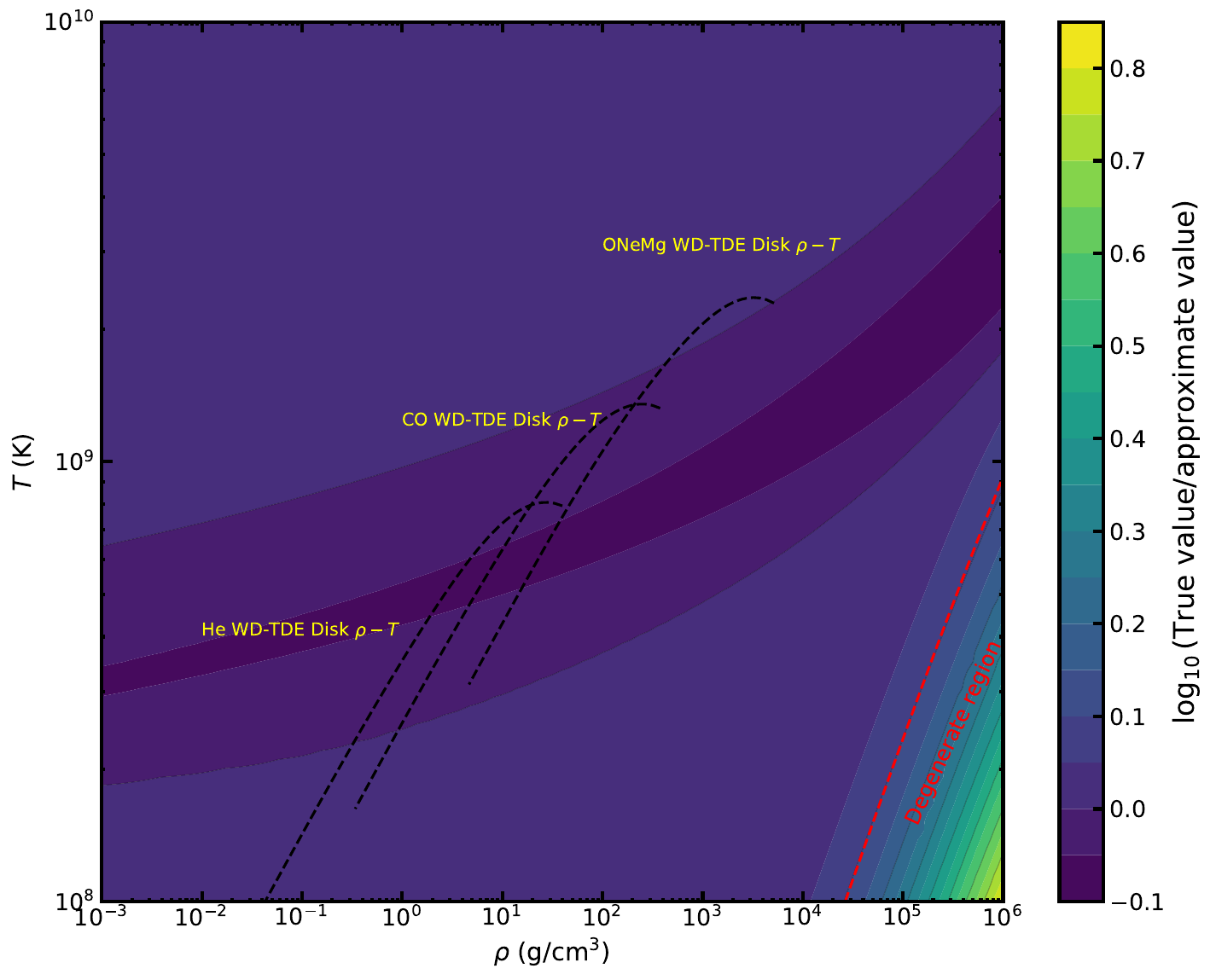}
    \caption{Ratio of the exact electron pressure (Equation~\ref{eq:true_pressure}) to the approximation in Equation~(\ref{eq:P-}). The approximation is accurate to better than $\sim 20\%$ over the relevant parameter range.}
\label{fig:Pressure_test}
\end{figure}

The electron/positron pressure is
\begin{equation} \label{eq:true_pressure}
    P_{e^{\pm}} = \frac{4\pi g_e}{3h^3} \int^{\infty}_0 \frac{vp^3\, dp}{\ue^{(E-\psi_{e^{\pm}})/k_{\rm b}T}+1},
\end{equation}
where $v$ is the particle velocity. In the non-degenerate regime relevant here, the ideal-gas relation applies,
\begin{equation}
    P_{e^{\pm}} = n_{e^{\pm}} k_{\rm b}T.
\end{equation}

Solving Equations~(\ref{eq:n_e}), (\ref{eq:ne_ne}), and (\ref{eq:true_pressure}) yields accurate number densities and pressures, but requires numerical integration. Instead, we use a simple decomposition in which the electron pressure includes both ionization electrons and pairs:
\begin{equation} \label{eq:P-}
    P_{e^-} \simeq (n_{e0}+n_{e,\psi=0})k_{\rm b} T.
\end{equation}
For positrons, we take
\begin{equation} \label{eq:P+}
    P_{e^+} \simeq n_{e,\psi=0}k_{\rm b} T.
\end{equation}
Figure~\ref{fig:Pressure_test} compares the exact electron pressure (Equation~\ref{eq:true_pressure}) with the approximation (Equation~\ref{eq:P-}); the agreement is better than $\sim 20\%$. The positron pressure is less accurate for $T\lesssim 5\times 10^8\,\rm K$ in the outer disk, which can overestimate $n_{e^+}$ and $P_{e^+}$. However, because $P_{e^+}$ is negligible compared to the total pressure, this has no practical impact.

\section{Neutrino production by pair annihilation}
\label{appendix:Neutrino_pair}
Here we compute neutrino production via $e^-e^+$ annihilation in WD TDE disks and provide a fitting formula that accurately reproduces the emissivity in both the non-relativistic and relativistic temperature regimes.

Accurate emissivities over a broad $\rho$--$T$ range were computed by \citet{itoh_neutrino_1989} using the Weinberg--Salam theory, including $s$-channel $Z^0$ exchange (all neutrino flavors) and $t$-channel $W^{\pm}$ exchange (electron-flavor neutrinos). The emissivity is highly temperature sensitive and is nearly density independent, except in the high-density, degenerate regime.

The pair-annihilation emissivity scales as
$\dot{q}_{\nu \overline{\nu}} \propto n_{e^-} n_{e^+} \langle\sigma_{\nu \overline{\nu}}v\rangle \langle E\rangle$,
where $\langle E \rangle$ is the characteristic pair energy and $\langle\sigma_{\nu \overline{\nu}}v\rangle$ is the thermally averaged cross section times velocity. In the relativistic limit, $\langle E \rangle \propto T$, $\langle\sigma_{\nu \overline{\nu}}v\rangle \propto T^2$, and $n_{e^{\pm}}\propto T^3$ (Equation~\ref{eq:n_epm}), giving $\dot{q}_{\nu \overline{\nu}} \propto T^9$. In the non-relativistic limit, $\langle\sigma_{\nu \overline{\nu}}v\rangle \propto T^{1/2}$ and $n_{e^{\pm}}\propto T^{3/2}$, yielding $\dot{q}_{\nu \overline{\nu}} \propto T^{4.5}$. Here $s$ denotes the squared center-of-mass energy of the pair.

To capture both limits with a single expression, we adopt the fitting form in Equation~(\ref{eq:q_ee}).

\begin{figure}
\centering
 \includegraphics[scale=0.5]{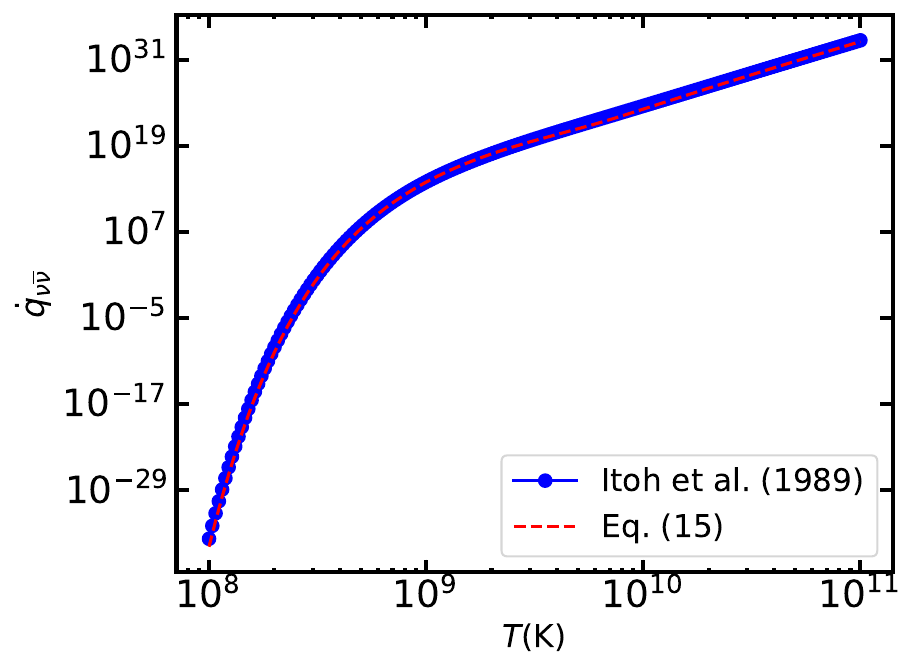}
    \caption{Comparison of the pair-annihilation emissivity $\dot{q}_{\nu \overline{\nu}}$ from \citet{itoh_neutrino_1989} (blue) with our fitting formula (Equation~\ref{eq:q_ee}; red dashed).}
\label{fig:q_vv_test}
\end{figure}

Our fitting formula agrees well with the numerical results of \citet{itoh_neutrino_1989} (see their Equation~18). Figure~\ref{fig:q_vv_test} compares their emissivity with our expression (Equation~\ref{eq:q_ee}) over the parameter range relevant here.

\section{Analytical calculation of an advection- and radiation-pressure-dominated disk}
\label{appendix:adv_disk}
For moderate super-Eddington accretion rates ($1\lesssim \dot{M}/\dot{M}_{\rm Edd}\lesssim 10^6$), the peak temperature typically remains below $10^8\,\rm K$. Nuclear burning and pair production are then unimportant, and cooling is dominated by advection with radiation pressure providing the main pressure support.

In this limit, an approximate analytic disk structure follows by setting the viscous heating rate equal to the advective cooling rate ($Q_{\rm vis}\simeq Q_{\rm adv}$) and taking $P\simeq aT^4/3$. The resulting temperature profile is
\begin{equation} \label{eq:T_adv}
    \begin{split}
    T &\simeq 4.4 \times10^6 \alpha^{-1/4} M_3^{1/8} \left(\frac{\dot{M}_{\rm in}}{10 \dot{M}_{\rm Edd}}\right)^{1/4} \\
    &\times \left(\frac{R}{10R_{\rm in}}\right)^{-5/8}f^{1/8}\ {\rm K}
    \end{split}
\end{equation}
with the corresponding surface density
\begin{equation} \label{eq:Sigma_adv}
    \Sigma \simeq 10^2 \alpha^{-1} M_3^{-1/2} \frac{\dot{M}_{\rm in}}{10 \dot{M}_{\rm Edd}} \left(\frac{R}{10R_{\rm in}}\right)^{-1/2}\ {\rm g\ cm^{-2}}.
\end{equation}

Including cooling by disk winds, we instead balance $Q_{\rm vis}\simeq Q_{\rm adv}+Q_{\rm w}$ and adopt a wind-driven inflow rate $\dot{M}_{\rm in} \simeq \dot{M}(R/R_{\rm out})^s$. The modified profiles become
\begin{equation} \label{eq:T_adv_w}
    \begin{split}
    T &\simeq 3.9 \times10^6 (1-\frac{sK}{3})^{-1/8} \alpha^{-1/4} M_3^{1/8} \left(\frac{\dot{M}}{10 \dot{M}_{\rm Edd}}\right)^{1/4} \\
    &\times \left(\frac{R}{10R_{\rm in}}\right)^{s/4-5/8} \left(\frac{100R_{\rm in}}{R_{\rm out}}\right)^{s/4} f^{1/8}\ {\rm K}
    \end{split}
\end{equation}
and
\begin{equation} \label{eq:Sigma_adv_w}
    \begin{split}
    \Sigma &\simeq 72 (1-\frac{sK}{3})^{-1} \alpha^{-1} M_3^{-1/2} \frac{\dot{M}}{10 \dot{M}_{\rm Edd}} \\
    &\times \left(\frac{R}{10R_{\rm in}}\right)^{s-1/2} \left(\frac{100R_{\rm in}}{R_{\rm out}}\right)^{s}\ {\rm g\ cm^{-2}}.
    \end{split}
\end{equation}

Substituting Equations~(\ref{eq:T_adv}--\ref{eq:Sigma_adv_w}) into Equation~(\ref{eq:dotE_rad}) and taking $\kappa_{\rm R}\simeq \kappa_{\rm es}\simeq 0.2\,\rm cm^2\,g^{-1}$ gives
\begin{equation} \label{eq:L_adv}
    L \simeq \frac{4\pi GM_{\rm h} m_p c \mu}{\sigma_{\rm T}} \ln{\frac{R_{\rm out}}{R_{\rm in}}} \simeq \mu \ln{\frac{R_{\rm out}}{R_{\rm in}}} L_{\rm Edd} \simeq 9L_{\rm Edd},
\end{equation}
consistent with standard slim-disk scalings \citep{watarai_radiative_1999}, for which the luminosity is only weakly dependent on $\dot{M}$ and $\alpha$. The normalization is higher by a factor $\mu\simeq 2$ compared to hydrogen-rich disks: for fully ionized WD debris, the Eddington luminosity is twice as high because the electron-scattering opacity is lower.

\section{Neutrino Flux Sensitivity Limit}
\label{sec:flux_limit_framework}

The differential flux sensitivity (quoted at a given confidence level, C.L.) is defined as the minimum incident neutrino flux in a narrow energy bin that would yield a statistically significant excess of events above background. For an MeV-scale detector, the expected number of signal events in the true-energy bin $\Delta E_i = [E_i - \Delta E/2,\, E_i + \Delta E/2]$ accumulated over an exposure time $T_{\rm expo}$ can be approximated as
\begin{equation}
N_{\mathrm{sig}, i} \approx T_{\rm expo}\, \bigl( \Phi_{\nu}|_{E_i}\, \Delta E \bigr)\, \sum_{c} \Bigl( N_{\mathrm{target}, c}\, \sigma_{c,i}\, \epsilon_{c,i} \Bigr),
\label{eq:simplified_signal}
\end{equation}
where $\Phi_{\nu}|_{E_i}$ is the differential neutrino flux in the $i$-th bin (in unit of $\mathrm{cm}^{-2}\,\mathrm{s}^{-1}\,\mathrm{MeV}^{-1}$), $N_{\mathrm{target},c}$ is the number of targets in the fiducial volume relevant for channel $c$ (free protons or electrons), $\sigma_{c,i}$ is the interaction cross section evaluated in bin $i$, and $\epsilon_{c,i}$ is the corresponding selection efficiency (trigger, reconstruction, and analysis cuts). In the MeV regime the dominant channels are inverse beta decay (IBD, $\bar{\nu}_e + p \to e^+ + n$) and elastic neutrino--electron scattering (ES, $\nu_x + e^- \to \nu_x + e^-$). We take JUNO and Hyper-K to be IBD-dominated, while for IceCube-Gen2 we include both IBD and ES. We adopt $\epsilon_c \simeq 0.73$, $0.6$, and $1$ for JUNO, Hyper-K, and IceCube-Gen2, respectively.

\subsection{Statistical Significance and Background Normalization}
The expected number of background events in bin $i$ over $T_{\rm expo}$ is
\begin{equation}
N_{\mathrm{bkg}, i} = T_{\rm expo}\, \mathcal{B}(E_{i})\, \Delta E,
\label{eq:background_integral}
\end{equation}
where $\mathcal{B}(E_i)$ is the differential background rate density. For a Poisson statistics, the minimum number of signal events required for a significance $Z$ (e.g., $Z=1.28$ for a one-sided $90\%$ C.L. upper limit) is
\begin{equation}
N_{\mathrm{sig}, i}^{\mathrm{min}} = Z\, \sqrt{N_{\mathrm{bkg}, i}}.
\label{eq:gaussian_limit}
\end{equation}
Substituting $N_{\mathrm{sig}, i}^{\mathrm{min}}$ into Eq.~(\ref{eq:simplified_signal}) yields the differential flux limit
\begin{equation}
\Phi_{\nu}|_{E_i}^{\mathrm{limit}} = \frac{N_{\mathrm{sig}, i}^{\mathrm{min}}}{T_{\rm expo}\, \Delta E\, \sum_{c} \bigl(N_{\mathrm{target}, c}\, \sigma_{c,i}\, \epsilon_{c,i}\bigr)}.
\label{eq:final_differential_limit}
\end{equation}

\subsection{Detector-Specific Parameterizations}
We evaluate Eq.~(\ref{eq:final_differential_limit}) for three representative next-generation detectors.

\subsubsection{Jiangmen Underground Neutrino Observatory (JUNO)}
JUNO is a $20\,\text{kt}$ liquid-scintillator detector optimized for precision measurements of reactor and geo-neutrinos \citep{an_neutrino_2016}. We consider IBD on free protons with $N_{\mathrm{target}, p} \approx 1.44 \times 10^{33}$, using the delayed-coincidence signature. The energy resolution ($\sim 3\%/\sqrt{E\,\text{(MeV)}}$) enables detailed spectral reconstruction. We model the low-energy background as reactor $\bar{\nu}_e$ plus cosmogenic $^{9}\text{Li}/^{8}\text{He}$ contributions,
\begin{equation}
\mathcal{B}_{\mathrm{JUNO}}(E) = 1.425\, \exp\left(-\frac{E}{3\ {\rm MeV}}\right) + 0.274.
\label{eq:juno_bkg}
\end{equation}

\subsubsection{Hyper-Kamiokande (Hyper-K)}
Hyper-K is well suited for MeV neutrino astronomy, combining a $187\,\text{kt}$ ultra-pure water volume (with $N_{\mathrm{target}, p} \approx 1.25 \times 10^{34}$ free protons) and high-fidelity 3D Cherenkov ring reconstruction, enabling sub-degree pointing for transient neutrinos \citep{proto-collaboration_hyper-kamiokande_2018}. For our MeV sensitivity estimate we treat IBD as the dominant channel. At low energies ($\lesssim 10\,\text{MeV}$) the sensitivity is limited primarily by radioactivity in detector materials (e.g., $^{214}\text{Bi}$), while at higher energies ($\gtrsim 15\,\text{MeV}$) it is limited by atmospheric ``invisible muon'' decay products. We parameterize the background as
\begin{equation}
\mathcal{B}_{\mathrm{HK}}(E) = 80\, \exp\left(-\frac{E}{3.5\ {\rm MeV}}\right) + 1.2.
\label{eq:hk_bkg}
\end{equation}

\subsubsection{IceCube-Gen2 (MeV continuous, full-asymmetry mode)}
For a transient, IceCube-Gen2 functions as a global rate detector: individual interactions are not reconstructed, but a collective excess in the single-photon hit rates is sought across $\sim 10^4$ multi-PMT modules (mDOMs) on top of a raw noise rate $R_{\mathrm{raw}} \approx 6.0 \times 10^6\,\text{Hz}$ \citep{collaboration_icecube-gen2_2021}. For an exposure $T_{\rm expo} \simeq 10\,\text{s}$ we assume a two-stage reduction: (i) a hardware local-coincidence (LC) requirement that reduces the noise to $N_{\mathrm{bkg,LC}} \simeq 4.4\times 10^4$ counts, and (ii) a directional asymmetry estimator with an effective asymmetry factor $\alpha \simeq 0.65$. The corresponding statistical fluctuation is $\sigma_D = \sqrt{N_{\mathrm{bkg,LC}}} \approx 210$ counts.

Requiring a one-sided $90\%$ C.L. excess ($Z=1.28$) gives $\alpha\, N_{\mathrm{sig}}^{\mathrm{min}} = Z\, \sigma_D$, i.e.
\begin{equation}
N_{\mathrm{sig}}^{\mathrm{min}} \simeq 4.13\times 10^{2}.
\label{eq:nsig_target_10s}
\end{equation}

Including both IBD and ES, we write
\begin{equation}
\sum_{c} \left[ N_{\mathrm{target}, c}\, \sigma_{c} \right] \equiv \left( N_{\mathrm{proton}}\, \sigma_{\mathrm{IBD}} \right) + \left( N_{\mathrm{electron}}\, \sigma_{\mathrm{ES}} \right).
\label{eq:gen2_combined}
\end{equation}
where $\sigma_{\mathrm{IBD}} \simeq 10^{-43} (E-1.293\ {\rm MeV})^2\ {\rm cm^{-2}}$ and $\sigma_{\mathrm{ES}} \simeq 9.2 \times 10^{-45} (E/{\rm MeV})\ {\rm cm^{-2}}$, $N_{\mathrm{proton}} \simeq 2.6 \times 10^{33}$ and $N_{\mathrm{electron}} = 5N_{\mathrm{proton}}$.

\begin{acknowledgments}
This work is supported by the National Natural Science Foundation of China (the NSFC Type C Youth Project 12503053),  the National Key R\&D Program of China (2025YFF0511100), the Hong Kong Research Grants Council (GRF 17314822). We also acknowledge the useful discussion with Weiming Yuan. We thank the participants of the TDE FORUM (Full-process Orbital to Radiative Unified Modeling) online seminar series for their inspiring discussions. 
\end{acknowledgments}

\bibliography{cited}{}
\bibliographystyle{aasjournalv7}

\end{CJK*}
\end{document}